\documentclass[aps,twocolumn,superscriptaddress,10pt,nofootinbib]{revtex4-2} 

\usepackage{graphicx}  	
\usepackage{dcolumn}   	
\usepackage{bm}        	
\usepackage{amssymb}   	
\usepackage{amsmath} 
\usepackage{epstopdf}
\usepackage[a4paper,margin=20mm]{geometry}
\usepackage{color}  
\usepackage{hyperref}
\usepackage{tabularx}

\newcommand{\bs}[1]{\boldsymbol{#1}}

\hypersetup{
    colorlinks=true, 
    citecolor=blue, 
    linktoc=all,     
    linkcolor=blue,  
}

\def \l {\left}
\def \r {\right}

\begin{document}

\title{Local polar order controls mechanical stress and triggers layer formation in developing {\it Myxococcus xanthus} colonies}

\author{Endao Han}
\email[E-mail:]{endao.han@ntu.edu.sg}
\altaffiliation{Current address: School of Physical and Mathematical Sciences, Nanyang Technological University, 637371, Singapore}
\affiliation{Joseph Henry Laboratories of Physics, Princeton University, Princeton, NJ 08544, USA.}

\author{Chenyi Fei}
\affiliation{Lewis-Sigler Institute for Integrative Genomics, Princeton University, Princeton, NJ 08544, USA.}
\affiliation{Department of Molecular Biology, Princeton University, Princeton, NJ 08544, USA.}

\author{Ricard Alert}
\affiliation{Max Planck Institute for the Physics of Complex Systems, N{\"o}thnitzerstra{\ss}e 38, 01187 Dresden, Germany. }
\affiliation{Center for Systems Biology Dresden, Pfotenhauerstra{\ss}e 108, 01307 Dresden, Germany. }
\affiliation{Cluster of Excellence Physics of Life, TU Dresden, 01062, Dresden, Germany. }

\author{Katherine Copenhagen}
\affiliation{Lewis-Sigler Institute for Integrative Genomics, Princeton University, Princeton, NJ 08544, USA.}

\author{Matthias D. Koch}
\affiliation{Lewis-Sigler Institute for Integrative Genomics, Princeton University, Princeton, NJ 08544, USA.}
\affiliation{Department of Molecular Biology, Princeton University, Princeton, NJ 08544, USA.}

\author{Ned S. Wingreen} 
\affiliation{Lewis-Sigler Institute for Integrative Genomics, Princeton University, Princeton, NJ 08544, USA.}
\affiliation{Department of Molecular Biology, Princeton University, Princeton, NJ 08544, USA.}

\author{Joshua W. Shaevitz} 
\email[E-mail:]{shaevitz@princeton.edu}
\affiliation{Joseph Henry Laboratories of Physics, Princeton University, Princeton, NJ 08544, USA.}
\affiliation{Lewis-Sigler Institute for Integrative Genomics, Princeton University, Princeton, NJ 08544, USA.}

\date{\today}

\begin{abstract}
Colonies of the social bacterium \textit{Myxococcus xanthus} go through a morphological transition from a thin colony of cells to three-dimensional droplet-like fruiting bodies as a strategy to survive starvation. 
The biological pathways that control the decision to form a fruiting body have been studied extensively. 
However, the mechanical events that trigger the creation of multiple cell layers and give rise to droplet formation remain poorly understood. 
By measuring cell orientation, velocity, polarity, and force with cell-scale resolution, we reveal a stochastic local polar order in addition to the more obvious nematic order. 
Average cell velocity and active force at topological defects agree with predictions from active nematic theory, but their fluctuations are anomalously large due to polar active forces generated by the self-propelled rod-shaped cells. 
We find that \textit{M. xanthus} cells adjust their reversal frequency to tune the magnitude of this local polar order, which in turn controls the mechanical stresses and triggers layer formation in the colonies. 
\end{abstract}

\maketitle

Biological cells often form densely-packed, two-dimensional monolayers that serve specific biological functions. 
Densely-packed cells with elongated shapes, from collectives of eukaryotes \cite{Gruler_1995,Saw_2017,Balasubramaniam_NM_2021,Comelles_elife_2021,Giomi_arxiv_2022,Silberzan_2017,Silberzan_2018,Kawaguchi_2017} to populations of bacteria \cite{Volfson_PNAS_2008,Poon_2018,Yaman_2019,Genkin_2017,Zhang_PNAS_2018,Meacock_2020,Copenhagen_NP2020}, typically align with each other and may behave as active nematic liquid crystals.
The constituents of such active nematics generate internal active stresses along the axis of alignment, which give rise to phenomena not found in their passive counterparts \cite{Dogic_2012,Yeomans_2018,Dogic_2020,Zhang_NRM}. 
A hallmark of these systems is the spontaneous creation of topological defects -- singularities in the orientation field that play an important role in apoptotic cell extrusion \cite{Saw_2017,Guillamat_NM_2022,Giomi_arxiv_2022}, the accumulation of neural progenitor cells \cite{Kawaguchi_2017}, tissue morphogenesis \cite{Keren_2021}, and pattern formation in bacterial colonies \cite{Poon_2018,Yaman_2019,Genkin_2017,Zhang_PNAS_2018,Meacock_2020,Copenhagen_NP2020}. 

However, recent work has challenged the completeness of this picture for many of these biological systems.
For example, epithelial cell layers can develop polar order, which drives flocking, morphogenesis at defects, and spreading \cite{Trepat_2019,Giomi_arxiv_2022,Guillamat_NM_2022}. 
Motile bacteria are also driven by polar forces produced by flagella, pili, or gliding motors. 
What role does this single-cell polarity have in the collective dynamics? 
This question has been addressed using the Self-Propelled Rod (SPR) model \cite{Ramaswamy_2002,Marchetti_RMP,Bar_SPR_review}, in which interactions between rods are apolar or only weakly polar. 
Some SPR systems show long-range nematic order only \cite{Ginelli_PRL_2010,Marchetti_2008} while others show a mixture of nematic and polar order \cite{Huber_science_2018,Denk_PNAS_2020}. 
What biological phenomena emerge from the coexistence of polar and nematic order? 

\begin{figure*}[!t]
    \vspace{-10pt}
    \begin{center}
        \includegraphics[width=\textwidth]{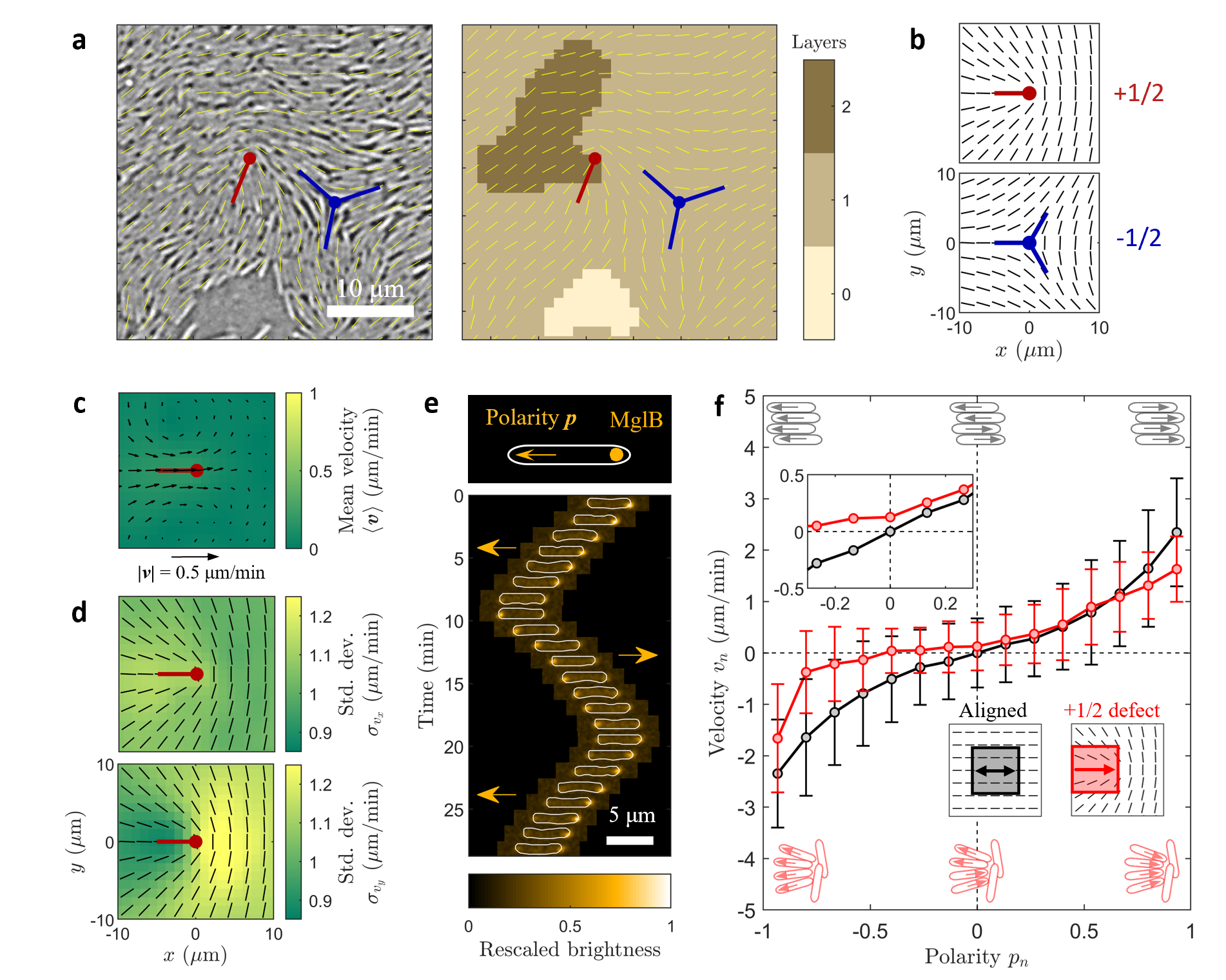}
    \end{center}
    \vspace{-10pt}
    \caption{ \label{fig:polarity} Nematic and polar order in thin \textit{M. xanthus} layers. 
    (a) An exemplary bright field image of cells on a solid surface and the corresponding number of cell layers (see SI Fig.~\ref{Efig:depth}). Yellow line segments indicate the director field (cell orientations) $\bm{\hat{n}}$, and there is a pair of +1/2 (red) and -1/2 (blue) defects in the view. The white scale bar is $10~\mu$m.  
    (b) Average director field $\l< \bm{\hat{n}} \r>$ around $\pm 1/2$ defects. 
    (c) Mean velocity of cell flow $\l< \bm{v} \r>$ around $+1/2$ defects; the black arrows show magnitude and direction and the color map shows the speed $\l| \l< \bm{v} \r> \r|$. 
    (d) Standard deviation $\sigma_{v_x}$ and $\sigma_{v_y}$ of the $x$ and $y$ components of the velocity field around $+1/2$ defects. The black line segments show $\l< \bm{\hat{n}} \r>$. 
    (e) MglB proteins localize to the rear of the cell and defines individual cell polarity (arrows). The kymograph shows the brightness of the fluorescent tag for one isolated moving cell (outlined by the white contours). The kymograph captures two reversal events. The white scale bar is $5~\mu$m. 
    (f) Local cell velocity $v_n$ is correlated with local polarity $p_n$. Black circles indicate measurements in nematically aligned regions (in the black square) while red circles are for measurements in the comet tail region of $+1/2$ defects (in the red square). The inset is a zoom-in of the region near $p_n = 0$. The positive direction for polarity and velocity are labeled by the thick arrows in the director field: it points toward the defect core near $+1/2$ defects and is left-right symmetric in aligned regions. } 
\end{figure*}

Here, we show that polarity fluctuations trigger the formation of new cell layers, which enables the starvation-induced development from monolayers to droplet-like fruiting bodies in the social bacterium \textit{Myxococcus xanthus} \cite{Kaiser_Review}.
Layers of \textit{M. xanthus} cells contain both comet-like defects of topological charge $+1/2$ and triangular defects of  charge $-1/2$ (Fig.~\ref{fig:polarity}a,b), indicating that the system is nematic. 
Previous work showed that the average cell flow around these defects is well-explained by an active nematic model \cite{Copenhagen_NP2020} (see Fig.~\ref{Efig:defect_mean_velocity}). 
Fig.~\ref{fig:polarity}c shows the mean cell velocity $\l< \bm{v} \r>$ near a $+1/2$ defect, which originates from the balance between the active nematic force density, 
\begin{equation}
    \bm{f}^\text{a}_\text{n} = - \zeta_\text{n} \nabla \cdot \bm{Q},
\label{eq:main:f_a}
\end{equation}
and an anisotropic cell-substrate friction, where $\bm{Q} = S(2\bm{\hat{n}}\bm{\hat{n}} - \bm{I})$ is the nematic order parameter tensor, $\bm{\hat{n}}$ is the director of the nematic order, $S$ is the scalar order parameter, and $\zeta_\text{n}$ is the activity coefficient \cite{Ramaswamy_2002,Yeomans_2018,Copenhagen_NP2020}. 
The velocity-based flux through a circular boundary $\mathcal{C}$ around a  $+1/2$ defect, $\Phi_{\bm v} = \oint_\mathcal{C} \l( \bm{v} \cdot \bm{\hat{r}} \r) ds$, is negative, indicating a net influx of cells toward the defect core.
Consequently, $+1/2$ defects promote cell accumulation and layer formation (Fig.~\ref{fig:polarity}a). 
Similarly, $-1/2$ defects produce a net outflux of cells and favor the formation of holes within the monolayer \cite{Copenhagen_NP2020}.
Thus, the nematic order is connected to colony morphology.

While the active nematic theory predicts a steady flow around $+1/2$ defects, the nucleation of new layers is rare, sudden, and stochastic. 
To study these stochastic events, we measured the velocity fluctuations. 
Strikingly, the standard deviations of $v_x$ and $v_y$, $\sigma_{v_x}$ and $\sigma_{v_y}$, are both several times larger than the mean speed $\l| \l< \bm{v} \r> \r|$ (Fig.~\ref{fig:polarity}d). 
We hypothesized that these velocity fluctuations are driven by fluctuations in local polar order around $+1/2$ defects. 
To test this hypothesis, we experimentally measured cell polarity and traction forces with cellular-scale resolution simultaneously with the cell velocity, nematic order, and thickness fields near defects.

\begin{figure*}
\vspace{-5pt}
  \begin{center}
    \includegraphics[width=1\textwidth]{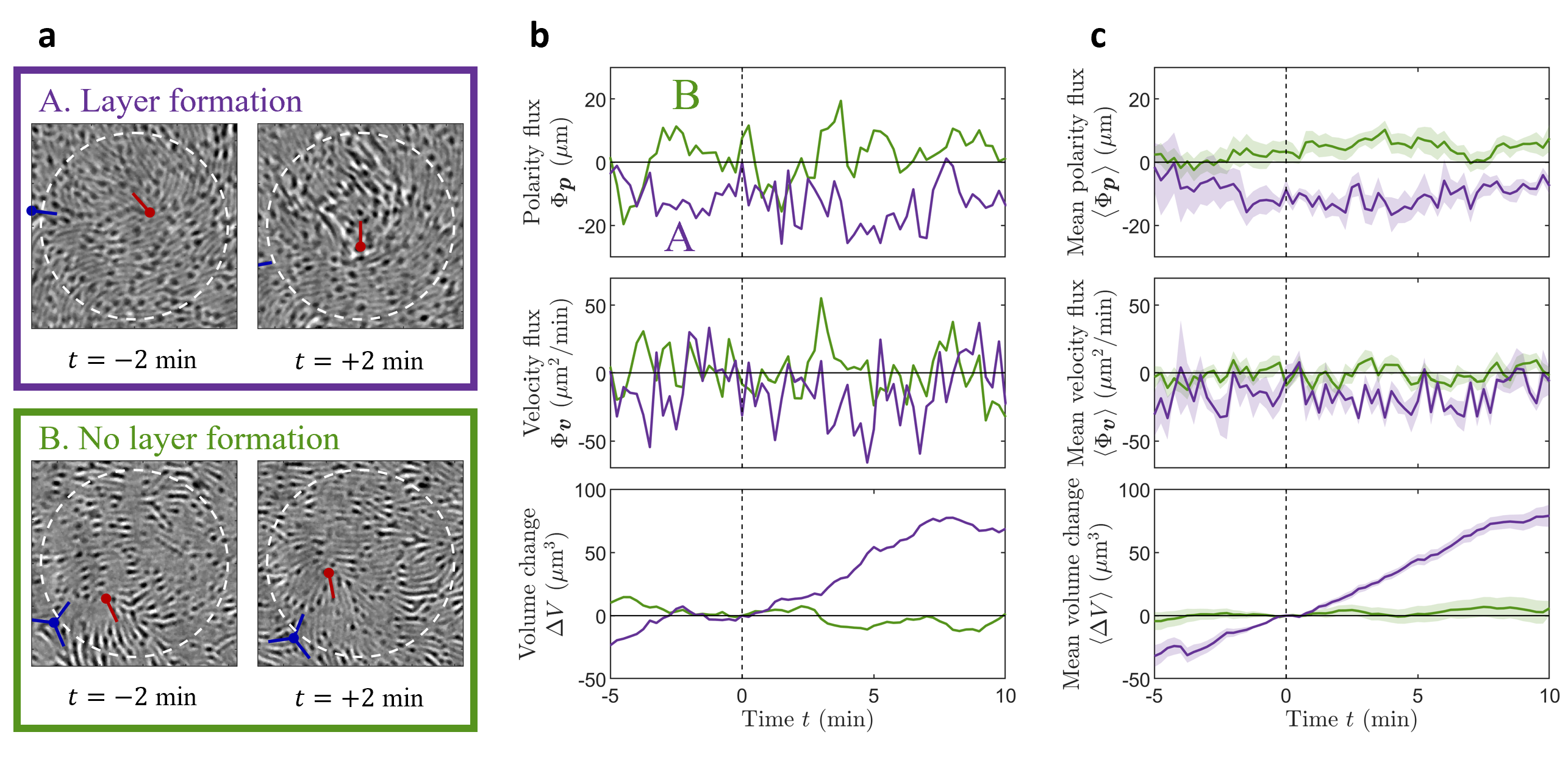}
  \end{center}
  \vspace{-10pt}
\caption{ \label{fig:layer} Instantaneous cell flux drives layer formation. 
(a) Two exemplary regions with (A, purple) and without (B, green) second layer formation. For region A, we set the time at which the second layer appeared as $t = 0~$min, while for B, the time point $t = 0~$min was chosen arbitrarily. The white circles with a radius of $l_p = 12~\mathrm{\mu}$m label the boundaries of the selected regions, and they each surround a $+1/2$ defect. 
(b) Polarity-based flux $\Phi_{\bm p}$, velocity-based flux $\Phi_{\bm v}$, and volume change $\Delta V$ in regions A (purple) and B (green). 
(c) Mean (curves) polarity-based flux $\l< \Phi_{\bm p} \r>$, velocity-based flux $\l< \Phi_{\bm v} \r>$, and volume change $\l< \Delta V \r>$ over multiple regions, each surrounding a $+1/2$ defect, with (purple) and without (green) second layer formation, and the corresponding standard errors (shaded areas). The definition of $t = 0~$min remains the same. }
\end{figure*}

\vspace{2mm}
\noindent {\bf Instantaneous cell polarity is the main driver of cell flow}\\
We measured cell polarity, $\bm{p}$, using the \textit{mglB::mVenus} strain of \textit{M. xanthus}. This strain expresses a fluorescent fusion to the MglB protein, which is localized to the rear pole of the cell (Fig.~\ref{fig:polarity}e) \cite{Lotte_2022}. 
By simultaneously imaging the cells and the localization of MglB, we defined the polarity of each cell within the population (see SI Sec.~\ref{sec:experiment}). 
We calculated the average local polarity $p_n$ and velocity $v_n$ in square regions with a side $l_p = 12~\mu$m, the length of two cells, where the subscript $n$ denotes the component projected along the director. 
To obtain $p_n$ and $v_n$, we reoriented the regions either with aligned cells or near a $+1/2$ defect as shown in Fig.~\ref{fig:polarity}f, and averaged the horizontal components of $\textbf{\textit p}$ and $\textbf{\textit v}$, respectively inside the boxes (see SI Fig.~\ref{Efig:polarity_velocity}). 
For the $+1/2$ defect, ``right'' was defined as the positive direction, while due to the left-right symmetry of the aligned region, $(p_n,v_n)$ and $(-p_n,-v_n)$ are equivalent. 
In the areas with aligned cells ($\nabla \cdot \bm{Q} \to 0$), $\bm{f}^\text{a}_\text{n}$ does not drive any consistent cell flow, and the cell flow is driven by instantaneous polarity. 
In these regions, $v_n$ is a monotonic function of $p_n$ (Fig.~\ref{fig:polarity}f, black), and the slope of $v_n$ versus $p_n$ increases as $\l| p_n \r| \to 1$, indicating that the increase of local polar order leads to higher cell speed. 
In contrast, near a $+1/2$ defect, both the nematic and polar order drive cell motion. 
Yet, instantaneous local polar order is still the major driver of the velocity field as seen in the monotonic $v_n$-$p_n$ relationship (Fig.~\ref{fig:polarity}f, red). 
However, compared to the aligned areas, this specific arrangement of cells around defects limits the local cell speed, leading to lower $v_n$ as $\l| p_n \r| \to 1$. 
Furthermore, at zero polarity $p_n=0$, the cell velocity is non-zero (Fig.~\ref{fig:polarity}f inset, red); this is the flow driven by the net active nematic force $\bm{f}^\text{a}_\text{n}$ (Fig.~\ref{fig:polarity}c). 
Our results now show that this average velocity due to nematic forces is small compared to the velocities produced by polarity fluctuations (Fig.~\ref{fig:polarity}f).

\vspace{2mm}
\noindent {\bf Polarity-driven cell influx drives layer formation}\\
The cell number was approximately constant as the cells did not grow or divide under our experimental conditions (see Methods). 
The formation of a second layer on top of a monolayer thus requires a local influx of cells via motility.
The change in volume $\Delta V$ within a region is given by the velocity-based flux, $\Delta V = - \int \Phi_{\bm v} h dt$, where $h$ is the thickness of the cell colony (see SI Fig.~\ref{Efig:volume}). 
As polarity fluctuations are one of the main drivers of cell flows around defects, they have a major contribution to the cell flux, and hence to layer formation.

Fig.~\ref{fig:layer}a shows two circular regions with radius $l_p$, each surrounding a $+1/2$ defect. 
A visible second layer appeared at $t = 0~$min in region $A$, while region $B$ remained as a monolayer. 
Fig.~\ref{fig:layer}b shows the polarity-based flux across the boundaries of these two regions $\Phi_{\bm p} = \oint_\mathcal{C} \l( \bm{p} \cdot \bm{\hat{r}} \r) ds$. 
For $A$, $\Phi_{\bm p}$ was consistently negative starting from several minutes before the out-of-plane cell motion, while $\Phi_{\bm p}$ for region $B$ fluctuated around $0$. 
The resulting velocity-based flux $\Phi_{\bm v}$ showed the same trends, with a net influx $(\Phi_{\bm v} < 0)$ for the region $A$ but not for $B$. 
We identified multiple regions around $+1/2$ defects with and without layer formation, and show the mean polarity-based flux $\l< \Phi_{\bm p} \r>$, velocity-based flux $\l< \Phi_{\bm v} \r>$, and volume change $\l< \Delta V \r>$ in Fig.~\ref{fig:layer}c. 
Their trends are consistent with the exemplary individual events in Fig.~\ref{fig:layer}b, where $\Phi_{\bm p}$ and $\Phi_{\bm v}$ are directly related to layer formation.

\begin{figure*}
\vspace{-5pt}
    \begin{center}
        \includegraphics[width=1\textwidth]{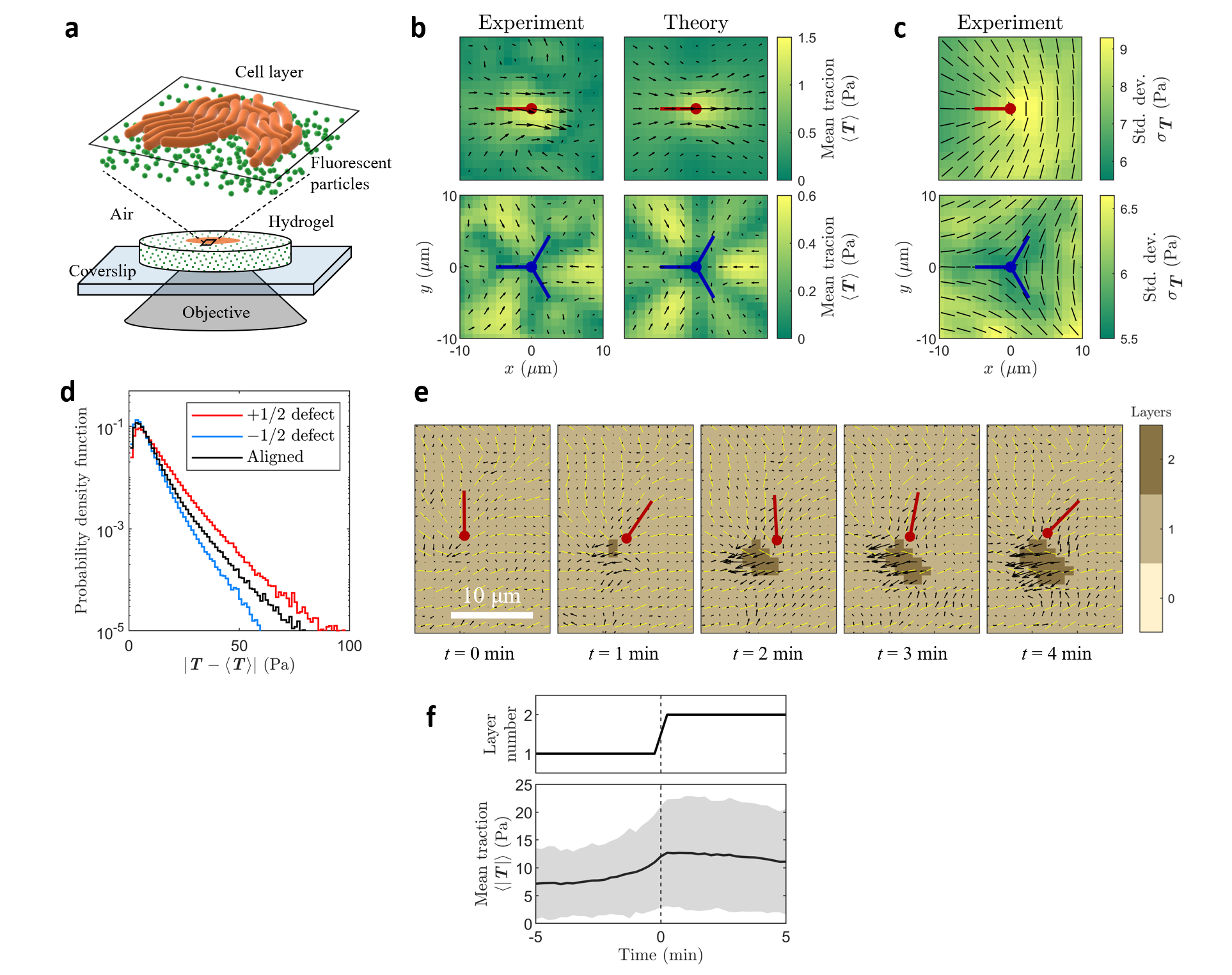}
    \end{center}
    \vspace{-10pt}
    \caption{ \label{fig:TFM} Traction force microscopy (TFM) reveals that large force fluctuations coincide with layer formation. 
    (a) Illustration of the TFM setup: \textit{M. xanthus} cells (orange) on a hydrogel surface with embedded fluorescent particles (green) imaged from below. 
    (b) Experimentally measured and theoretically calculated mean traction field $\l< \bm{T} \r>$ near $\pm 1/2$ defects. The color maps show magnitudes, and the arrows indicate magnitude and direction. 
    (c) Experimentally measured standard deviation of traction $\sigma_{\bm T}$ near $\pm 1/2$ defects. The black lines indicate the mean director field. 
    (d) Distributions of traction fluctuations $\l| \bm{T}-\l< \bm{T} \r> \r|$ within $5~\mu$m from the centers of $+1/2$ (red) and $-1/2$ (blue) defects, and in regions where the cells were aligned (black). 
    (e) Traction variation when a second layer formed near a +1/2 defect. The black arrows indicate traction and the yellow lines directors. The white scale bar is $10~\mu$m. 
    (f) Local traction variation when a second cell layer formed around a $+1/2$ defect. At each location, we shifted the time series of traction magnitude $|\bm{T}(t)|$ such that the formation of the second layer occurred at time $t = 0$~min. Then we calculated its mean (black curve) and standard deviation (gray shade). } 
\end{figure*}

Besides the average instantaneous flux $\l< \Phi_{\bm v} \r>$, we calculated the flux based on the mean velocity $\l< \bm{v} \r>$ near $+1/2$ defects (Fig.~\ref{fig:polarity}c), and obtained $\Phi_{\l< \bm{v} \r>} = -5.1~\mu\text{m}^2/\text{min}$. 
In $\Phi_{\l< \bm{v} \r>}$, $\l< \bm{v} \r>$ was averaged over all the $+1/2$ defects, with or without second layer formation, while in reporting $\l< \Phi_{\bm v} \r>$, we selected two separate subsets: those with second layer formation and those without. 
Note that $\Phi_{\l< \bm{v} \r>}$ is significantly weaker than the value of $\l< \Phi_{\bm v} \r>$ leading to layer formation (Fig.~\ref{fig:layer}c, purple). 
This shows that most of the influx at $+1/2$ defects is due to the strong velocity fluctuation induced by the local polar order, which explains why out-of-plane cell motion is not deterministic in our system. 
Instead of accumulating cells at a steady rate, when a second layer forms, it forms fast. 
Moreover, $\Phi_{\bm v}$ and $\Phi_{\bm p}$ do not always have a strong positive correlation, because the velocity of a cell in the colony is not directly determined by its polarity but also depends on its mechanical interactions with neighboring cells. 
This involves the active propelling forces generated by each individual cell, local cell-substrate friction, and cell-cell contact interactions, which introduce further stochasticity to the system.

\vspace{2mm}
\noindent{\bf Cellular tractions and layer formation}\\
To probe cellular forces in \textit{M. xanthus} colonies, we used traction force microscopy (TFM) \cite{Sabass_Chapter, Sabass_review, Sabass_PNAS2017} to measure the shear stress exerted on the solid substrate (the $x$-$y$ plane) by the cells (Fig.~\ref{fig:TFM}a). 
By tracking the embedded fluorescent particles, we measured the displacements of the substrate tangential to its surface, and in turn, reconstructed the traction $\bm{T}$ (see SI Sec.~\ref{sec:TFM_method}). 
According to Eq.~\ref{eq:main:f_a}, the active nematic force $\bm{f}^\text{a}_\text{n}$ is due to bending and splay in the nematic order, given by $\l| \nabla \cdot \bm{Q} \r| > 0$. 
Consequently, $\bm{f}^\text{a}_\text{n}$ should be large near the defects, where $\l< \bf{\hat{n}} \r>$ exhibits strong distortions (Fig.~\ref{fig:polarity}b). 
This is confirmed by our measurements of the average traction field $\l< \bm{T} \r>$ around $\pm 1/2$ defects (Fig.~\ref{fig:TFM}b). 
We then set out to explain these measurements theoretically. 
Since the cells glide on both the substrate and on neighboring cells, the total active force density $\bm{f}^\text{a}$ has a cell-substrate component $\bm{f}^\text{a}_\text{s}$ and a cell-cell component $\bm{f}^\text{a}_\text{c}$. 
The mean traction reflects the cell-cell interactions: $\l< \bm{T} \r> = \l< \bm{f}^\text{a}_\text{c} \r> - \nabla \l< P \r>$, where $P$ is the pressure within the cell layer (see SI Sec.~\ref{sec:theory}). 
Our measurements agree with the theoretical predictions (Fig.~\ref{fig:TFM}b), assuming the cell layer is an active nematic system and the average cell-cell active force $\l< \bm{f}^\text{a}_\text{c} \r>$ is given by $\bm{f}^\text{a}_\text{n}$ in Eq.~\ref{eq:main:f_a} with $\zeta_\text{n} > 0$, which corresponds to extensile active stresses.

Our experiments showed that, similar to $\bm{v}$, the instantaneous traction $\bm{T}(t)$ had a standard deviation $\sigma_{\bm T}$ about an order of magnitude larger than the mean $\l| \l< \bm{T} \r> \r|$ (Fig.~\ref{fig:TFM}c). 
Unlike $\l< \bm{T} \r>$, the fluctuations were not controlled by $\bm{Q}$. 
In regions where the cells were aligned ($\nabla \cdot \bm{Q} \to 0$), there were still consistent, strong force fluctuations (Fig.~\ref{fig:TFM}d). 
Moreover, traction fluctuations were stronger at $+1/2$ defects and weaker at $-1/2$ defects (Fig.~\ref{fig:TFM}c,d), although both types of defects had large $|\nabla \cdot \bm{Q}|$ near their cores (see SI Fig.~\ref{Efig:divQ}). 
Polarity can contribute to these stress fluctuations via $\bm{f}^\text{a}_\text{s} = \zeta_\text{p} \bm{p}$, where $\zeta_\text{p}$ represents the polar active force a cell generates while gliding on the substrate.

The fluctuations in polar active forces (in $\zeta_\text{p}$ or $\bm{p}$) will produce fluctuations in the pressure field $P$ in the cell colony. 
The local build-up of pressure can then trigger the formation of new cell layers. 
Fig.~\ref{fig:TFM}e shows an example of a second layer emerging near a $+1/2$ defect, simultaneously with a strong increase in $\bm{T}$. 
We identified multiple such monolayer to double-layer transitions and found that $\l| \bm{T}(t) \r|$ increases at the reference time $t = 0$, when the second layer forms (Fig.~\ref{fig:TFM}f). 
The increase in $\l< \l| \bm{T}(t) \r| \r>$ upon second layer formation was about 5~Pa, which significantly exceeds $\l| \l< \bm{T} \r> \r| \sim 1$ Pa around $+1/2$ defects (Fig.~\ref{fig:TFM}b). 
The traction did not relax immediately after second layer formation. 
Similar to $\Phi_{\bm p}$, $\l| \bm{T}(t) \r|$ evolved slowly, on the time scale of minutes.

\begin{figure*}
\vspace{-10pt}
  \begin{center}
    \includegraphics[width=\textwidth]{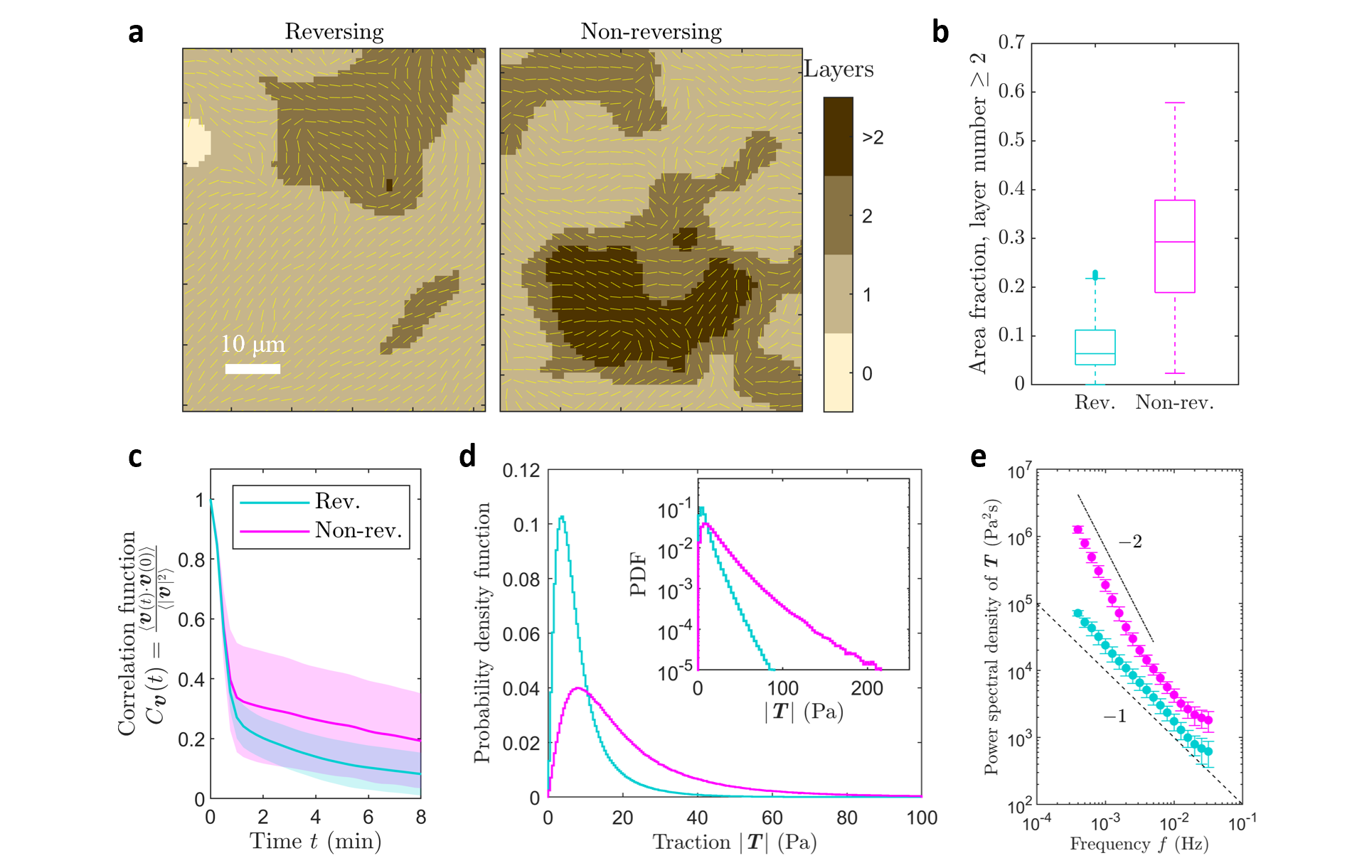}
  \end{center}
  \vspace{-20pt}
  \caption{ \label{fig:mutants} Turning off cell reversal enhances traction fluctuation and layer formation. 
(a) Representative examples of layer-number fields for colonies of reversing (left) and non-reversing (right) cells. Yellow lines indicate the director field $\bm{ \hat{n} }$. White scale bar is 10~$\mu$m. 
(b) Distributions of area fraction of regions with layer number $\ge 2$ for reversing and non-reversing cells.  
(c) Temporal auto-correlation functions of velocity $C_{\bm{v}}(t)$ for reversing (turquoise) and non-reversing (magenta) cells. The same colors are used in the following panels comparing these two strains. 
(d) Distributions of traction magnitude $|\bm{T}|$. Inset shows the same data in log-linear scale. 
(e) Power spectral density of traction. The dashed line has slope $-1$ and the dot-dashed line has slope $-2$. 
} 
\end{figure*}

\vspace{2mm}
\noindent{\bf Cell reversals control local polar order and layer formation}\\
Our results so far show that cell polarity produces strong fluctuations in traction and cell flux, which triggers layer formation. 
How do cells control local polar order in the system? 
During locomotion, \textit{M. xanthus} cells can reverse their direction of motion on the minute time scale \cite{frz_book}. 
Cells control this reversal frequency in response to starvation to induce layer formation. 
In a nutrient-rich environment, the average revresal time is $\tau \approx 10$~min \cite{Zusman_reversal,Thutupalli_2015} and the cells spread into a thin layer on a solid surface (Fig.~\ref{fig:polarity}a). 
As nutrients become scarce, the cells increase $\tau$, which leads to layer formation and ultimately to three-dimensional droplets called fruiting bodies \cite{Kaiser_Review, Perez_Review, Liu_PRL2019}. 
To understand the effect of cellular reversal on surface traction, we probed the non-reversing mutant $\Delta$\textit{frzE} while keeping the nutrient-rich environment invariant. 
Fig.~\ref{fig:mutants}a shows exemplary maps of layer thickness for reversing and non-reversing cells. 
Across multiple measurements, the non-reversing mutant generated more multi-layer regions than the reversing ones (Fig.~\ref{fig:mutants}b). 

Longer reversal time $\tau$ enhances the local polar order in the system. 
With the polarity assay (see SI Sec.~\ref{sec:experiment}), we measured the temporal autocorrelation functions of $\bm{p}$ and $\bm{v}$, $C_{\bm p}(t)$ and $C_{\bm v}(t)$, respectively, and showed that the correlation times of polarity $\tau_{\bm p}$ and velocity $\tau_{\bm v}$ were approximately equivalent. 
More importantly, they both increased along with the local polarity $p_n$ (see SI Fig.~\ref{Efig:polarity_correlation}). 
Similarly, we measured $C_{\bm v}(t)$ for reversing and non-reversing strains (Fig.~\ref{fig:mutants}c), and found that the correlation time increased from $0.8$~min with reversals to $3.7$~min without, implying stronger local polar order in the non-reversing cell colonies. 
As a result, although the speeds of reversing and non-reversing cells were similar, the non-reversing populations generated more persistent flows, leading to enhanced aggregation.

As expected, the increased local polar order in the non-reversing cells leads to stronger stress fluctuations, as shown by the longer tail in the traction distribution (Fig.~\ref{fig:mutants}d). 
This increase in traction fluctuations happens predominantly at low frequencies (Fig.~\ref{fig:mutants}e). 
The power spectral densities (PSDs) show two different power laws: traction generated by reversing cells follows a $f^{-1.1}$ power law across almost two decades in frequency $f$, while the PSD for non-reversing cells approaches $f^{-2}$ at low frequencies (Fig.~\ref{fig:mutants}e). 
Understanding the origin of these power laws requires further theoretical investigations.

\vspace{2mm}
In summary, our experiments reveal stochastic local polar order in thin \textit{M. xanthus} colonies, which not only leads to strong fluctuations in cell velocity and mechanical stress but also triggers layer formation. 
Although active nematic theory explains the average flows generated around topological defects and the cell accumulation that promotes layer formation \cite{Copenhagen_NP2020}, it needs to be extended to capture the large stress fluctuations we measure or the stochasticity in layer formation.
Our results show that the polarity fluctuations generated by collectives of self-propelled rod-shaped cells produce stronger forces and flows than the active nematic stresses around topological defects.
We further show that \textit{M. xanthus} colonies have found a simple knob \textemdash the cell reversal time \textemdash that they can use to control internal mechanical stress and colony morphology via tuning the local polar order. 
This control mechanism enables the colony to spread on a surface when nutrients are abundant and then initiate aggregation when food is scarce.

\vspace{2mm}
\noindent \textbf{Acknowledgments}: We thank the Confocal Imaging Facility, a Nikon Center of Excellence, in the Department of Molecular Biology at Princeton University for instrument use and technical advice.
We thank Roy Welch and Lotte Sogaard-Andersen for providing \textit{M. xanthus} strains. 
We thank Matthew Black, Aaron Bourque, Benjamin Bratton, Zemer Gitai, Guannan Liu, Howard Stone, Nan Xue, Cassidy Yang, and Rui Zhang for their assistance in the research and many useful discussions. 
This work was supported by the NSF through awards PHY-1806501 and PHY-2210346, and the Center for the Physics of Biological Function (PHY-1734030). 
NSW acknowledges National Institutes of Health Grant R01 GM082938. 

\vspace{2mm}
\noindent \textbf{Author contributions}:
E.H. and J.W.S. conceived the research. E.H. performed the experiments and data analysis. C.F. and R.A. derived the theoretical model. K.C. and M.D.K. provided key data processing and experimental techniques, respectively. N.S.W. and J.W.S. supervised the project. All authors designed the research and interpreted the results. E.H., C.F., R.A., N.S.W., and J.W.S. wrote the paper with input from the other authors. 

\let\oldaddcontentsline\addcontentsline 
\renewcommand{\addcontentsline}[3]{} 
\bibliographystyle{unsrt}
\bibliography{TFM_monolayer}

\begin{thebibliography}{10}

\bibitem{Gruler_1995}
Hans Gruler, Manfred Schienbein, Kurt Franke, and Anne~de Boisfleury-chevance.
\newblock Migrating {Cells}: {Living} {Liquid} {Crystals}.
\newblock {\em Molecular Crystals and Liquid Crystals Science and Technology.
  Section A. Molecular Crystals and Liquid Crystals}, 260(1):565--574, February
  1995.

\bibitem{Saw_2017}
T.~B. Saw, A.~Doostmohammadi, V.~Nier, L.~Kocgozlu, S.~Thampi, Y.~Toyama,
  P.~Marcq, C.~T. Lim, J.~M. Yeomans, and B.~Ladoux.
\newblock Topological defects in epithelia govern cell death and extrusion.
\newblock {\em Nature}, 544(7649):212--216, 2017.

\bibitem{Balasubramaniam_NM_2021}
Lakshmi Balasubramaniam, Amin Doostmohammadi, Thuan~Beng Saw, Gautham Hari
  Narayana~Sankara Narayana, Romain Mueller, Tien Dang, Minnah Thomas, Shafali
  Gupta, Surabhi Sonam, Alpha~S. Yap, Yusuke Toyama, René-Marc Mège, Julia~M.
  Yeomans, and Benoît Ladoux.
\newblock Investigating the nature of active forces in tissues reveals how
  contractile cells can form extensile monolayers.
\newblock {\em Nature Materials}, 20(8):1156--1166, August 2021.

\bibitem{Comelles_elife_2021}
Jordi Comelles, Soumya Ss, Linjie Lu, Emilie Le~Maout, S~Anvitha, Guillaume
  Salbreux, Frank Jülicher, Mandar~M Inamdar, and Daniel Riveline.
\newblock Epithelial colonies in vitro elongate through collective effects.
\newblock {\em eLife}, 10:e57730, January 2021.

\bibitem{Giomi_arxiv_2022}
Josep-Maria Armengol-Collado, Livio~Nicola Carenza, Julia Eckert, Dimitrios
  Krommydas, and Luca Giomi.
\newblock Epithelia are multiscale active liquid crystals, 2022.

\bibitem{Silberzan_2017}
Guillaume Duclos, Christoph Erlenkämper, Jean-François Joanny, and Pascal
  Silberzan.
\newblock Topological defects in confined populations of spindle-shaped cells.
\newblock {\em Nature Physics}, 13(1):58--62, January 2017.

\bibitem{Silberzan_2018}
C.~Blanch-Mercader, V.~Yashunsky, S.~Garcia, G.~Duclos, L.~Giomi, and
  P.~Silberzan.
\newblock Turbulent {Dynamics} of {Epithelial} {Cell} {Cultures}.
\newblock {\em Physical Review Letters}, 120(20):208101, May 2018.

\bibitem{Kawaguchi_2017}
K.~Kawaguchi, R.~Kageyama, and M.~Sano.
\newblock Topological defects control collective dynamics in neural progenitor
  cell cultures.
\newblock {\em Nature}, 545(7654):327--331, 2017.

\bibitem{Volfson_PNAS_2008}
Dmitri Volfson, Scott Cookson, Jeff Hasty, and Lev~S. Tsimring.
\newblock Biomechanical ordering of dense cell populations.
\newblock {\em Proceedings of the National Academy of Sciences},
  105(40):15346--15351, October 2008.

\bibitem{Poon_2018}
D.~Dell'Arciprete, M.~L. Blow, A.~T. Brown, F.~D.~C. Farrell, J.~S. Lintuvuori,
  A.~F. McVey, D.~Marenduzzo, and W.~C.~K. Poon.
\newblock A growing bacterial colony in two dimensions as an active nematic.
\newblock {\em Nat Commun}, 9(1):4190, 2018.

\bibitem{Yaman_2019}
Y.~I. Yaman, E.~Demir, R.~Vetter, and A.~Kocabas.
\newblock Emergence of active nematics in chaining bacterial biofilms.
\newblock {\em Nat Commun}, 10(1):2285, 2019.

\bibitem{Genkin_2017}
M.~M. Genkin, A.~Sokolov, O.~D. Lavrentovich, and I.~S. Aranson.
\newblock Topological defects in a living nematic ensnare swimming bacteria.
\newblock {\em Physical Review X}, 7(1), 2017.

\bibitem{Zhang_PNAS_2018}
H.~Li, X.~Shi, M.~Huang, X.~Chen, M.~Xiao, C.~Liu, H.~Chat\'e, and H.~P. Zhang.
\newblock Data-driven quantitative modeling of bacterial active nematics.
\newblock {\em Proceedings of the National Academy of Sciences},
  116(3):777--785, 2019.

\bibitem{Meacock_2020}
O.~J. Meacock, A.~Doostmohammadi, K.~R. Foster, J.~M. Yeomans, and W.~M.
  Durham.
\newblock Bacteria solve the problem of crowding by moving slowly.
\newblock {\em Nature Physics}, 17(2):205--210, 2020.

\bibitem{Copenhagen_NP2020}
K.~Copenhagen, R.~Alert, N.~S. Wingreen, and J.~W. Shaevitz.
\newblock Topological defects promote layer formation in myxococcus xanthus
  colonies.
\newblock {\em Nature Physics}, 17(2):211--215, 2021.

\bibitem{Dogic_2012}
T.~Sanchez, D.~T. Chen, S.~J. DeCamp, M.~Heymann, and Z.~Dogic.
\newblock Spontaneous motion in hierarchically assembled active matter.
\newblock {\em Nature}, 491(7424):431--4, 2012.

\bibitem{Yeomans_2018}
A.~Doostmohammadi, J.~Ignes-Mullol, J.~M. Yeomans, and F.~Sagues.
\newblock Active nematics.
\newblock {\em Nat Commun}, 9(1):3246, 2018.

\bibitem{Dogic_2020}
G.~Duclos, R.~Adkins, D.~Banerjee, M.~S.~E. Peterson, M.~Varghese, I.~Kolvin,
  A.~Baskaran, R.~A. Pelcovits, T.~R. Powers, A.~Baskaran, F.~Toschi, M.~F.
  Hagan, S.~J. Streichan, V.~Vitelli, D.~A. Beller, and Z.~Dogic.
\newblock Topological structure and dynamics of three-dimensional active
  nematics.
\newblock {\em Science}, 367(6482):1120--1124, 2020.

\bibitem{Zhang_NRM}
R.~Zhang, A.~Mozaffari, and J.~J. de~Pablo.
\newblock Autonomous materials systems from active liquid crystals.
\newblock {\em Nature Reviews Materials}, 6(5):437--453, 2021.

\bibitem{Guillamat_NM_2022}
Pau Guillamat, Carles Blanch-Mercader, Guillaume Pernollet, Karsten Kruse, and
  Aurélien Roux.
\newblock Integer topological defects organize stresses driving tissue
  morphogenesis.
\newblock {\em Nature Materials}, 21(5):588--597, May 2022.

\bibitem{Keren_2021}
Y.~Maroudas-Sacks, L.~Garion, L.~Shani-Zerbib, A.~Livshits, E.~Braun, and
  K.~Keren.
\newblock Topological defects in the nematic order of actin fibres as
  organization centres of hydra morphogenesis.
\newblock {\em Nature Physics}, 17(2):251--259, 2020.

\bibitem{Trepat_2019}
Carlos Pérez-González, Ricard Alert, Carles Blanch-Mercader, Manuel
  Gómez-González, Tomasz Kolodziej, Elsa Bazellieres, Jaume Casademunt, and
  Xavier Trepat.
\newblock Active wetting of epithelial tissues.
\newblock {\em Nature Physics}, 15(1):79--88, January 2019.

\bibitem{Ramaswamy_2002}
R.~Aditi~Simha and S.~Ramaswamy.
\newblock Hydrodynamic fluctuations and instabilities in ordered suspensions of
  self-propelled particles.
\newblock {\em Phys Rev Lett}, 89(5):058101, 2002.

\bibitem{Marchetti_RMP}
M.~C. Marchetti, J.~F. Joanny, S.~Ramaswamy, T.~B. Liverpool, J.~Prost, Madan
  Rao, and R.~Aditi Simha.
\newblock Hydrodynamics of soft active matter.
\newblock {\em Reviews of Modern Physics}, 85(3):1143--1189, 2013.

\bibitem{Bar_SPR_review}
M.~B\"ar, R.~Gro{\ss}mann, S.~Heidenreich, and F.~Peruani.
\newblock Self-propelled rods: Insights and perspectives for active matter.
\newblock {\em Annual Review of Condensed Matter Physics}, 11(1):441--466,
  2020.

\bibitem{Ginelli_PRL_2010}
F.~Ginelli, F.~Peruani, M.~Bar, and H.~Chate.
\newblock Large-scale collective properties of self-propelled rods.
\newblock {\em Phys Rev Lett}, 104(18):184502, 2010.

\bibitem{Marchetti_2008}
A.~Baskaran and M.~C. Marchetti.
\newblock Hydrodynamics of self-propelled hard rods.
\newblock {\em Phys Rev E Stat Nonlin Soft Matter Phys}, 77(1 Pt 1):011920,
  2008.

\bibitem{Huber_science_2018}
L.~Huber, R.~Suzuki, T.~Krüger, E.~Frey, and A.~R. Bausch.
\newblock Emergence of coexisting ordered states in active matter systems.
\newblock {\em Science}, 361(6399):255--258, 2018.

\bibitem{Denk_PNAS_2020}
J.~Denk and E.~Frey.
\newblock Pattern-induced local symmetry breaking in active-matter systems.
\newblock {\em Proc Natl Acad Sci U S A}, 117(50):31623--31630, 2020.

\bibitem{Kaiser_Review}
D.~Kaiser.
\newblock Coupling cell movement to multicellular development in myxobacteria.
\newblock {\em Nat Rev Microbiol}, 1(1):45--54, 2003.

\bibitem{Lotte_2022}
D.~Szadkowski, L.~A.~M. Carreira, and Lotte Søgaard-Andersen.
\newblock A bipartite, low-affinity roadblock domain-containing gap complex
  regulates bacterial front-rear polarity.
\newblock {\em bioRxiv}, page 2022.03.17.484758, 2022.

\bibitem{Sabass_Chapter}
S.~V. Plotnikov, B.~Sabass, U.~S. Schwarz, and C.~M. Waterman.
\newblock High-resolution traction force microscopy.
\newblock {\em Methods Cell Biol}, 123:367--94, 2014.

\bibitem{Sabass_review}
B.~Sabass, M.~L. Gardel, C.~M. Waterman, and U.~S. Schwarz.
\newblock High resolution traction force microscopy based on experimental and
  computational advances.
\newblock {\em Biophys J}, 94(1):207--20, 2008.

\bibitem{Sabass_PNAS2017}
B.~Sabass, M.~D. Koch, G.~Liu, H.~A. Stone, and J.~W. Shaevitz.
\newblock Force generation by groups of migrating bacteria.
\newblock {\em Proc Natl Acad Sci U S A}, 114(28):7266--7271, 2017.

\bibitem{frz_book}
D.R. Zusman, Y.F. Inclan, and T.~Mignot.
\newblock {\em Chapter 7, The Frz Chemosensory System of Myxococcus xanthus. In
  Myxobacteria - Multicellularity and Differentiation, D.E. Whitworth (Ed.)}.
\newblock American Society for Microbiology (ASM), 2008.

\bibitem{Zusman_reversal}
W~Shi, F~K Ngok, and D~R Zusman.
\newblock Cell density regulates cellular reversal frequency in myxococcus
  xanthus.
\newblock {\em Proceedings of the National Academy of Sciences},
  93(9):4142--4146, 1996.

\bibitem{Thutupalli_2015}
S.~Thutupalli, M.~Sun, F.~Bunyak, K.~Palaniappan, and J.~W. Shaevitz.
\newblock Directional reversals enable myxococcus xanthus cells to produce
  collective one-dimensional streams during fruiting-body formation.
\newblock {\em J R Soc Interface}, 12(109):20150049, 2015.

\bibitem{Perez_Review}
J.~Munoz-Dorado, F.~J. Marcos-Torres, E.~Garcia-Bravo, A.~Moraleda-Munoz, and
  J.~Perez.
\newblock Myxobacteria: Moving, killing, feeding, and surviving together.
\newblock {\em Front Microbiol}, 7:781, 2016.

\bibitem{Liu_PRL2019}
G.~Liu, A.~Patch, F.~Bahar, D.~Yllanes, R.~D. Welch, M.~C. Marchetti,
  S.~Thutupalli, and J.~W. Shaevitz.
\newblock Self-driven phase transitions drive myxococcus xanthus fruiting body
  formation.
\newblock {\em Phys Rev Lett}, 122(24):248102, 2019.

\bibitem{Black_2021}
M.~E. Black and J.~W. Shaevitz.
\newblock Rheological dynamics of active myxococcus xanthus populations during
  development, 2021.

\bibitem{zhu1999coarsening}
J.~Zhu, L.~Chen, J.~Shen, and V.~Tikare.
\newblock Coarsening kinetics from a variable-mobility cahn-hilliard equation:
  Application of a semi-implicit fourier spectral method.
\newblock {\em Physical Review E}, 60(4):3564, 1999.

\bibitem{Qin_Science}
B.~Qin, C.~Fei, A.~A. Bridges, A.~A. Mashruwala, H.~A. Stone, N.~S. Wingreen,
  and B.~L. Bassler.
\newblock Cell position fates and collective fountain flow in bacterial
  biofilms revealed by light-sheet microscopy.
\newblock {\em Science}, 369(6499):71--77, 2020.

\bibitem{Kaiser_TFP}
Daniel Wall and Dale Kaiser.
\newblock Type iv pili and cell motility.
\newblock {\em Molecular Microbiology}, 32(1):01--10, 1999.

\bibitem{Zusman_gliding}
E.~M. Mauriello, T.~Mignot, Z.~Yang, and D.~R. Zusman.
\newblock Gliding motility revisited: how do the myxobacteria move without
  flagella?
\newblock {\em Microbiol Mol Biol Rev}, 74(2):229--49, 2010.

\bibitem{frzE}
C.~Kaimer and D.~R. Zusman.
\newblock Regulation of cell reversal frequency in myxococcus xanthus requires
  the balanced activity of chey-like domains in frze and frzz.
\newblock {\em Mol Microbiol}, 100(2):379--95, 2016.

\bibitem{Herrick_3TPM}
W.~G. Herrick, T.~V. Nguyen, M.~Sleiman, S.~McRae, T.~S. Emrick, and S.~R.
  Peyton.
\newblock Peg-phosphorylcholine hydrogels as tunable and versatile platforms
  for mechanobiology.
\newblock {\em Biomacromolecules}, 14(7):2294--304, 2013.

\bibitem{Li_PNAS2019}
H.~Li, X.~Q. Shi, M.~Huang, X.~Chen, M.~Xiao, C.~Liu, H.~Chate, and H.~P.
  Zhang.
\newblock Data-driven quantitative modeling of bacterial active nematics.
\newblock {\em Proc Natl Acad Sci U S A}, 116(3):777--785, 2019.

\bibitem{Vromans_2016}
A.~J. Vromans and L.~Giomi.
\newblock Orientational properties of nematic disclinations.
\newblock {\em Soft Matter}, 12(30):6490--5, 2016.

\bibitem{PTV}
D.~Blair and Dufresne E.
\newblock The matlab particle tracking code repository.
\newblock {\em Particle-tracking code available at
  site.physics.georgetown.edu/matlab/}, 2018.

\bibitem{Deen_PIV}
N.~G. Deen, P.~Willems, M.~van Sint~Annaland, J.~A.~M. Kuipers, R.~G.~H.
  Lammertink, A.~J.~B. Kemperman, M.~Wessling, and W.~G.~J. van~der Meer.
\newblock On image pre-processing for piv of single- and two-phase flows over
  reflecting objects.
\newblock {\em Experiments in Fluids}, 49(2):525--530, 2010.

\bibitem{Schwarz_TFM}
U.~S. Schwarz and J.~R. Soine.
\newblock Traction force microscopy on soft elastic substrates: A guide to
  recent computational advances.
\newblock {\em Biochim Biophys Acta}, 1853(11 Pt B):3095--104, 2015.

\bibitem{Landau_book}
L.D. Landau and E.M. Lifshitz.
\newblock {\em Theory of Elasticity (Second Edition)}.
\newblock Pergamon Press, 1970.

\end{thebibliography}
\let\addcontentsline\oldaddcontentsline 

\clearpage
\renewcommand{\theequation}{S\arabic{equation}}
\renewcommand{\thefigure}{S\arabic{figure}}
\renewcommand{\thetable}{S\arabic{table}}
\setcounter{equation}{0}
\setcounter{figure}{0}
\setcounter{table}{0}

\onecolumngrid
\begin{center}
{\LARGE Supplemental Information} 
\vspace{5mm}
{\Large Spontaneous polar order controls morphology of {\it Myxococcus xanthus} colonies }
\end{center}
\vspace{5mm}
\tableofcontents
\clearpage
%
%
\clearpage
\section{Extended data}

\begin{figure*}[h]
  \begin{center}
    \includegraphics[width=\textwidth]{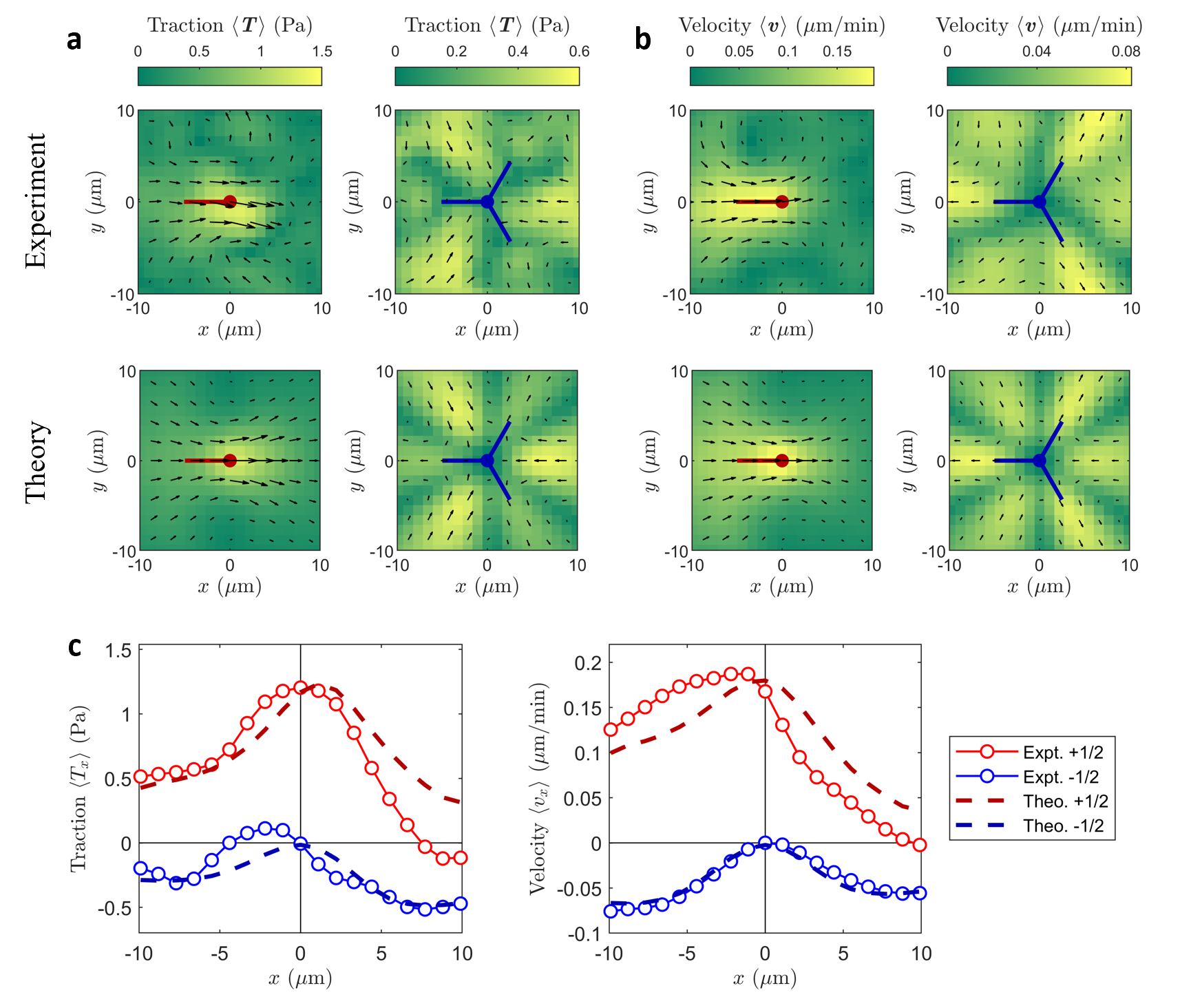}
  \end{center}
  \caption{ \label{Efig:defect_mean_velocity} Comparison between experimentally measured mean traction $\l< \bm{T} \r>$ and mean velocity $\l< \bm{v} \r>$, and the corresponding theoretical predictions. 
(a) Mean traction $\l< \bm{T} \r>$ near $+1/2$ (red symbols) and $-1/2$ (blue symbols) defects. 
(b) Mean velocity of the cell flow $\l< \bm{v} \r>$ near $+1/2$ (red symbols) and $-1/2$ (blue symbols) defects. 
In both panels, the top row is experimental and the bottom row is theoretical. 
The experimental measurements were obtained with the TFM assay (SI Sec.~\ref{sec:experiment}). 
Both $\l< \bm{v} \r>$ and $\l< \bm{T} \r>$ were calculated with 7354 frames of $+1/2$ defects and 6640 frames of $-1/2$ defects identified in 11 replicated experiments. 
Details on how the theoretical velocity and traction distributions were calculated are in Sec.~\ref{sec:theory}. 
(c) Compare experimental (circles) and theoretical (dashed lines) $\l< T_x(x) \r>$ and $\l< v_x(x) \r>$ at $y = 0~\mu$m near the $+1/2$ (red) and $-1/2$ (blue) defects. 
The experimental data shown here are calculated with 7354 frames for the $+1/2$ defects and 6640 frames for the $-1/2$ defects. These defects are identified and tracked in 11 replicated experiments. 
} 
\end{figure*}

\begin{figure*}
  \begin{center}
    \includegraphics[width=1\textwidth]{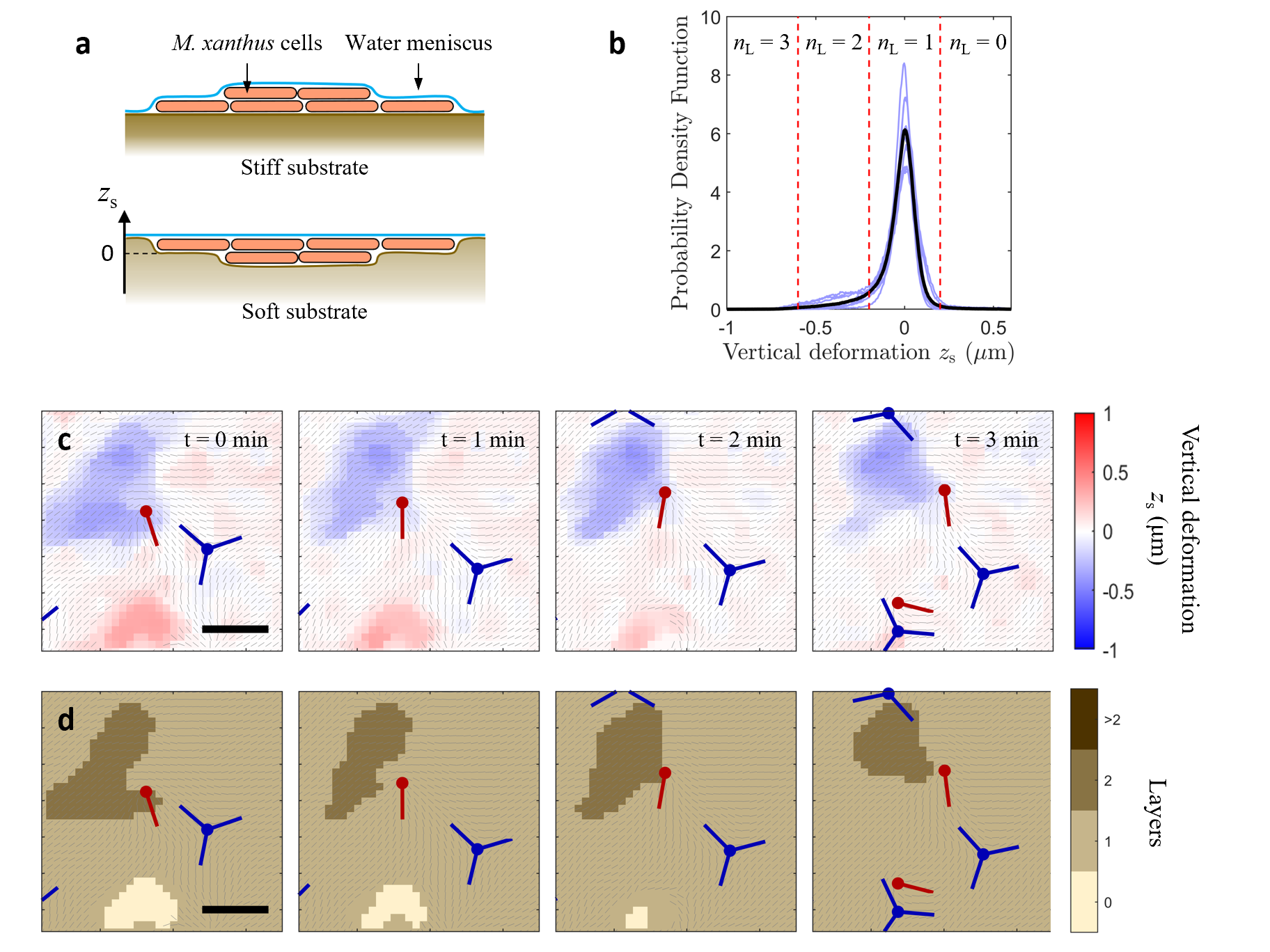}
  \end{center}
  \caption{ \label{Efig:depth} Layer thickness obtained from the measurement of surface deformation $z_\mathrm{s}$. The way we measured $z_\mathrm{s}$ is discussed in SI Sec.~\ref{sec:normal_deformation}. 
(a) Illustration of \textit{M. xanthus} layers on a rigid (top) and soft (bottom) substrate. The orange rods represent the cells, the cyan lines are air-water interfaces, and the brown lines represent the surfaces of the solid substrates. On a solid substrate, the water-air interface is deformed to conform to the shape of the cell layer. However, on a soft substrate, such as in our TFM assay, the substrate was so soft that deforming the substrate was easier than deforming the water-air interface. As a result, the cells were actually mostly embedded into the gel. We define $z_\text{s} = 0~\mu$m as the position just below a cell monolayer, so a double layer has $z_\text{s} < 0$ and regions on the substrate without any cell have $z_\text{s} > 0$. 
(b) Probability distribution function (PDF) of $z_\text{s}$ obtained with $\Delta$\textit{pilA} cells and the TFM assay. Each blue curve was obtained from a video in the experiments. The black curve shows the PDF of all the data. The thickness of each cell layer was about $0.4~\mu$m, thus we chose the red dashed lines as thresholds that turned $z_\text{s}$ into integer cell layer thickness $n_\text{L}$.  
(c) Exemplary surface deformation $z_\text{s}$ measured with the TFM assay. The color map shows $z_\text{s}$ and the gray lines show the local director field $\bm{\hat{n}}$. The scale bar is $10~\mu$m. 
(d) Layer thickness of the cell colony $n_\text{L}$ based on $z_\text{s}$. The color map shows $n_\text{L}$ and the gray lines show the local director field $\bm{\hat{n}}$. The scale bar is $10~\mu$m.
} 
\end{figure*}

\begin{figure*}
  \begin{center}
    \includegraphics[width=\textwidth]{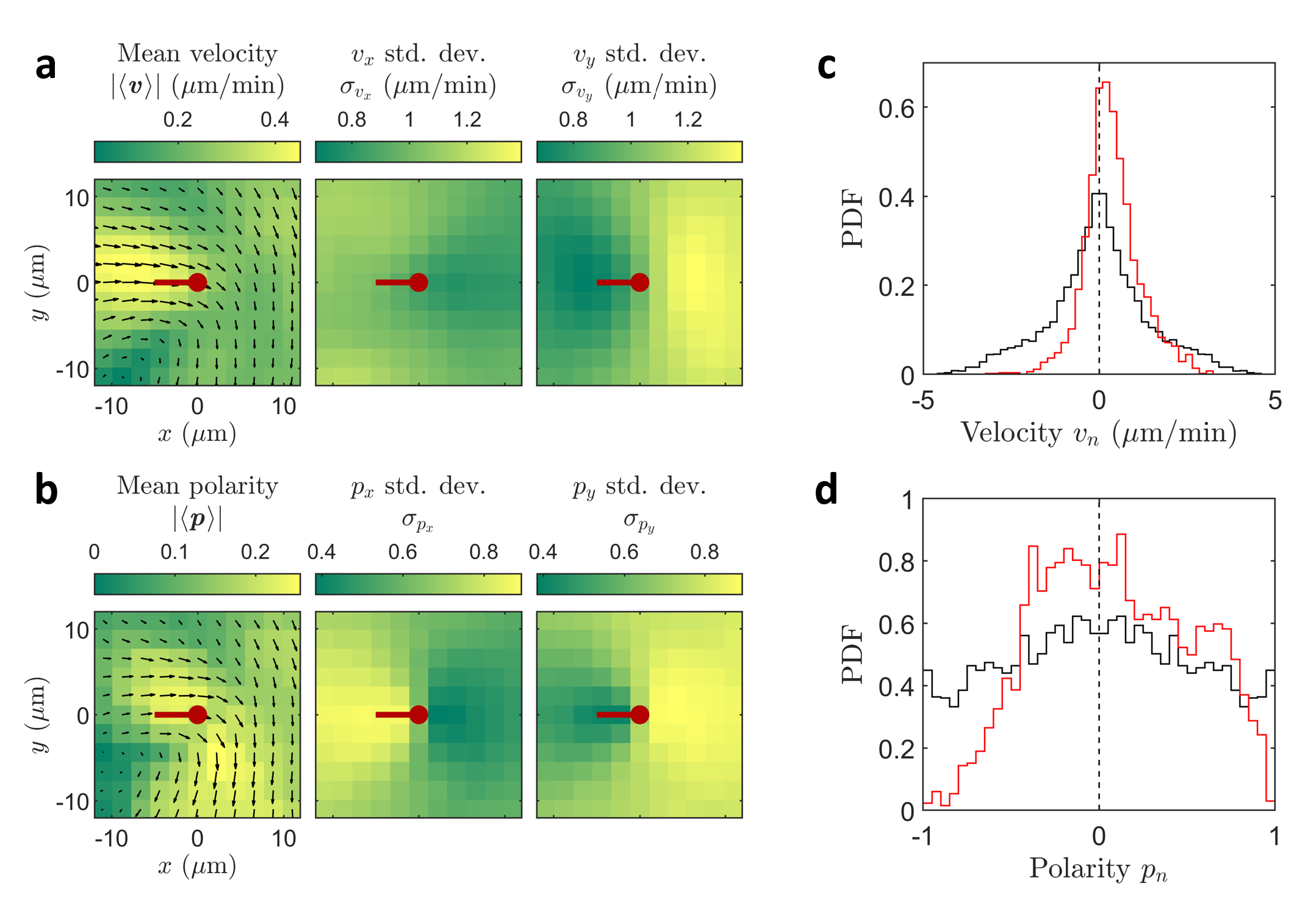}
  \end{center}
  \caption{ \label{Efig:polarity_velocity} Polarity and velocity in ordered regions and near $+1/2$ defects obtained with the polarity assay. 
(a) Mean velocity of cell flow $\l< \bm{v} \r>$ and standard deviations of $v_x$ and $v_y$ ($\sigma_{v_x}$ and $\sigma_{v_y}$) near $+1/2$ defects. The black arrows show the magnitude and direction of $\l< \bm{v} \r>$. The color maps show $\l| \l< \bm{v} \r> \r|$, $\sigma_{v_x}$, and $\sigma_{v_y}$, respectively. Here $\l<~\r>$ denote the temporal average across all the frames. 
(b) Mean cell polarity $\l< \bm{p} \r>$ and standard deviations of $p_x$ and $p_y$ ($\sigma_{p_x}$ and $\sigma_{p_y}$) near $+1/2$ defects. The black arrows show the magnitude and direction of $\l< \bm{p} \r>$. The color maps show $\l| \l< \bm{p} \r> \r|$, $\sigma_{p_x}$, and $\sigma_{p_y}$, respectively. 
(c) Probability density functions (PDF) of $v_n$ in regions with aligned cells (black) and in the tail region of $+1/2$ defects (red). 
(d) PDF of $p_n$ in regions with aligned cells (black) and in the tail region of $+1/2$ defects (red). 
In (c) and (d), $v_n$ and $p_n$ are calculated as described in SI Sec.~\ref{sec:polarity}. 
The positive and negative directions are defined as shown by the orange arrows in Fig.~\ref{SI:polarity_velocity}. 
} 
\end{figure*}

\begin{figure*}
  \begin{center}
    \includegraphics[width=1\textwidth]{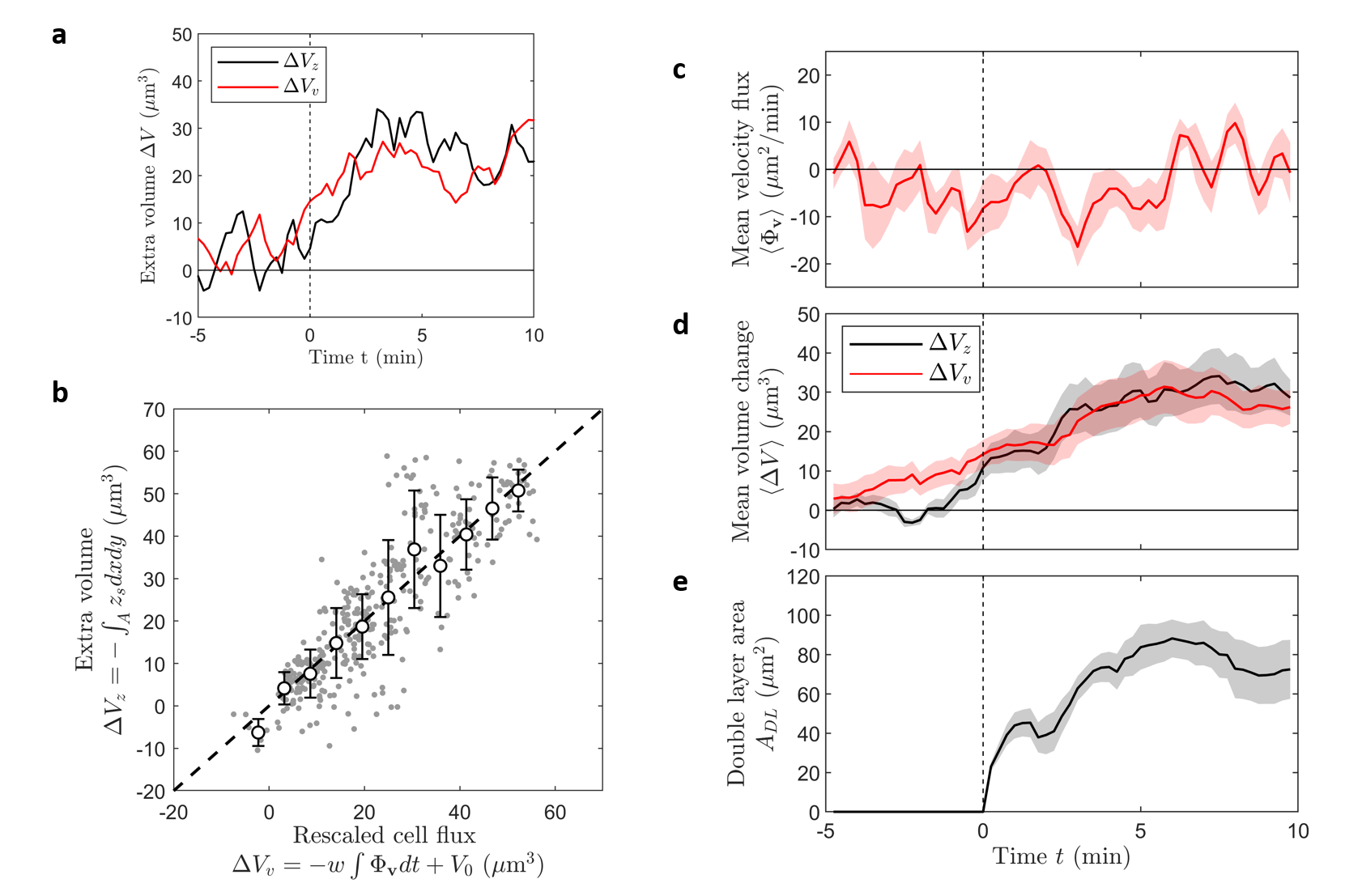}
  \end{center}
  \caption{ \label{Efig:volume} Equivalence and difference of locally accumulated colony volume obtained with substrate surface deformation $z_\text{s}$ and cell velocity $\bm{v}$. The minus surface deformation $-z_\text{s}$ is equivalent to the thickness of the cell layer $h$: $-z_
  \text{s} = h$. 
(a) An exemplary double-layer formation event where the local colony volume increased. The extra volume $\Delta V$ was calculated in two ways: (1) using $z_\text{s}$, $\Delta V_z = - \int_A z_\text{s} \text{d}x \text{d}y$, where $A$ is the area of a circular region of interest with a radius of $l_p = 12~\mu$m; and (2) using $\bm{v}$, $\Delta V_v = - w \int \Phi_v \text{d}t + V_0$, where $w$ is the thickness of the colony at the boundary of this circular region $\mathcal{C}$, $\Phi_v = \oint_\mathcal{C} \l( \bm{v} \cdot \bm{\hat{r}} \r) ds$ is the cell flux out of this region, and $V_0$ represents $\Delta V_v(t = 0)$. We calculated $\Delta V_z$ and $\int \Phi_v \text{d}t$ directly, and used $w$ and $V_0$ as fitting parameters to match $\Delta V_z$ and $\Delta V_v$. The time $t = 0~$min was when a double cell layer appeared. 
(b) Direct comparison between $\Delta V_z$ and $\Delta V_v$. The gray data points are from seven different regions like what (a) shows. The black circles show the mean and the error bars show the standard deviation. The black dashed line is $\Delta V_z = \Delta V_v$. The layer thickness obtained was $\l< w \r> = 0.46 \pm 0.18~\mu$m, which agrees with the thickness of one cell layer. 
(c) Mean cell flux $\l< \Phi_v \r>$ (red line) and its standard error (shaded area) calculated using the same seven regions as in (b). 
(d) Volume change $\Delta V_z$ (red) and $\Delta V_v$ (black). The solid lines show the mean and the shaded areas show the corresponding standard errors. 
(e) Area of the double layer $A_{\text{DL}}$. The solid line shows the mean and the shaded area show the corresponding standard error. A double layer was defined as regions with $z_\text{s} < -0.2~\mu$m. The time $t = 0~$min was when a double layer appeared in panels (c), (d), and (e). 
We can see that $\Delta V_z$ started to increase before a visible double layer actually appeared, because of the threshold in defining double layers. 
More interestingly, $\Delta V_v$ started increasing even earlier, which we think is due to the compression of cell colony locally. In other words, the local cell concentration increased first due to the cell influx, then the surface of the substrate started showing visible deformation, and eventually a visible double layer formed. 
} 
\end{figure*}

\begin{figure*}[h]
  \begin{center}
    \includegraphics[width=1\textwidth]{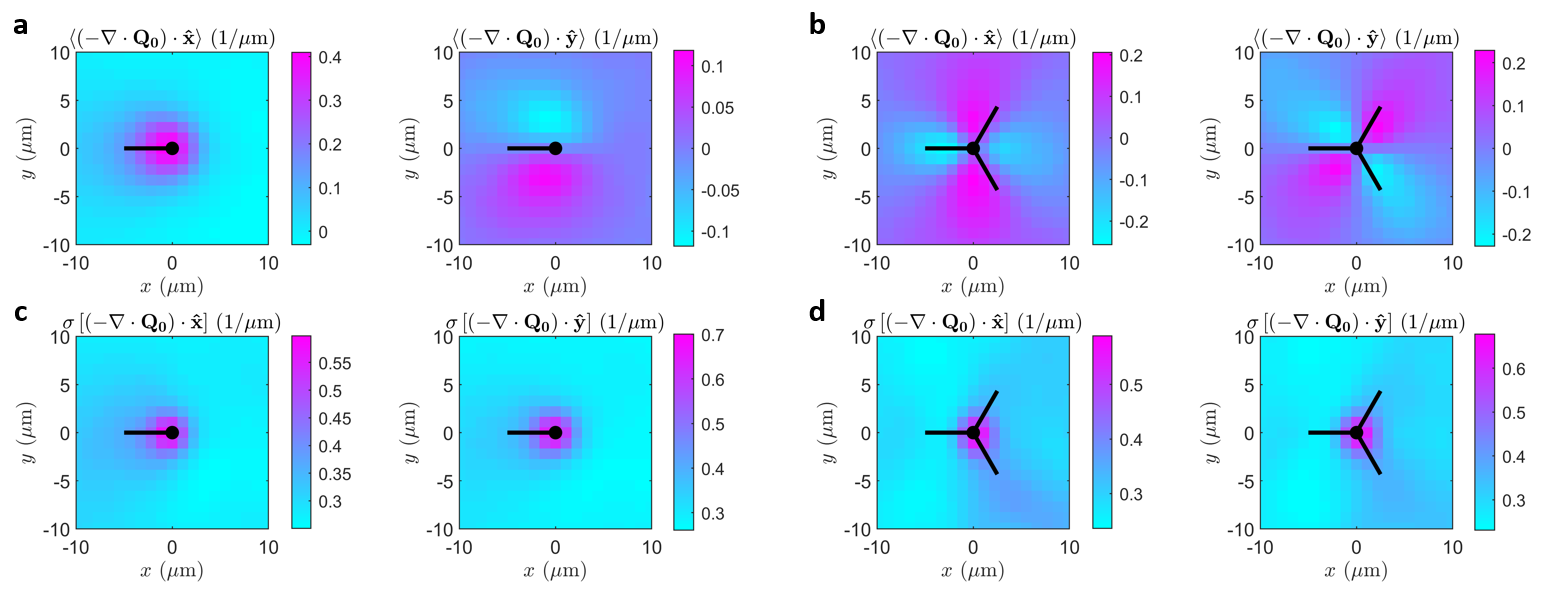}
  \end{center}
  \caption{ \label{Efig:divQ} Mean and fluctuation of $-\nabla \cdot \bm{Q}_0$, where $\bm{Q}_0 = \begin{pmatrix}
        \cos(2 \l< \theta \r>) & \sin(2 \l< \theta \r>) \\
        \sin(2 \l< \theta \r>) & -\cos(2 \l< \theta \r>)
    \end{pmatrix}$ and $\l< \theta \r>$ is the local average director angle. 
(a) Mean components of $-\nabla \cdot \bm{Q}_0$ near $+1/2$ defects in the $x$ and $y$ directions, $\l< (-\nabla \cdot \bm{Q}_0) \cdot \bm{\hat{x}} \r>$ and $\l< (-\nabla \cdot \bm{Q}_0) \cdot \bm{\hat{y}} \r>$, where $\bm{\hat{x}}$ and $\bm{\hat{y}}$ are unit vectors in the $x$ and $y$ directions, respectively. 
(b) Mean $x$ and $y$ components of $-\nabla \cdot \bm{Q}_0$ near $-1/2$ defects, $\l< (-\nabla \cdot \bm{Q}_0) \cdot \bm{\hat{x}} \r>$ and $\l< (-\nabla \cdot \bm{Q}_0) \cdot \bm{\hat{y}} \r>$. 
(c) Standard deviations of the $x$ and $y$ components of $-\nabla \cdot \bm{Q}_0$ near $+1/2$ defects, $\sigma \l[ (-\nabla \cdot \bm{Q}_0) \cdot \bm{\hat{x}} \r]$ and $\sigma \l[ (-\nabla \cdot \bm{Q}_0) \cdot \bm{\hat{y}} \r]$. 
(d) Standard deviations of the $x$ and $y$ components of $-\nabla \cdot \bm{Q}_0$ near $-1/2$ defects, $\sigma \l[ (-\nabla \cdot \bm{Q}_0) \cdot \bm{\hat{x}} \r]$ and $\sigma \l[ (-\nabla \cdot \bm{Q}_0) \cdot \bm{\hat{y}} \r]$. 
Within all the components of $\l< -\nabla \cdot \bm{Q}_0 \r>$ near $\pm 1/2$ defects, the most significant one was $\l< (-\nabla \cdot \bm{Q}_0) \cdot \bm{\hat{x}} \r>$ for $+1/2$ defects. 
All the standard deviations in c and d are of the same order of magnitude. 
As a result, the fluctuation in the director field alone does not explain the huge fluctuations in the velocity and mechanical stress in the system, which were both about one order of magnitude stronger than the mean. 
} 
\end{figure*}

\begin{figure*}
  \begin{center}
    \includegraphics[width=0.9\textwidth]{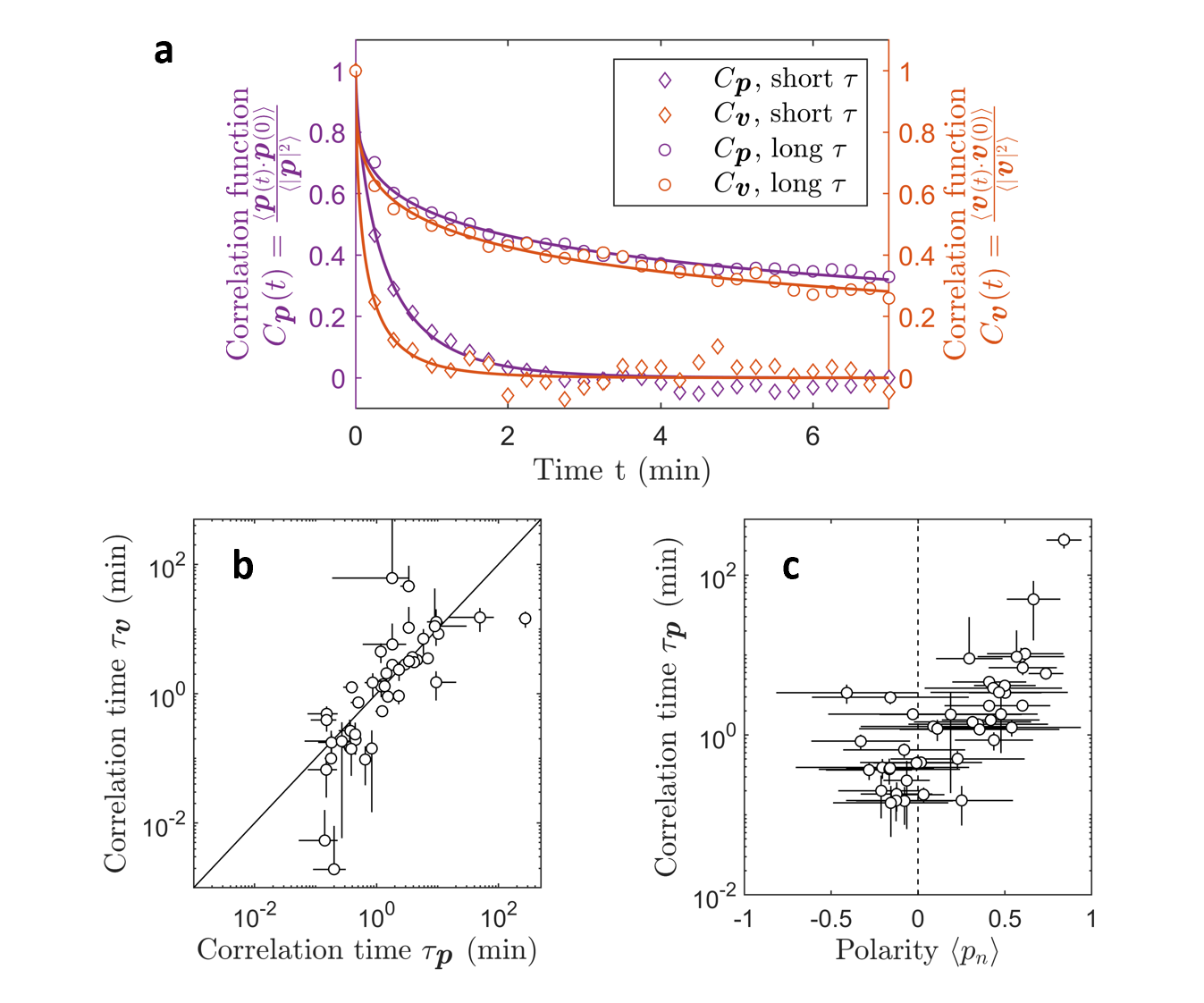}
  \end{center}
  \caption{ \label{Efig:polarity_correlation} Correlation functions and correlation times of polarity $\bm{p}$ and velocity $\bm{v}$ in the tail regions of $+1/2$ defects. 
(a) Exemplary correlation functions of polarity $C_p(t) = \dfrac{\l< \bm{p}(t) \cdot \bm{p}(0) \r> }{ \l< | \bm{p} |^2 \r> }$ and velocity $C_v(t) = \dfrac{\l< \bm{v}(t) \cdot \bm{v}(0) \r> }{ \l< | \bm{v} |^2 \r> }$. The data were fitted with a stretched exponential function $C(t) = \text{exp} \l[ -(t/\tau)^\beta \r]$, where $\tau$ is the correlation time and $\beta$ is a fitting parameter. A pair of $C_p(t)$ and $C_v(t)$ obtained from the data of one defect with relatively short $\tau$ (diamonds) and another pair from another defect with relatively long $\tau$ (circles) are shown in the plot. The solid curves show the corresponding best fits. 
(b) Compare correlation times obtained with polarity $\tau_{\bm{p}}$ and velocity $\tau_{\bm{v}}$. They were approximately equivalent across two orders of magnitude. 
(c) Relationship between the correlation time $\tau_{\bm{p}}$ and the average local polarity $\l< p_n \r>$. Higher polarity leads to a longer correlation time. 
Each data point in (b) and (c) corresponds to one video of a $+1/2$ defect. 
} 
\end{figure*} 


\begin{figure*}
  \begin{center}
    \includegraphics[width=\textwidth]{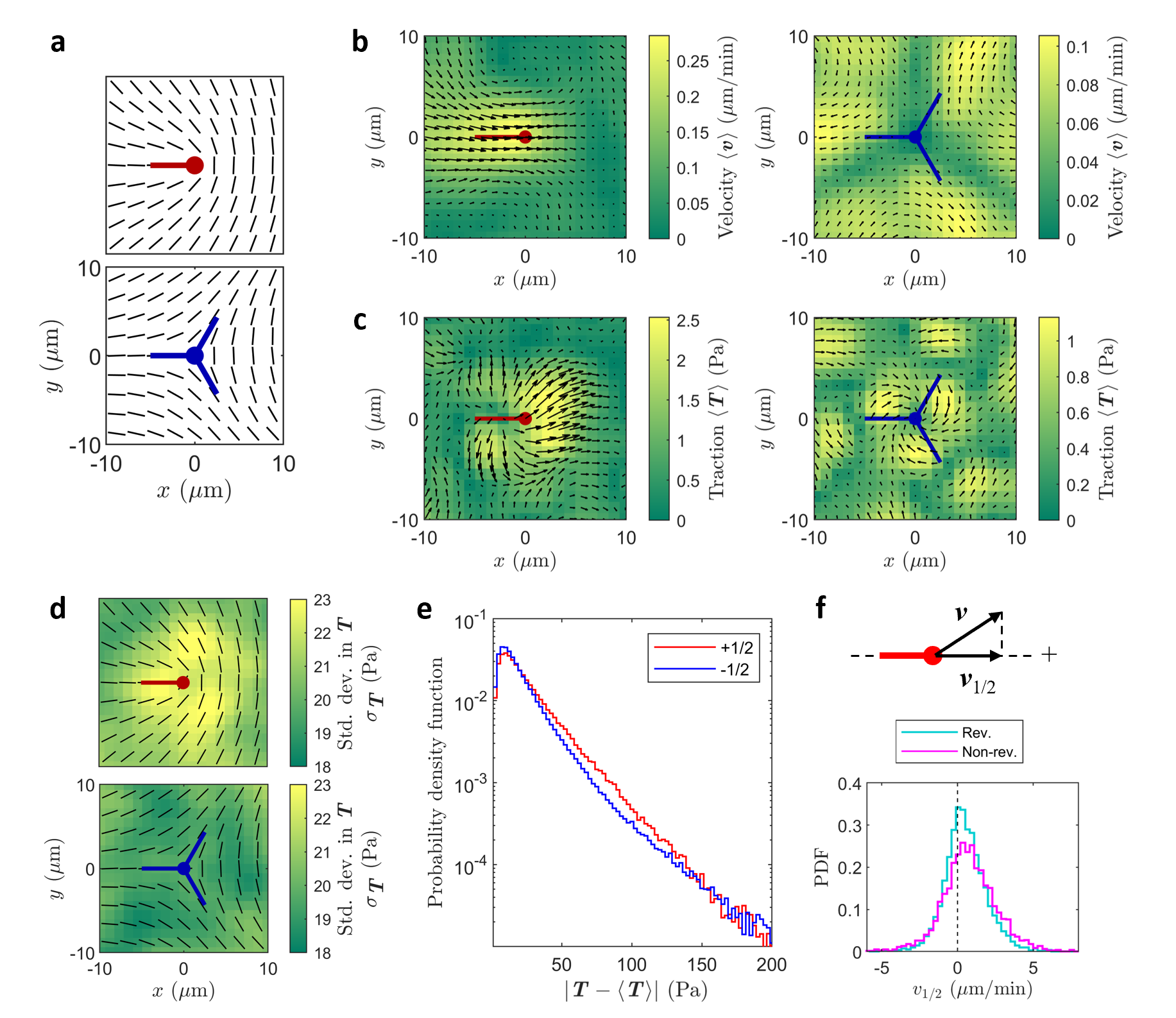}
  \end{center}
  \caption{ \label{Efig:frzE} Properties of the $\pm 1/2$ defects in thin colonies of $\Delta$\textit{frzE} cells (without reversal). 
(a) Experimentally measured mean directors $\l< \bm{\hat{n}} \r>$ near the $\pm 1/2$ defects. 
(b) Experimentally measured mean velocity of cell flow $\l< \bm{v} \r>$ near the $\pm 1/2$ defect. The black arrows show its magnitude and direction and the color map shows the speed $\l| \l< \bm{v} \r> \r|$. 
(c) Experimentally measured mean traction $\l< \bm{T} \r>$ near $\pm 1/2$ defects. The color maps show their magnitudes, and the arrows label their magnitude and direction. 
(d) Experimentally measured standard deviation of traction $\sigma_{\bm T}$ near $\pm 1/2$ defects. The black lines show $\l< \bm{\hat{n}} \r>$. 
(e) Distributions of traction fluctuations $\l| \bm{T}-\l< \bm{T} \r> \r|$ within $5~\mu$m from the centers of $+1/2$ (red) and $-1/2$ (blue) defects. 
(f) Distributions of the forward moving velocity of $+1/2$ defects $v_{1/2}$ for the $\Delta$\textit{pilA} (reversing) and $\Delta$\textit{frzE} (non-reversing) cells. The sketch above shows how we defined the defect velocity $v_{1/2}$. The right-hand side is the positive direction. 
} 
\end{figure*}

\clearpage

\clearpage
\section{Theories on traction and cell flow near defects}
\label{sec:theory}

\subsection{Landau-de Gennes theory}

We use the Landau-de Gennes theory to describe the nematic state of dense, planar bacterial colonies. 
The tensorial order parameter is defined as
\begin{gather}
    \bm{Q} = 2S\Big(\bm{\hat{n}}\bm{\hat{n}} - \frac{1}{2}\bm{I}\Big), \label{eq_Q_tensor}
\end{gather}
where the director $\bm{\hat{n}} \equiv (\cos\theta,\sin\theta)$ denotes the local axis along which cells align, $\bm{I}$ is the unit matrix, and $S$ is the scalar order parameter. 
The 2D Laudau-de Gennes energy functional takes the form
\begin{gather}
    \mathcal{F}(\bm{Q},\nabla\bm{Q}) = \int d^2\bs{x} \Big[-\frac{a}{2}\mathrm{Tr}(\bm{Q}^T\bm{Q}) + \frac{b}{4}\Big(\mathrm{Tr}(\bm{Q}^T\bm{Q})\Big)^2 + \frac{L}{2}(\nabla \bm{Q}):(\nabla \bm{Q})\Big], \label{eq_F_deg}
\end{gather}
where $a,b>0$ and $L$ denotes the orientation elastic modulus. 
Note that here we do not include the $\partial_j Q_{ij} \partial_k Q_{ik}$ invariant because in 2D there is no twisting mode of the distortion, which corresponds to the Frank elastic energy $\frac{K_3}{2}|\bm{\hat{n}}\times\nabla\times \bm{\hat{n}} |^2$.

Introducing Eq. (\ref{eq_Q_tensor}) into Eq. (\ref{eq_F_deg}), we obtain the following nematic energy:
\begin{gather}
    \mathcal{F}(S,\theta,\nabla S,\nabla \theta) = \int d^2\bs{x} \Big[ -a S^2 + b S^4 + L|\nabla S|^2 + 4LS^2 |\nabla \theta|^2\Big]. \label{eq_F_defect}
\end{gather}
In the uniformly ordered phase, the order parameter strength is given by $S_0^2 = a / 2b$. 
Around a defect of charge $q$, where $\theta = q \phi$, and $S = S_0 \Phi(r)$, we obtain $\nabla\theta = q/r~\bm{e}_\phi$ and $\nabla S = S_0 \Phi^\prime(r)~\bm{e}_r$. 
Introducing these relations into Eq. (\ref{eq_F_defect}), we obtain, for $\pm 1/2$ defects ($q=\pm1/2$), 
\begin{gather}
    \mathcal{F}[\Phi] = L S_0^2 \int 2\pi \tilde{r}d\tilde{r}\Big[-\Phi^2 + \frac{1}{2} \Phi^4 + \Big(\frac{d\Phi}{d\tilde{r}}\Big)^2 + \frac{\Phi^2}{\tilde{r}^2}\Big],
\end{gather}
where we defined a dimensionless radial coordinate as $\tilde{r} = r/\ell$, in which $\ell  = \sqrt{L/a}$ denotes the nematic correlation length. 
Minimizing the energy functional yields the saturating coefficient of the nematic order $\Phi(\tilde{r})$, which satisfies $\Phi(0) = 0$ and $\Phi(\infty) = 1$ and the ODE, 
\begin{equation}
    -\Phi + \Phi^3 = \Phi'' + \Phi'/\tilde{r} - \Phi/\tilde{r}^2, 
\end{equation}
where $\Phi^\prime$ denotes $d\Phi/d\tilde{r}$. Although the exact expression for $\Phi$ cannot be solved analytically, it can be approximated by the Pad\'e approximant:
\begin{equation}
    \Phi(\tilde{r}) \approx \sqrt{\frac{0.07 \tilde{r}^4 + 0.34 \tilde{r}^2}{1 + 0.41 \tilde{r}^2 + 0.07 \tilde{r}^4}}. 
    \label{eqn:pade}
\end{equation}

\begin{figure*}[!t]
\vspace{-10pt}
  \begin{center}
    \includegraphics[width=1\textwidth]{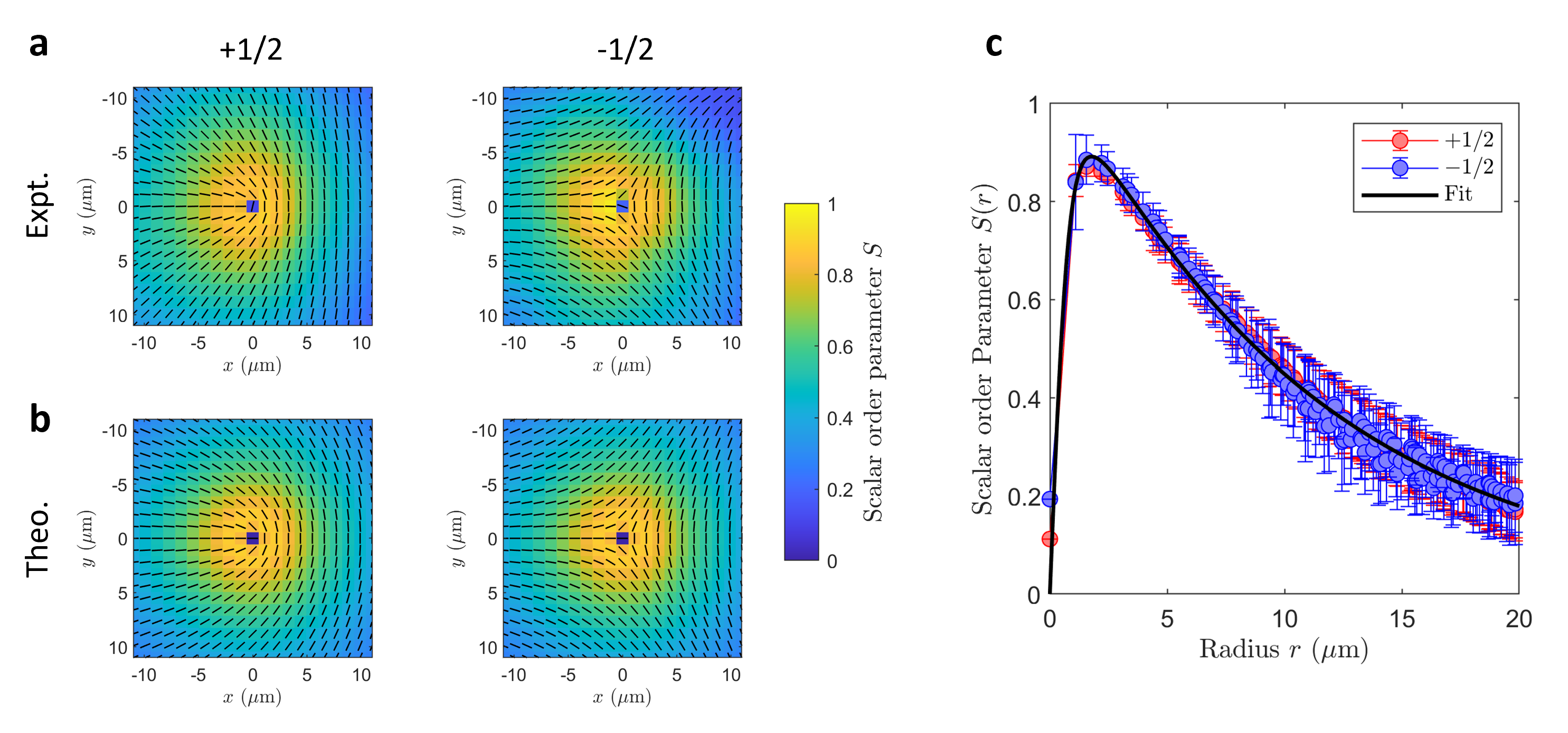}
  \end{center}
  \vspace{-20pt}
  \caption{ Director field and scalar order parameter $S$ around $\pm 1/2$ defects. 
  (a) Experimentally measured mean director fields around $+1/2$ (left) and $-1/2$ (right) defects. The color map shows the scalar order parameter $S$ and the black lines label the local directors. 
  (b) Theoretically calculated director fields around $+1/2$ (left) and $-1/2$ (right) defects. Labels identical to (a). 
  (c) Radial distribution of the scalar order parameter $S$ for $+1/2$ (red) and $-1/2$ (blue) defects. The black curve is the best fit using Eq.~\ref{eq:Sr}.  } 
  \label{fig:defect_director}
\end{figure*}

The alignment order parameter $S$ measured by experiments can be fitted by 
\begin{equation}
    S(r)=S_0 \Phi(r/\ell)\exp(-r/R), 
    \label{eq:Sr}
\end{equation}
where $S_0$, $\ell$, and $R$ are fitting parameters, and $\Phi$ is given by Eq.~\ref{eqn:pade}. 
The exponential decay is purely phenomenological. 
In each frame, in the regions away from the defect cores, the cells are normally well aligned. 
However, in different frames (at different times), they may align in different directions.  
Consequently, when averaged across all frames, the mean orientational order is weak away from the defect cores. 
This is why we introduce an exponential term here in Eq.~\ref{eq:Sr}. 
The measured $S(r)$ are similar between $+1/2$ defects and $-1/2$ defects (Fig.~\ref{fig:defect_director}). Thus, we use the same parameters to fit both data, yielding $S_0=1.2,~\ell = 0.8~{\mu m},~\mathrm{and}~R = 10~\mathrm{\mu m}$. 
Note that as a fitting parameter, $S_0$ is allowed to exceed 1, while the order parameters $S$ must be in the range of $[0,1]$. 
For regions around the topological defects, the complete form of $\bm{Q}$ is given by 
\begin{equation}
    \bm{Q}(r,\phi) = S(r) 
    \begin{pmatrix}
        \cos(2q\phi) & \sin(2q\phi) \\
        \sin(2q\phi) & -\cos(2q\phi)
    \end{pmatrix},     
\end{equation}
where $S(r)$ is given by Eq.\ (\ref{eq:Sr}).

\subsection{Hydrodynamics of two-dimensional incompressible cell monolayers}

Here, we consider the flow in a two-dimensional nematic system driven by the active stress and dampened by drag-like friction between cells and substrate. 
In such a microscopic system, inertia plays no role, so the force balance equation is
\begin{equation}
    \bm{f}^\text{a}_\text{s} + \bm{f}^\text{a}_\text{c} + \bm{f}^\text{f}_\text{s} + \bm{f}^\text{f}_\text{c} - \nabla P + \nabla \cdot \bm{\sigma}^\text{el} + \nabla \cdot \bm{\sigma}^\mu = 0. 
    \label{eq:EoM}
\end{equation}
The meanings of the terms are described in Table~\ref{tab:traction_symbols}. 
\begin{table}[!t]
    \centering
    \begin{tabular}{|c|l|}
    \hline
    Term & Description \\
    \hline
     $\bm{f}^\text{a}_\text{s}$ & Active force density due to cells gliding on substrate \\ 
     $\bm{f}^\text{a}_\text{c}$ & Active force density due to cells gliding on other cells \\
     $\bm{f}^\text{f}_\text{s}$ & Density of friction between cells and substrate \\ 
     $\bm{f}^\text{f}_\text{c}$ & Density of friction between cells \\ 
     $\nabla P$ & Pressure gradient \\
     $\nabla \cdot \bm{\sigma}^\text{el}$ & Force density arising from nematic elasticity \\
     $\nabla \cdot \bm{\sigma}^{\mu}$ & Viscous force density \\
     \hline
    \end{tabular}
    \caption{Different force densities in the system. Their units are all Pa. The pressure $P$ and stresses $\sigma^{\text{el}}$ and $\sigma^{\mu}$ are all linear force densities with the unit Pa$\cdot$m. }
    \label{tab:traction_symbols}
\end{table}
The basic idea is that cells self-propel by exerting active forces both on the substrate $\bm{f}^\text{a}_\text{s}$ and on other cells $\bm{f}^\text{a}_\text{c}$, and they experience frictions applied by the substrate $\bm{f}^\text{f}_\text{s}$ and by other cells $\bm{f}^\text{f}_\text{c}$. 
The pressure $P$ ensures that the velocity field $\bm{v}$ satisfies any imposed velocity-divergence conditions. 
Lastly, there is an elastic term $\nabla \cdot \bm{\sigma}^\text{el}$ due to distortions of the nematic director, and a viscous term $\nabla \cdot \bm{\sigma}^\mu$ due to cell-cell friction. 
Following previous work \cite{Copenhagen_NP2020}, we neglected these last two terms in our calculation.  

What is the traction $\bm{T}$ measured in our TFM experiments? 
It reflects the total force applied on the substrate by the cells, thus corresponding to the sum of cell-substrate forces, $\bm{f}^\text{a}_\text{s} + \bm{f}^\text{f}_\text{s}$. 
By virtue of the force balance Eq.~\ref{eq:EoM}, the traction can also be expressed as the sum of the internal forces in the cell layer: 
\begin{subequations}
\begin{align}
    \bm{T} &= - \bm{f}^\text{a}_\text{s} - \bm{f}^\text{f}_\text{s}, 
    \label{eq:trac_1} \\ 
    \bm{T} &= \bm{f}^\text{a}_\text{c} + \bm{f}^\text{f}_\text{c} - \nabla P. 
    \label{eq:trac_2}
\end{align}
\label{eq:trac_both}
\end{subequations}
The active force due to cells gliding on the substrate, $\bm{f}^\text{a}_s$ is controlled by the instantaneous polarity $\bm{p}$ of the cells: each cell pushes the substrate in the direction opposite to its polarity, so we can write  
\begin{equation}
    \bm{f}^\text{a}_\text{s} = \zeta_p \bm{p}. 
\end{equation}
Following \cite{Copenhagen_NP2020}, we model the friction force as an anisotropic viscous force that depends on the orientation field of the bacterial cells:  
\begin{equation}
    \bs{f}^\text{f}_\text{s} = - \xi_0 (\bm{I} - \epsilon \bm{Q}) \cdot \bs{v}, 
\end{equation}
where $\xi_0$ is the isotropic contribution to the friction coefficient, $\epsilon$ is the friction anisotropy, and $\bs{v}$ denotes the velocity of the cells. 
The cell-cell active force $\bm{f}^\text{a}_\text{c}$ results from force dipoles generated between cell pairs. 
This force is, therefore, the standard active force in active nematics, which emerges from the active stress $\bm{\sigma}^\text{a}_\text{c} = \zeta_\text{c} \bm{Q}$, with $\bm{f}^\text{a}_\text{c} = - \nabla \cdot \bm{\sigma}^\text{a}_\text{c}$, so we get  
\begin{equation}
	\bm{f}^\text{a}_\text{c} = - \zeta_\text{c} \nabla \cdot \bm{Q}.    
	\label{eq:f_c_a}
\end{equation} 
Our system is described by extensile active stresses, with a positive active-stress coefficient $\zeta_\text{c} > 0$.
Lastly, $\bm{f}^\text{f}_\text{c}$ is controlled by the relative velocity between adjacent cells, and it is similar to the viscous dissipation term in the Navier-Stokes equations: 
\begin{equation}
    \bm{f}^\text{f}_\text{c} = h \mu_\text{c} \nabla^2 \bm{v}, 
    \label{eq:f_c_r}
\end{equation}
where $h$ is the thickness of the cell layer and $\mu_\text{c}$ is the effective viscosity of the cell colony.

\begin{figure*}
  \begin{center}
    \includegraphics[width=1\textwidth]{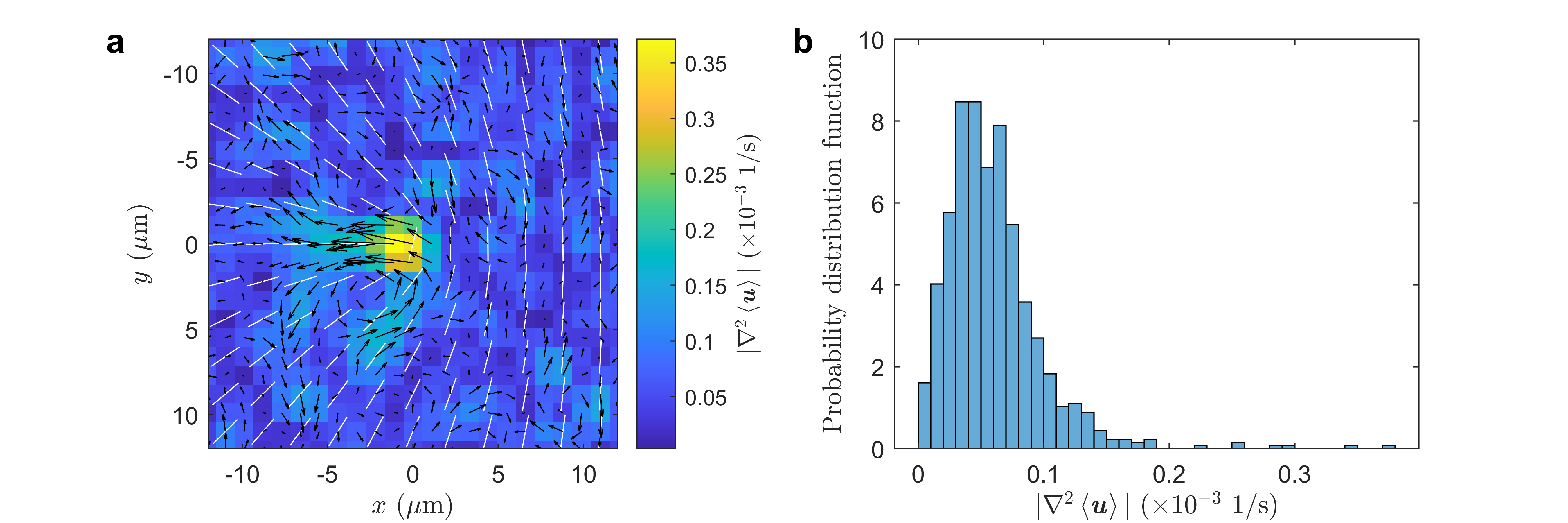}
  \end{center}
  \caption{ \label{SI_viscosity} Distribution of $\nabla^2 \l< \textbf{\textit{u}} \r>$ near a $+1/2$ defect. 
  (a) The color map shows the magnitude of $\l| \nabla^2 \l< \textbf{\textit{u}} \r> \r|$. The black arrows indicate the magnitude and direction of $\nabla^2 \l< \textbf{\textit{u}} \r>$. The white bars show the average director field. 
  (b) Distribution of $\l| \nabla^2 \l< \textbf{\textit{u}} \r> \r|$ in (a). 
} 
\end{figure*}

In our experiments, we measured the average traction near the defects. 
According to Eq.~\ref{eq:trac_both}, we have 
\begin{subequations}
\begin{align}
    \l< \bm{T} \r> &= - \l< \bm{f}^\text{a}_\text{s} \r> - \l< \bm{f}^\text{f}_\text{s} \r>, \\ 
    \l< \bm{T} \r> &= \l< \bm{f}^\text{a}_\text{c} \r> + \l< \bm{f}^\text{f}_\text{c} \r> - \l< \nabla P \r>. 
\end{align}
\label{eq:trac_mean_both}
\end{subequations}
Within all these terms, $\l< \bm{f}^\text{f}_\text{c} \r>$ is negligible. 
Given Eq.~\ref{eq:f_c_r}, we obtain 
\begin{equation}
    \l< \bm{f}^\text{f}_\text{c} \r> = h \l< \mu_\text{c} \nabla^2 \bm{v}\r> = h \mu_\text{c} \nabla^2 \l< \bm{v}\r>.  
\end{equation}
The spatial distribution of $\nabla^2 \l< \bm{v}\r>$ is shown in Fig.~\ref{SI_viscosity}. 
Except for the small region very close to the core of the defect, $\l| \nabla^2 \l< \bm{v}\r> \r| < 2 \times 10^{-4}~$s$^{-1}$.  
According to a recent measurement \cite{Black_2021}, the loss modulus $G''$ of a 12-hour \textit{M. xanthus} fruiting body is $G'' \approx 0.2~$kPa at frequency $\omega = 0.1~$Hz and $G'' \approx 1~$kPa at frequency $\omega = 1~$Hz. 
Considering the cell speed of the order of microns per minute in our system, 1~Hz is in the high-frequency regime. 
As a result, we estimate that the upper limit of the dynamic viscosity in our thin cell layers is about $\mu_\text{c} = G''/\omega \approx 1~$kPa$\cdot$s, which is $10^6$ times more viscous than water. 
Furthermore, as a fruiting body ages, its viscosity increases \cite{Black_2021}. 
In our cell monolayer, which is at a much earlier stage than a 12-hour fruiting body, we expect an even smaller viscosity. 
Using $\mu_\text{c} = 1~$kPa$\cdot$s and $h = 0.5~\mathrm{\mu}$m to estimate the stress due to cell-cell drag, we get 
\begin{equation}
    \l< \bm{f}^\text{f}_\text{c} \r> < 0.1~\text{Pa} 
\end{equation}
in most regions around a $+1/2$ defect. 
The actual value of $\l< \bm{f}^\text{f}_\text{c} \r>$ could be even smaller given the way we chose the parameters' values. 
Thus in the following analysis, we assume this term is negligible. 
Note that here we neglected the anisotropy in the cell layer, but the estimated order of magnitude should hold. 

Now we can introduce all the other terms above to Eqs.~\ref{eq:trac_mean_both}:  
\begin{subequations}
\begin{align}
    \l< \bm{T} \r> &= - \zeta_p \l< \bm{p} \r> + \xi_0 \l< (\bm{I} - \epsilon \bm{Q}) \cdot \bs{v} \r>, 
    \label{eq:trac_mean_1} \\ 
    \l< \bm{T} \r> &= - \zeta_\text{c} \nabla \cdot \l< \bm{Q} \r> - \nabla \l< P \r>.  
    \label{eq:trac_mean_2}
\end{align}
\end{subequations}
In the experiments, we measured $\l< \bm{T} \r>$, $\l< \bm{Q} \r>$ (based on $\theta$), and $\l< \bm{v} \r>$. 
The unknown fields are $\l< \bm{p} \r>$ and $\l< P \r>$ and the unknown parameters are $\zeta_p$, $\xi_0$, $\epsilon$, and $\zeta_\text{c}$. 
Note that all the force densities here are areal force densities, which have the unit of Pa, and the pressure $P$ has the unit of Pa$\cdot$m. 
Correspondingly, $\zeta_\text{c}$ has the unit of Pa$\cdot$m, $\zeta_p$ has the unit of Pa, and $\xi_0$ has the unit of Pa$\cdot$s/m.

\subsection{Velocity and traction around topological defects}
\label{sec:theory_defect}

\begin{figure*}
  \begin{center}
    \includegraphics[width=1\textwidth]{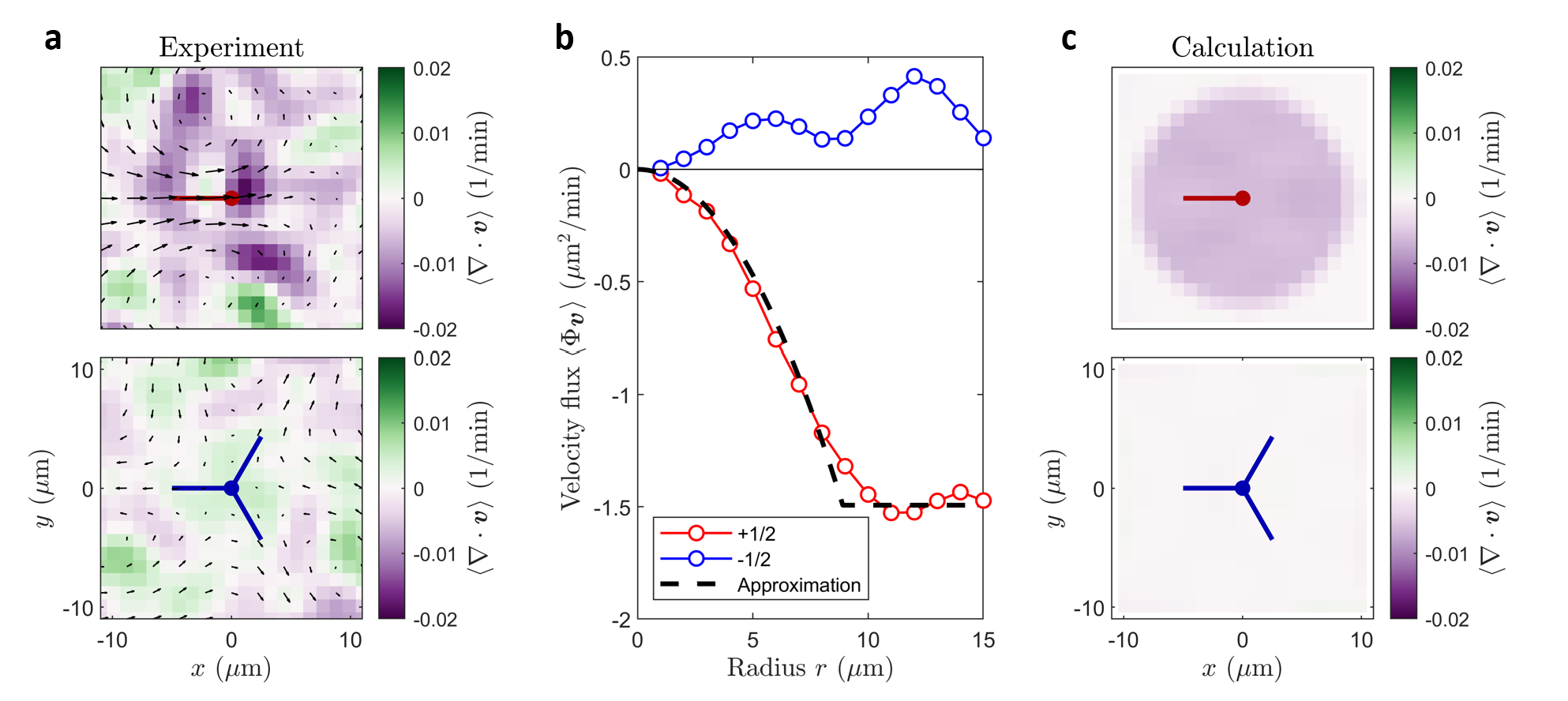}
  \end{center}
  \caption{ \label{Efig:divV} Compressibility of the cell layer near $\pm 1/2$ defects. 
(a) Divergence of the mean cell velocity $\l< \nabla \cdot \bm{v} \r>$ near $+1/2$ (red symbols) and $-1/2$ (blue symbols) defects. 
The color maps show $\l< \nabla \cdot \bm{v} \r>$, with green being positive and purple being negative, and the black arrows show the magnitude and direction of local velocity $\l< \bm{v} \r>$. 
The experimental measurements were obtained with the TFM assay (SI Sec.~\ref{sec:experiment}). 
The mean velocities were calculated with 7354 frames of $+1/2$ defects and 6640 frames of $-1/2$ defects identified in 11 replicated experiments. 
We tracked the motion of these defects, aligned the velocity fields around them in every frame based on the defects' locations and orientations, and then calculated the mean. 
Note that even though the velocity fields were aligned in the comoving frame of the defects, what we show here and in the main text are the cell velocities with respect to the substrate (in the lab frame). 
(b) Velocity flux $\Phi_v$ across a circular region with radius $r$ and centered at $+1/2$ (red) and $-1/2$ (blue) defects, $\Phi_v = \oint_\mathcal{C} \l( \bm{v} \cdot \bm{\hat{r}} \r) \text{d}s = \iint_\mathcal{A} \l( \nabla \cdot \bm{v} \r) \text{d}x \text{d}y$, where $\mathcal{C}$ is the boundary of this circular region and $\mathcal{A}$ is the area it covers. 
The flux near the $+1/2$ defect was approximated by a parabolic function at small $r$ ($r < 8.9~\mu$m) and a constant where $r > 8.9~\mu$m, as the black dashed line shows. 
This means that in the parabolic regime ($r < 8.9~\mu$m) $\l< \nabla \cdot \bm{v} \r>$ is approximately a constant, and $\l< \nabla \cdot \bm{v} \r> = 0$ as $r > 8.9~\mu$m
(c) In the theoretical calculations, we used the approximated spatial distribution of $\l< \nabla \cdot \bm{v} \r>$ as a constraint. We assumed $\l< \nabla \cdot \bm{v} \r> = \text{constant}$ as $r < 8.9~\mu$m and $0$ otherwise near the $+1/2$ defects, and we assumed the cell flow near the $-1/2$ defects was incompressible ($\l< \nabla \cdot \bm{v} \r> = 0$).  
Details are described in SI Sec.~\ref{sec:theory_defect}. 
} 
\end{figure*}

Combining Eq.~\ref{eq:trac_mean_1} and Eq.~\ref{eq:trac_mean_2} and assuming $\l< \bm{p} \r>=0$, the balance of average forces near defects is given by
\begin{gather}
    \l< \bm{T} \r> = \xi_0 (\bm{I} - \epsilon \l< \bm{Q} \r>) \cdot \l< \bs{v} \r> = - \zeta_\text{c} \nabla \cdot \l< \bm{Q} \r> - \nabla \l< P \r>.
    \label{eq:hydro}
\end{gather}
The isotropic pressure $\l< p \r>$ ensures that $\l< \bs{v} \r>$ satisfies the constraint $\nabla \cdot \l< \bs{v} \r> = j(\bs{r})$, where $j$ is the flux of cells across cell layers. For $-1/2$ defects, we impose the incompressibility condition $j(\bs{r}) \equiv 0$. To capture the asymmetric cell flows around the $+1/2$ defects (Fig.\ \ref{Efig:defect_mean_velocity}), we set
\begin{gather}
j (\bs{r}) = 
\begin{cases}
- j_0 & \mathrm{for~} r \leqslant R_0,\\
0 & \mathrm{for~} r > R_0,
\end{cases}
\label{eq: div_V_inflow}
\end{gather}
where $r$ denotes the distance from the defect center, and $j_0 = 6\times 10^{-3}~\mathrm{min^{-1}}$ and $R_0 = 8.9 \mathrm{~\mu m}$ are determined from experimental measurements (Fig.\ \ref{Efig:divV}).
Since $\zeta_\mathrm{c}$ sets the scale for the stress/pressure, we rewrite Eq.~(\ref{eq:hydro}) with scaled quantities $\tilde{p} = p/\zeta_\text{c}$ and $\tilde{\xi} = \xi_0/\zeta_\mathrm{c}$, which yields 
\begin{gather}
\tilde{\xi}(\bm{I} - \epsilon \l< \bm{Q} \r>) \cdot \l< \bs{v} \r> = - \nabla \cdot \l< \bm{Q} \r> - \nabla \l< \tilde{p} \r>.\label{eq:hydro-scaled}
\end{gather}
We solve Eq.~(\ref{eq:hydro-scaled}) numerically to obtain the velocity field $\l< \bm{v} \r>$ as described below. The traction field $\l< \bm{T} \r>$ is then computed by $\l< \bm{T} \r> =\xi_0 (\bm{I} - \epsilon \l< \bm{Q} \r>) \cdot \l< \bs{v} \r>$.

\noindent{\bf Numerical scheme}: To solve Eq.~(\ref{eq:hydro-scaled}) with the constraint $\nabla \cdot \l< \bs{v} \r> = j(\bs{r})$, we formally separate $\l< \bm{v} \r>$ into $\l< \bm{v} \r>=\l< \bm{u} \r> + \l< \bm{w} \r>$, where $\l< \bm{w} \r>$ is an arbitrary velocity field that satisfies $\nabla \cdot \l< \bm{w} \r> = j$. Specifically, we set $\l< \bm{w} \r> = \bs{0}$ for $-1/2$ defects, and $\l< \bm{w} \r> = w(r) \hat{\bs{r}}$ for $+1/2$ defects, where $w(r) = -j_0 r/2$ for $r\leqslant R_0$ and $w(r)=-j_0R_0^2/(2r)$ for $r > R_0$.
Next, the governing equation for $\l< \bm{u} \r>$ becomes
\begin{gather}
\tilde{\xi}(\bm{I} - \epsilon \l< \bm{Q} \r>) \cdot \l< \bs{u} \r> = - \tilde{\xi}(\bm{I} - \epsilon \l< \bm{Q} \r>) \cdot \l< \bs{w} \r> - \nabla \cdot \l< \bm{Q} \r> - \nabla \l< \tilde{P} \r>, ~\mathrm{and}~ \nabla \cdot \l< \bs{u} \r>=0.\label{eq:hydro-separate}
\end{gather}
Eq.~\ref{eq:hydro-separate} is solved using a semi-implicit Fourier spectral method as previously described \cite{zhu1999coarsening,Qin_Science}.
Briefly, we formally rewrite Eq.~(\ref{eq:hydro-separate}) as 
\begin{gather}
a \l< {\bs{u}} \r> - b \nabla^2 \l< {\bs{u}} \r> =  a \l< {\bs{u}} \r>  - b \nabla^2 \l< {\bs{u}} \r> -\tilde{\xi}(\bm{I} - \epsilon \l< \bm{Q} \r>) \cdot (\l< {\bs{u}} \r> + \l< {\bs{w}} \r>)- \nabla \cdot \l< \bm{Q} \r> - \nabla \l< \tilde{P} \r>\equiv -\nabla \l< \tilde{P} \r> + \bm{F}(\l< {\bm{u}} \r>),\label{eq:hydro-num}
\end{gather}
where $a$ and $b$ are introduced to stabilize the numerical scheme, and we have grouped all the non-pressure terms into $\bm{F}(\l< {\bm{u}} \r>) = a \l< {\bs{u}} \r>  - b \nabla^2 \l< {\bs{u}} \r> -\tilde{\xi}(\bm{I} - \epsilon \l< \bm{Q} \r>) \cdot (\l< {\bs{u}} \r> + \l< {\bs{w}} \r>)- \nabla \cdot \l< \bm{Q} \r>$. Taking the divergence of Eq.~(\ref{eq:hydro-num}) and using the incompressibility condition $\nabla \cdot \l< \bs{u} \r>= 0$, we obtain 
\begin{gather}
-\nabla^2 \l< \tilde{P} \r>  + \nabla\cdot\bs{F}= 0. \label{eq:laplacian-p}
\end{gather}
We denote by $[f]_{\bs{k}} = \int f(\bs{x}) e^{-i\bs{k}\cdot \bs{x}}d^2\bs{x}$ the Fourier transform of an arbitrary function $f(\bs{x})$. The Fourier transform of Eq.~(\ref{eq:laplacian-p}) leads to $|\bs{k}|^2[ \l< \tilde{P} \r>  ]_{\bs{k}}+ i \bs{k}\cdot[\bs{F}]_{\bs{k}}=0$. Introducing this relation into the Fourier transform of Eq.~(\ref{eq:hydro-num}), we obtain that $(a+b|\bs{k}|^2) [\l< {\bm{u}} \r>]_{\bs{k}} = (\bm{I}-\frac{\bm{k}\bm{k}}{|\bs{k}|^2}) [\bm{F}(\l< \bm{u} \r>)]_{\bm{k}} $. Thus, $\l< \bm{u} \r>$ can be solved using the following recursion relation
\begin{gather}
[\l< {\bm{u}} \r>^{(m+1)}]_{\bs{k}} = \frac{1}{a+b|\bs{k}|^2} (\bm{I}-\frac{\bm{k}\bm{k}}{|\bs{k}|^2}) [\bm{F}(\l< {\bm{u}} \r>^{(m)})]_{\bm{k}},
\end{gather}
where the superscript $^{(m)}$ denotes the expression evaluated at the $m^\mathrm{th}$ iteration step. The iteration ends when $\max |\l< {\bm{u}} \r>^{(m+1)} - \l< {\bm{u}} \r>^{(m)}| < 10^{-3} \max \big[\l< {\bm{u}} \r>^{(m+1)} \big]$. We set $a=b=3\tilde{\xi}$, and implemented the above numerical scheme in MATLAB.

\noindent{\bf Fitting to experimental data}: We follow previous work \cite{Copenhagen_NP2020} and use $\l< \bm{Q} \r> = \Phi(r/\ell) \begin{pmatrix}
        \cos(2q\phi) & \sin(2q\phi) \\
        \sin(2q\phi) & -\cos(2q\phi)
    \end{pmatrix}$ in Eq.~(\ref{eq:hydro-scaled}) to solve for $\l< {\bm{v}} \r>$.
We treat $\ell$, $\xi_0$,$\epsilon$, and $\zeta_\mathrm{c}$ as fitting parameters. We assume that $\pm1/2$ defects have the same $\ell$, $\xi_0$, and $\epsilon$, but can have different $\zeta_\mathrm{c}$. The values of these fitting parameters are determined by minimizing the root-mean-square deviation between the model and the experiments, which yields $\ell = 2.0~\mathrm{\mu m}$, $\epsilon = 0.35$, $\xi_0 = 6.5~\mathrm{Pa\cdot min/\mu m}$, and $\zeta_\text{c} = 3.8~\mathrm{Pa}\cdot\mathrm{\mu m}$ for $+1/2$ defects and $\zeta_\text{c} = 4.8~\mathrm{Pa}\cdot\mathrm{\mu m}$ for $-1/2$ defects.

\section{Experimental methods}
\label{sec:experiment}

\subsection{Bacterial strains and culture}

\begin{figure*}
  \begin{center}
    \includegraphics[width=1\textwidth]{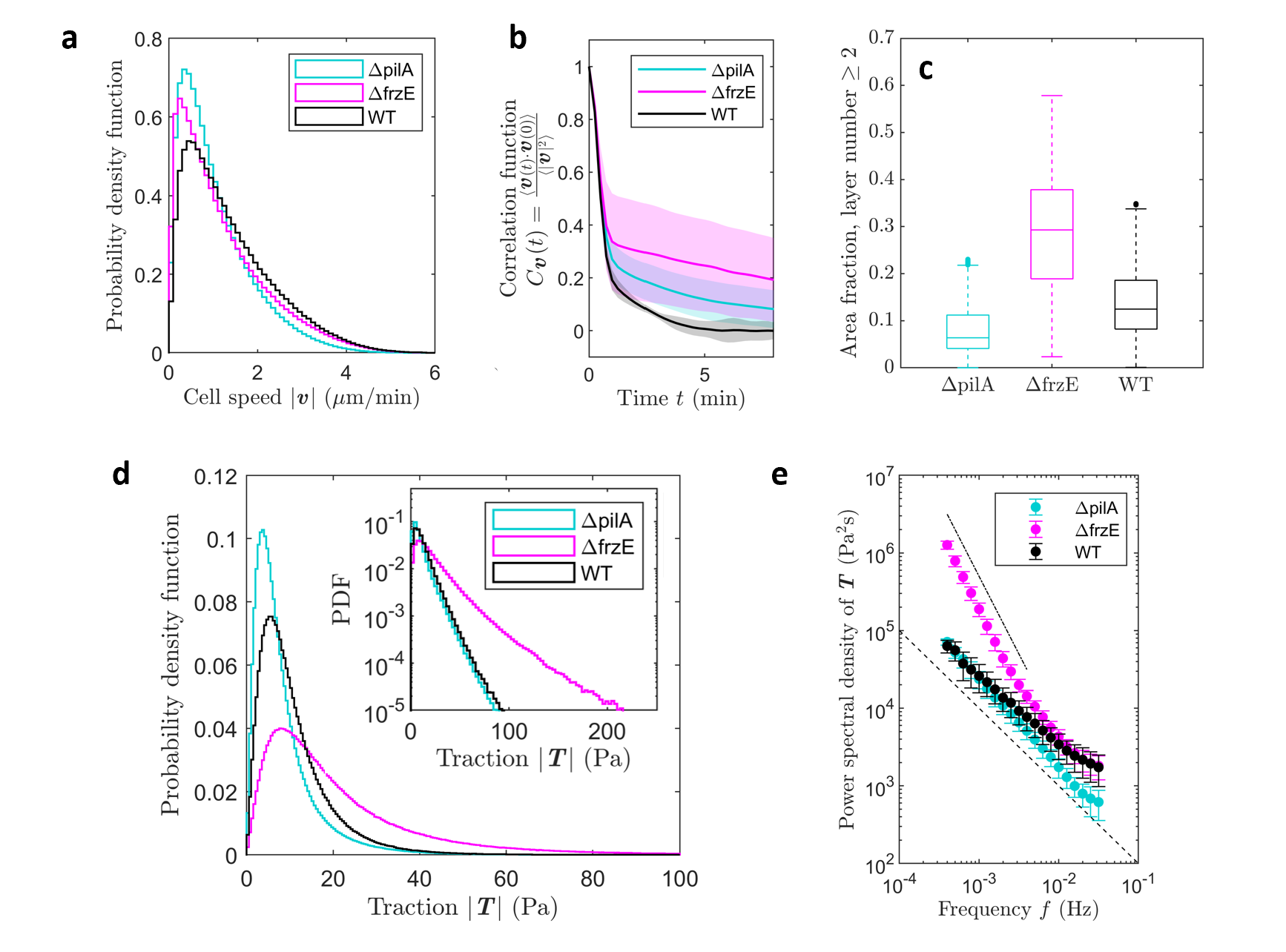}
  \end{center}
  \caption{ \label{Efig:mutants} Compare three different strains: WT, $\Delta$\textit{pilA}, and $\Delta$\textit{frzE}. 
  The WT and $\Delta$\textit{pilA} cells reverse, but the $\Delta$\textit{frzE} cells do not. 
  The WT and $\Delta$\textit{frzE} cells have type-IV pili, but the $\Delta$\textit{pilA} cells do not. 
  In all the experiments, the cells stayed in a nutrient rich environment. 
(a) Probability distribution functions of cell speed $\l| \bm{v} \r|$ in thin colonies. The difference between these three strains was minor. 
(b) Temporal correlation functions of cell velocity $C_v(t) = \dfrac{\l< \bm{v}(t) \cdot \bm{v}(0) \r> }{ \l< | \bm{v} |^2 \r> }$. The solid curves show the means and the shaded areas show the standard deviations. Similar to Fig.~\ref{Efig:polarity_correlation}, we fit each $C_{\bm v}(t)$ to a stretched exponential function $C(t) = \text{exp} \l[ -(t/\tau)^\beta \r]$. The WT cells had the minimum correlation time $\l< \tau \r>$, then it was the $\Delta$\textit{pilA} cells, and they both had shorter $\l< \tau \r>$ than the $\Delta$\textit{frzE} cells. 
(c) Areal ratio of regions where the cell colony was thicker than a monolayer ($n_\text{L} \ge 2$) in the whole field of view of the videos. The $\Delta$\textit{frzE} cells created more areas where the cells aggregated into multiple layers compared to the WT and $\Delta$\textit{pilA} cells. 
(d) Probability distribution functions of traction magnitudes $\l| \bm{T} \r|$ for the three strains. The inset shows the same data in log-linear scales. The difference between the WT and $\Delta$\textit{pilA} cells was marginal, but they both had significantly narrower distributions than the $\Delta$\textit{frzE} strain. 
(e) Power spectral densities (PSD) of traction $\bm{T}$ for all three strains. The PSD of WT approached that of $\Delta$\textit{pilA} in the low-frequency regime, while it approached that of $\Delta$\textit{frzE} in the high-frequency regime. This suggests that the cell reversal controls the low-frequency component of the mechanical stress in the colonies and the pili only affect the high-frequency regime. } 
\end{figure*}

Wild type (WT) \textit{M. xanthus} cells have two types of motility: social (S) motility and adventurous (A) motility. 
The former is driven by the type IV pili \cite{Kaiser_TFP} and the latter by the gliding motors \cite{Zusman_gliding}. 
A strain with both S-motility and A-motility is called A$^+$S$^+$. 
Besides WT, we used various genetically modified strains in the experiments for different purposes. 
To simplify the intercellular and cell-substrate interactions, we used a $\Delta$\textit{pilA} mutant, which does not make any type IV pili and thus has no S-motility (A$^+$S$^-$). 
This mutant has simpler cell-cell and cell-substrate interactions compared to WT -- active force is only generated via direct contact and each cell can only interact with the substrate or its neighbors. 
In general, A$^+$S$^-$ cells' behaviors in a thin layer are very similar to WT (see Fig.~\ref{Efig:mutants}), so we used an A$^+$S$^-$ strain to obtain the data in Figs.~1, 2, 4, and 5 (reversing) of the main text. 
To test the effect of cell reversal in Fig.~5 (main text), as the non-reversing cells we used a $\Delta$\textit{frzE} mutant (A$^+$S$^+$), which reverses the direction of motion much less frequently compared to WT and $\Delta$\textit{pilA} cells in a nutrient-rich environment. 
Figure~\ref{Efig:mutants} shows a comparison of the WT, $\Delta$\textit{pilA}, and $\Delta$\textit{frzE} strains, demonstrating that the increased polarity, layering, and stress we see in the non-reversing mutant is not a function of the presence of pilus-driven S-motility in the $\Delta$\textit{frzE} strain.
To measure the cell polarity in Figs.~3 and 4 (main text), we used a \textit{mglB}$::$\textit{mVenus} strain with fluorescent labels on the MglB protein, which localizes to the lagging pole of the cell. 
These \textit{mglB}$::$\textit{mVenus} cells are A$^+$S$^+$ and their behaviors are very close to WT.  
The strains and their properties are summed up in Table~\ref{tab:cells}. 

\begin{table*}[!ht]
\begin{center}
 \begin{tabular}{|c|c|c|c|c|c|} 
 \hline
 Strain & ~Motility~ & ~With pilus?~ & ~Reversal?~ & ~Fluorescent label~ & ~Reference \\ 
 \hline
 WT (DK1622) & A$^+$S$^+$ & Yes & Yes & No & \cite{Copenhagen_NP2020} \\ 
 \hline
 $\Delta$\textit{pilA} & A$^+$S$^-$ & No & Yes & No & \cite{Sabass_PNAS2017} \\ 
 \hline
 $\Delta$\textit{frzE} & A$^+$S$^+$ & Yes & No & No & \cite{frzE} \\ 
  \hline
 \textit{mglB}$::$\textit{mVenus} & A$^+$S$^+$ & Yes & Yes & MglB & \cite{Lotte_2022} \\ 
 \hline
 
\end{tabular}
\end{center}
\caption{ \textit{M. xanthus} strains used in the experiments. }
\label{tab:cells}
\end{table*}

To grow the cells from frozen stock, we plated them onto $1.5\%$ agar pads in CTTYE ($1\%$ casitone, 10~mM Tris–HCl, 1~mM KH$_2$PO$_4$, 8~mM MgSO$_4$, and adjusted its pH to 7.6) and kept the plates in a $32^\circ$C incubator for at least three days. 
Then we picked some cells from a colony on the agar plate and transferred them into 10~ml CTTYE solution in a flask to make a liquid culture. 
The flask was kept at $32^\circ$C overnight with shaking. 
The optical density (OD) of the liquid culture was measured before each experiment, and we only used cell cultures in the exponential phase. 


\subsection{Preparation of samples for traction force microscopy (TFM assay)}
The samples for traction force microscopy (TFM) experiments were prepared using 35~mm diameter Petri dishes with a 0.17~mm thick glass bottom (Thermo Fisher). 
The surface of the glass inside the Petri dish was treated with 3-(Trimethoxysilyl)propyl methacrylate (TMSPMA) so that it combined tightly to the polyacrylamide (PAA) hydrogel \cite{Herrick_3TPM}. 
The protocol is: 
\begin{enumerate}
\item Plasma clean the glass surface. 
\item Mix $2\%$ (volume fraction) TMSPMA with $98\%$ (volume fraction) $95\%$ ethanol and adjust its pH to 5.0 with glacial acetic acid. 
\item Add 1~ml TMSPMA solution in each Petri dish and remove it after soaking for 2 minutes. 
\item Wash each Petri dish three times with pure ethanol and dry it at room temperature for 15 minutes. 
\end{enumerate}

The formulas for making PAA gel at three different stiffnesses are listed in Table~\ref{tab:PAA_solution}. 
We always prepared our gel in two steps: first, we made a PAA stock following the formula in Table~\ref{tab:PAA_stock}, and then we made PAA solution using the corresponding stock. 
In the main text, all the data were obtained with PAA substrates with a 430~Pa shear modulus. 
Fluorescent beads (FluoSpheres Carboxylate-Modified Microspheres, yellow-green fluorescent 505/515, 2$\%$ solids) with a 110~nm diameter were distributed uniformly in the PAA solution to measure the horizontal and vertical deformations at the gel surface. 
When making the PAA substrate, we put 15~$\mu$l PAA solution in the middle of each Petri dish. 
Then we put a glass coverslip (Thermo Fisher 12CIR-1, 12 mm diameter) on top of the droplet and waited half an hour for the PAA to gelate. 
The 12CIR-1 coverslips were treated with water repellent in advance to make their surfaces more hydrophobic. 

\begin{table*}[!ht]
\begin{center}
 \begin{tabular}{|c|c|c|c|c|c|} 
 \hline
 Shear modulus (Pa) & ~PAA Stock (ml)~ & ~Water (ml)~ & ~ Bead suspension ($\mu$l) ~ & ~$10\%$ APS ($\mu$l)~ & ~Temed ($\mu$l)~ \\ 
 \hline
 230 & 0.03 & 0.2185 & 10 & 1.25 & 0.375 \\ 
 \hline
 430 & 0.05 & 0.1985 & 10 & 1.25 & 0.375 \\ 
 \hline
 1500 & 0.075 & 0.1735 & 10 & 1.25 & 0.375 \\ 
 \hline
\end{tabular}
\end{center}
\caption{Formula for making 0.25~ml PAA hydrogels using the PAA stocks. APS stands for ammonium persulfate solution in water. }
\label{tab:PAA_solution}
\end{table*}

\begin{table*}[!ht]
\begin{center}
 \begin{tabular}{|c|c|c|c|} 
 \hline
 Shear modulus (Pa) & ~$40\%$ Acrylamide (ml)~ & ~$2\%$ Bis (ml)~ & ~Water (ml)~ \\ 
 \hline
 230 & 3.13 & 1.25 & 0.63 \\ 
 \hline
 430 & 3.13 & 0.63 & 1.25 \\ 
 \hline
 1500 & 3.13 & 0.42 & 1.46 \\ 
 \hline
\end{tabular}
\end{center}
\caption{Formula for making 5~ml PAA stocks. }
\label{tab:PAA_stock}
\end{table*}

After gelation, we removed the coverslip and submerged the PAA pad in $66~\mu$g/ml chitosan solution for at least 45 minutes.
The chitosan solution for coating was prepared following these steps: 
\begin{enumerate}
\item Make 0.2~M Acetic acid (0.12~ml glacial acetic acid to 9.88~ml DI water). 
\item Dissolve 10~mg chitosan in 3~ml 0.2~M acetic acid. 
\item Make a 1 to 50 dilution of the chitosan solution prepared in Step 2. 
\end{enumerate}
Then we removed the chitosan solution, washed the PAA pad once with sufficient DI water, soaked the pad in sufficient CTTYE for about 10 minutes, removed the CTTYE, and soaked the pad in fresh CTTYE for another 10 minutes. 
After soaking, we removed CTTYE and dried the excess liquid with paper tissue.


To make the cell monolayer, we used overnight liquid cultures of \textit{M. xanthus} in CTTYE shaken at 32$^\circ$C. 
We separated the cells by centrifuging the liquid culture and resuspended them in DI water so that the cell concentration was $8 \times$ the concentration of OD $= 1$ (550~nm wavelength). 
The cells were dispersed using a pipette for at least 1 minute and then shaken on a vortex for at least 30 seconds. 
Lastly, we added a 1.5~$\mu l$ droplet of this concentrated cell suspension onto the surface of the PAA gel, and let it settle for several minutes. 

To prevent dehydration of the gel during imaging, we put a 2~mm thick acrylic spacer (cut by a laser cutter) in each Petri dish and put a $22 \times 22$~mm cover glass on top of the spacer to create a chamber above the substrate and cells. 
The glass did not touch the gel, so at the air-gel interface, it was gel, cells, and air from bottom up. 
Outside the chamber, the edges of the spacer and the cover glass were sealed with Valap ($1/3$ vaseline, $1/3$ lanolin, and $1/3$ paraffin by mass). 
After preparation, we kept the sealed samples in a $32^\circ$C incubator for about two hours, took them out, and kept them all at room temperature during the imaging session.

\subsection{Preparation of samples for polarity measurement (Polarity assay)}
To measure cell polarity, we used the \textit{mglB}$::$\textit{mVenus} strain, which has fluorescent MglB proteins at the lagging cell pole. 
Due to the small size of these fluorescent spots, we needed the cell layer to be very flat so that we could capture the polarity of as many cells as possible. 
As a result, to get images with high enough quality for the analysis, we measured cell polarity using a $1.5\%$ agarose substrate (shear modulus $\approx 50$~kPa), which was much stiffer than the 430~Pa PAA in the TFM assay. 
Each sample sat on a 25~mm $\times$ 75~mm $\times$ 1~mm glass slide. 
We laser cut 1~mm thick rubber spacers with an elliptical hole in the middle and put one on each glass slide. 
We then dissolved 300~mg agarose powder in 20~ml CTTYE by heating it up, filled each hole with the solution, and put a 22~mm $\times$ 40~mm $\times$ 0.15~mm cover glass on top to flatten the surface. 
It took several minutes for the agarose to cool down and solidify. 
Then we removed the cover glass, and on each pad, put 10~$\mu$l liquid culture of \textit{mglB}$::$\textit{mVenus} cells at $2.5 \times$ the concentration of OD $= 1$. 
We waited several minutes until there was no visible liquid on the gel surface and put a 22~mm $\times$ 40~mm $\times$ 0.17~mm cover glass on top of the spacer and the gel. 
We then sealed the edges with Valap and incubated the samples for two hours at 32$^\circ$C.

\subsection{Imaging}

The images were taken with a commercial Nikon Ti-E microscope with the Perfect Focus System (PFS) and Yokogawa CSU-21 spinning disk confocal. 
We used a $60 \times$ Plan Apochromat Water Immersion objective (Nikon, NA 1.20, working distance 0.27~mm) and an Andor Zyla camera. 
The samples were placed in a humidifying chamber, and we kept the temperature $25^\circ$C. 
For TFM, the objective touched the bottom of the Petri dish and we imaged the gel-cell-air interface from the bottom up, through the substrate. 
The incident laser beam (488~nm) went through the sample from below while the white light was emitted above the sample. 
With each sample, we took both time series and $z$-stack images. 
For the time series, we placed the imaging plane right at the surface of the substrate, so that we saw the cells and the fluorescent particles simultaneously. 
In each acquisition, we took one bright field image of the cells with white light and one fluorescence image of the fluorescent particles. 
The time between adjacent acquisitions was 15~s. 
For the $z$-stack, we moved the focal plane in the $z$ direction (perpendicular to the gel surface) from below the gel surface to above. 
At each $z$, we took one bright field image and one fluorescence image. 
In most experiments, the step size between adjacent $z$ slices was $0.2~\mu$m. 
In the others, we used a $2~\mu$m step size to scan across a larger range of $z$. 

For the polarity measurement, the images were taken through the 0.17~mm thick cover glass. 
We used white light to take bright field images of the cells and 488~nm laser to image the fluorescently labeled MglB proteins. 
We define the $z$ position of the imaging plane that cuts through the cell bodies as $z \equiv 0~\mu$m. 
In each acquisition, we took one fluorescence image at $z = 0~\mu$m and three bright field images at $z = -0.9~\mu$m, $0~\mu$m, and $0.9~\mu$m. 
The time between adjacent acquisitions was 15~s.

\vspace{10mm}

\section{Measuring director and velocity fields from images of cells}
\label{sec:velocity}

\subsection{Processing bright field images of the cells}
We used the bright field images to obtain the director field and cell flow velocity. 
We started with some basic processing steps, including removing the slowly varying background and adjusting the contrast. 
We also removed the slow global drift measured using the fluorescence images as described in Section~\ref{sec:laser_process}. 
An example of the processed bright field images is shown in Fig.~\ref{SI_findHoles}a.

\subsection{Detecting holes in cell layers}

\begin{figure}
    \begin{center}
        \includegraphics[width = 1\textwidth]{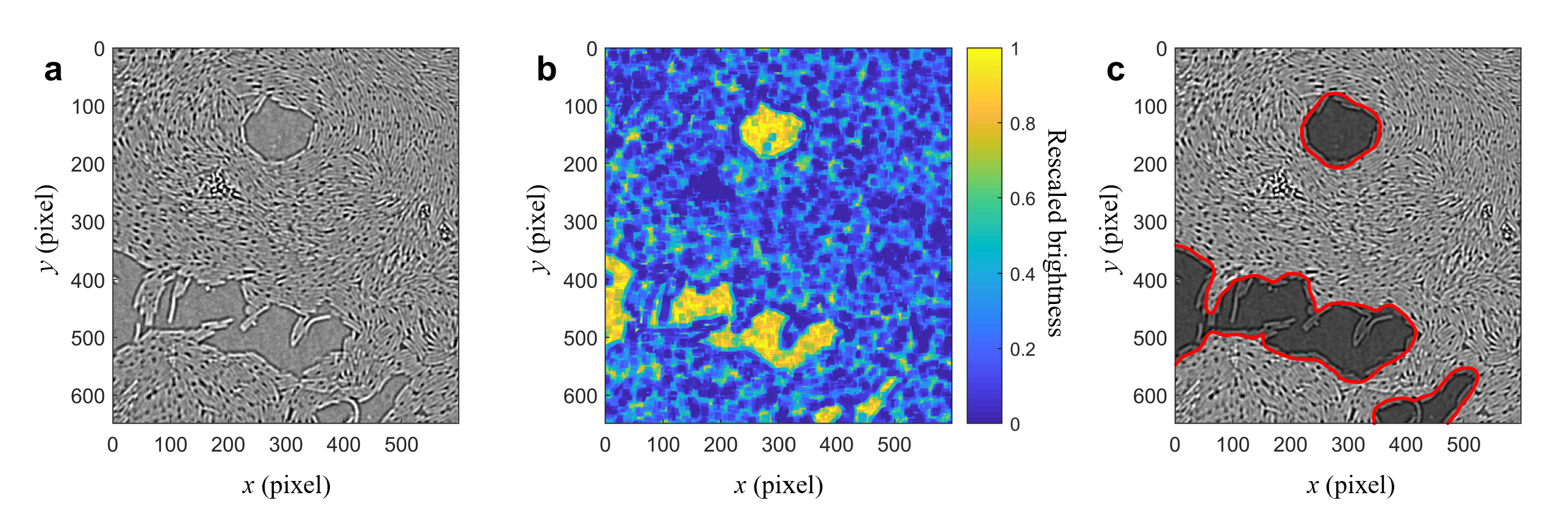}
    \end{center}
\caption{ Detecting holes in the bacterial colony. This example was imaged at $60 \times$ magnification, and the length scale is 0.11~$\mu$m per pixel. (a) Bright field images of the cells. (b) Rescaled brightness $I_\text{diff}$ (Eq.~\ref{eq:I_diff}). (c) Final result: the holes are highlighted with darker gray and red edges. 
}
\label{SI_findHoles}
\end{figure}

We used the local maximum difference in brightness to separate holes (bare gel surface) from the cell layer. 
In a region with cells, this difference is more substantial than for the bare gel surface (holes in the cell layer), where the brightness is more uniform. 
For each pixel, we calculated the maximum and minimum brightness, $I_\text{local, max}$ and $I_\text{local, min}$, respectively, within a $10 \times 10$~pixel box centered at this pixel. 
Then we rescaled the brightness difference using 
\begin{equation}
    I_\text{diff} = 1-\text{normalize}(I_\text{local, max} - I_\text{local, min}), 
    \label{eq:I_diff}
\end{equation}
where ``normalize'' means adjusting the brightness of the image $(I_\text{local, max} - I_\text{local, min})$ so that it ranges between 0 and 1. 
The rescaled image $I_\text{diff}$ enhances the difference between cell layers and holes as shown in Fig.~\ref{SI_findHoles}b. 
We then set a threshold and binarized $I_\text{diff}$ to distinguish holes from cell layers. 

In this project, we care more about the regions where there is a densely packed cell layer and the $\pm 1/2$ topological defects in these regions. 
However, sometimes our algorithm recognized some features inside or on the edges of the holes as $\pm 1/2$ defects, and such defects were to be excluded from our analysis. 
As a result, we dilated the regions recognized as holes so that their boundaries slightly extended beyond the actual edges of the holes. 
This allowed us to only keep the $\pm 1/2$ defects located inside the cell layer. 
We also excluded the holes with an area smaller than 900 pixels (about 11~$\mu \text{m}^2$). 
Consequently, when the cell layer was slightly cracked open in a small region due to a fluctuation in cell concentration, it was still counted as a cell layer. 
Lastly, we smoothed the edges of the holes both in space and time using a Gaussian filter. 
The detected holes are labeled in Fig.~\ref{SI_findHoles}c.

\subsection{Nematic order}

\begin{figure}[!t]
    \begin{center}
        \includegraphics[width=0.9\textwidth]{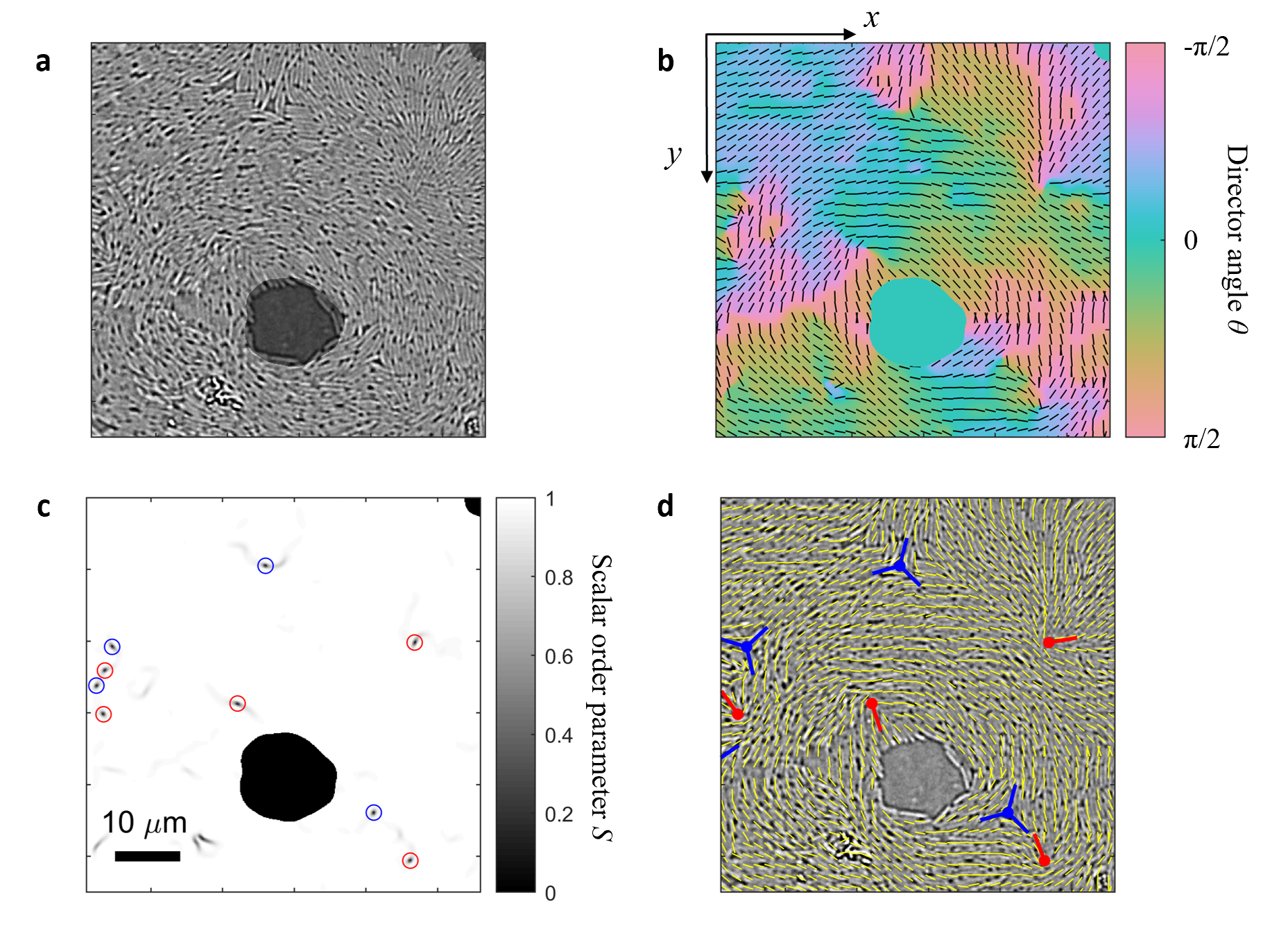}
    \end{center}
\caption{ Measuring director field $\theta$ and detecting $\pm 1/2$ defects. 
(a) Bright field image of the cells, with the holes shaded in a darker color. 
(b) Director field $\theta$ showing the orientation of the cells labeled by both the color map and the short black lines. The director is horizontal when $\theta = 0$ and $\theta$ increases when the line rotates clockwise. 
(c) Scalar order parameter $S$. We label the regions recognized as defects with circles: red circles represent $+1/2$ defects and blue circles represent $-1/2$ defects. The holes are excluded. 
(d) Directors (yellow lines) overlaid on the bright field image of the cells. The locations and orientations of the topological defects are labeled using the same color as in (c). 
The scale bar in (c) is 10~$\mu$m and applies to all panels.
}
\label{SI_findDefect}
\end{figure}

To characterize the nematic order within a cell layer, we first measured the cell orientation angle $\theta_{ij}$ following \cite{Li_PNAS2019, Copenhagen_NP2020} using the pre-processed bright field image. 
In an image, the brightness of the pixel in row $i$ and column $j$ is $I_{ij}$. 
We calculated the Hessian matrix $H_{ij}$ pixel by pixel, where  
\begin{equation}
    H_{ij} = 
    \left( 
    \begin{array}{cc} \left( \partial I_{ij} / \partial x \right)^2 & \left( \partial I_{ij} / \partial x \right)\left( \partial I_{ij} / \partial y \right) \\ \left( \partial I_{ij} / \partial x \right)\left( \partial I_{ij} / \partial y \right) & \left( \partial I_{ij} / \partial y \right)^2 \end{array} 
    \right), 
    \label{eq:structure_tensor}
\end{equation}
and smoothed each element in the matrix by means of a Gaussian filter with standard deviation $\sigma = 10$~pixels (1.1~$\mu$m for $60 \times$ magnification). 
The eigenvector associated with the smallest eigenvalue of $H_{ij}$ gives the local direction of the smallest brightness gradient, which was taken to represent the cell orientation.
An exemplary director field is shown in Fig.~\ref{SI_findDefect}b, based on the bright field image in Fig.~\ref{SI_findDefect}a.

\subsection{Detection and tracking of topological defects}

From the measured $\theta_{ij}$, we obtained the scalar order parameter $S$: 
\begin{equation}
    S_{ij} = 
    \sqrt{ \l< \text{cos}(2\theta_{ij}) \r>_R^2 + \l< \text{sin}(2\theta_{ij}) \r>_R^2 }, 
    \label{eq:strength_field}
\end{equation}
where $\left<~\right>_R$ represents averaging within a disk of radius $R = 5$~pixels (0.55~$\mu$m for $60 \times$ magnification) centerd at the pixel $(i,j)$. 
As shown in Fig.~\ref{SI_findDefect}c, $S_{ij}$ vanishes near the cores of defects and is approximately 1 everywhere else.
For each point identified as a potential defect core, we calculated the topological charge $q$ by applying its definition:
\begin{equation}
    q = \frac{1}{2 \pi} \oint_\mathcal{C} \text{d}\theta, 
    \label{eq:charge}
\end{equation}
where $\mathcal{C}$ is a circular circuit of radius $R_c = 6$~pixels ($0.66~\mu$m at $60 \times$) around the defect core. 
We ignored the candidate points with $q=0$. 
For each defect, we identified its axes of symmetry using the method in Refs.~\cite{Copenhagen_NP2020,Vromans_2016}.
Specifically, the axis of a $+1/2$ defect is given by 
$\hat{p}_\alpha = \partial_\beta Q_{\alpha \beta} / |\partial_\beta Q_{\alpha \beta}|$, where $Q$ is the nematic order parameter tensor with components 
\begin{equation}
    Q_{\alpha \beta} = S \left[ 2 \hat{n}_\alpha \hat{n}_\beta - \delta_{\alpha \beta} \right], 
\end{equation}
where $\hat{\textbf{n}} = (\text{cos} \theta,~\text{sin} \theta)$.
Similarly, for $-1/2$ defects, we first defined the opposite nematic angle $\theta' = -\theta$, obtained the corresponding $Q'$ tensor, and then calculated $\hat{p}'_\alpha = \partial_\beta Q'_{\alpha \beta} / |\partial_\beta Q'_{\alpha \beta}|$. 
Finally, defining $\hat{\textbf{p}}' = (\text{sin} \psi',~\text{cos} \psi')$, one symmetry axis of a $-1/2$ defect is given by the angle $\psi = -\psi'/3$. 
The other two symmetry axes of a $-1/2$ defect were then found by its three-fold symmetry.

We tracked the defects' motion using Blair and Dufresne's particle tracking velocimetry (PTV) code \cite{PTV}. 
We ignored defects that exist for shorter than eight frames (2 minutes). 
We also ignored defects in holes, and those within 50 pixels of the edges of the images. 
Figure~\ref{SI_findDefect}d shows the director field obtained from the bright field image, and the locations and orientations of the detected $+1/2$ (red) and $-1/2$ (blue) defects. 
One can see that some of the singular points circled out in Fig.~\ref{SI_findDefect}c are not shown in Fig.~\ref{SI_findDefect}d, because they did not last long enough and thus were excluded.

\subsection{Measuring velocity and cell flows}
\begin{figure}
    \begin{center}
        \includegraphics[width=\textwidth]{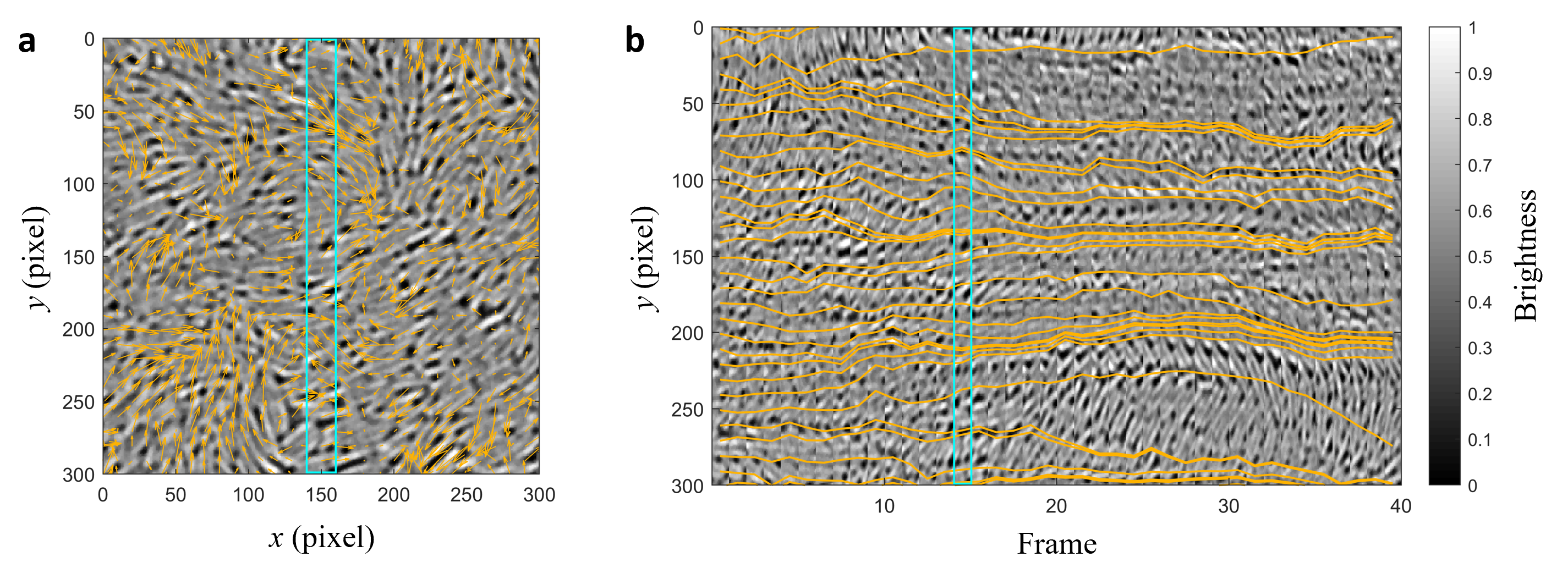}
    \end{center}
\caption{ Using optical flow to measure the velocity of cell flow. (a) Processed bright field image of the cells with the orange velocity vectors overlaid. (b) Kymograph obtained by stitching together the same rectangular region (surrounded by lines in cyan in (a)) in the bright field image series in chronological sequence. This selected region is 300 pixels tall and 10 pixels wide shown by the cyan rectangle in (a) and (b). The cyan rectangular regions in (a) and (b) are identical. The overlaid orange curves are the time integral of the $y$ velocity measured at different $y$ positions. }
\label{SI_OpticalFlow}
\end{figure}

We used the method of optical flow to measure cell velocity. 
After processing the original bright field images of the cells, we used the MATLAB (R2019a) function opticalFlowFarneback() to obtain the velocity field. 
Figure~\ref{SI_OpticalFlow}(a) shows an exemplary image of the cells with the measured velocity vectors overlaid. 
The parameters for the optical flow were calibrated using images of a \textit{frzS}$::$\textit{GFP} strain. 
The fluorescently labeled FrzS proteins generated bright spots at both poles of the cells, which allowed us to measure the same cell flow fields with two different methods: using PIV for the fluorescent images and optical flow for the bright field images. 
The two methods provided the same measurements in most regions. 
In the end, we calculated the displacement by integrating the velocity over time and overlaying the displacement curves on top of the kymograph made from the bright field images (Fig.~\ref{SI_OpticalFlow}(b)). 
These orange displacement curves nicely follow the up-and-down movement of the speckles in the kymograph.

\vspace{10mm}

\section{Measuring cell polarity}
\label{sec:polarity}

\begin{figure}
    \begin{center}
        \includegraphics[width=1\textwidth]{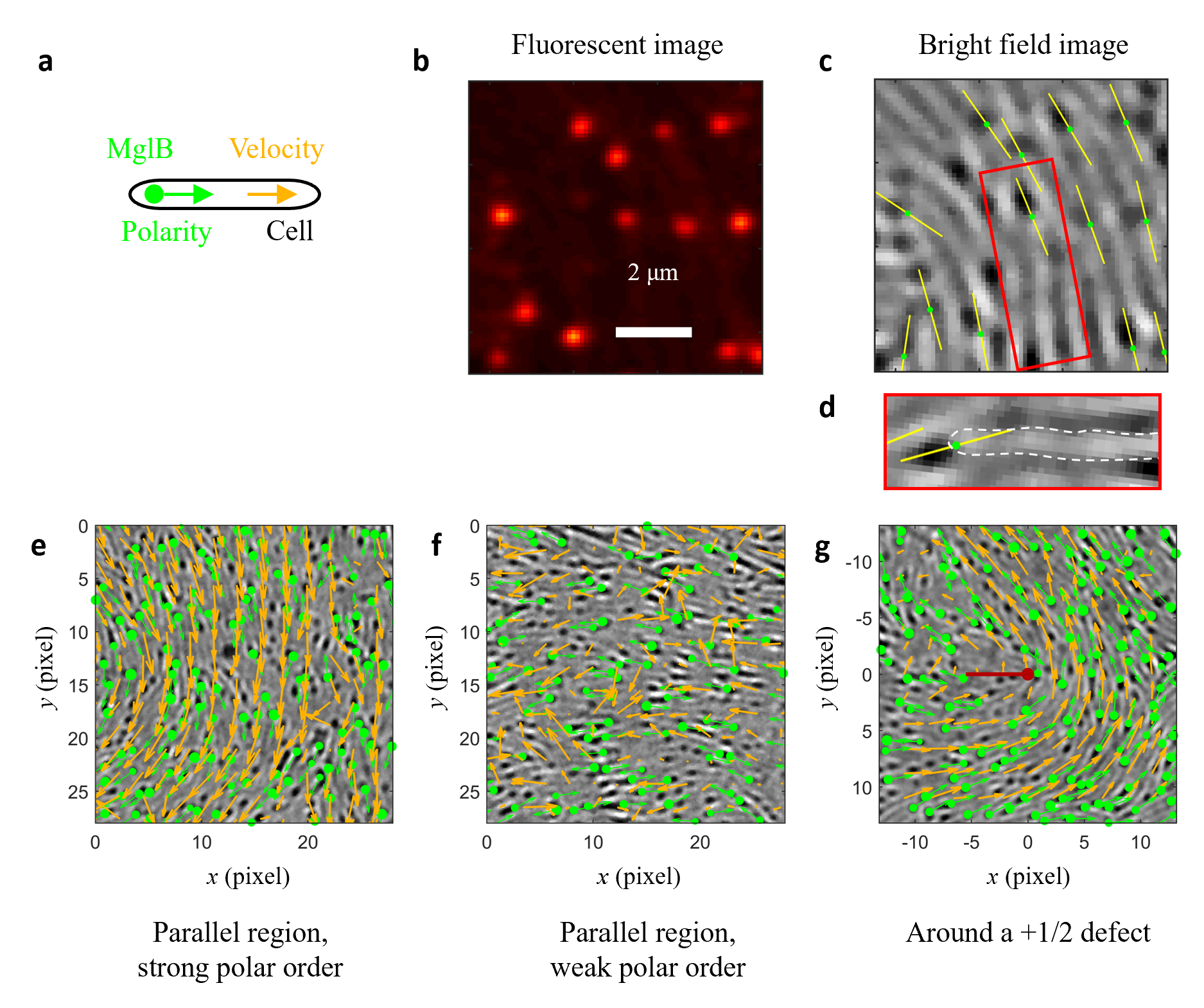}
    \end{center}
\caption{ Cell polarity measurement. 
(a) A schematic illustration of the relationship between cell velocity and the location of the fluorescently labeled MglB protein for a single cell. 
(b) Fluorescence image of the labeled MglB proteins. The scale bar represents $2~\mu$m. 
(c) Bright-field image of the cells in (b) with the positions of the MglB labels (green dots) overlaid. The yellow lines show the local directors $\bm{\hat{n}}$. 
(d) Reoriented region inside the red box in (c) with one cell body outlined by the white dashed line. 
(e-g) Polarity (green) and velocity (orange) vectors overlaid on bright-field images. The green dots show the MglB labels and their sizes represent the sizes of the bright MglB spots in the corresponding laser images. The Three types of flow: (e) ordered region with cells moving in the same direction; (f) ordered region with cells moving in opposite directions; (g) near a $+1/2$ defect. 
}
\label{SI_Polarity_Velocity}
\end{figure}

\subsection{Single-cell polarity vs. velocity}
The \textit{mglB}$::$\textit{mVenus} strain that we used was generated and characterized by Szadkowski \textit{et al.} \cite{Lotte_2022}. 
We tested the strain using samples with sparse cells on the surface, and indeed, the direction of motion of a single cell was highly correlated with the position of the MglB protein: MglB localized to the lagging pole of the cell (Fig.~\ref{SI_Polarity_Velocity}a).

\subsection{Data processing for cell layers}
Combining the fluorescence images of the fluorescent labels and the bright-field images of the cells, we measured cell polarity $\bm{p}$. 
One way to achieve this is to perform cell segmentation and determine the location of MglB within each cell. 
However, this was difficult with our bright-field images, so we used another method that measured the local polarity without cell segmentation.  
First, we processed the fluorescence images with a band-pass filter to denoise and enhanced their contrast, then we located the centers of the fluorescent MglB labels (Fig.~\ref{SI_Polarity_Velocity}b). 
Then, we measured the local directors at the locations of these centers, as shown by the yellow lines in Fig.~\ref{SI_Polarity_Velocity}c. 
As Fig.~\ref{SI_Polarity_Velocity}d shows, since MglB always appeared close to a cell pole, for each MglB focus (green dot), we compared the brightness in the bright-field image along the corresponding yellow line. 
Because the gap between adjacent cells had lower brightness, the brighter side indicated the inside of the cell, and the darker side indicated the outside. 
Since MglB localized to the lagging pole of the cell, we obtained the polarity vector $\bm{p}$ pointing along the direction of the yellow line toward the brighter side.

As shown in Fig.~\ref{SI_Polarity_Velocity}e-g, in each frame, we obtained the cell polarity (green arrows) and velocity (orange arrows, see SI Sec.~\ref{sec:velocity}) simultaneously. 
When comparing cell polarity and velocity, we focused on two special cases: ordered regions where the cells were approximately parallel to each other (Fig.~\ref{SI_Polarity_Velocity}e and f) and regions around $+1/2$ defects (Fig.~\ref{SI_Polarity_Velocity}g). 
In ordered regions, we observed both polar flows and nematic flows, meaning the cells moving in the same direction (Fig.~\ref{SI_Polarity_Velocity}e) and opposite directions (Fig.~\ref{SI_Polarity_Velocity}f), respectively. 
To quantify the local polar order, we reoriented the cells to become horizontal, as shown in Fig.~\ref{Efig:polarity_velocity}a. 
We first calculated the mean director $\l< \bm{\hat{n}} \r>$, and then rotated the image such that $\l< \bm{\hat{n}} \r>$ became horizontal. 
Then we took the horizontal components of the reoriented polarity $p_\text{h}$ and velocity $v_\text{h}$ inside a $12 \times 12~\mu$m$^2$ square boxes (two cell lengths) and obtained $p_n = \l< p_\text{h} \r>$ and $v_n = \l< v_\text{h} \r>$, where $\l<~\r>$ denote spatial average inside the box. 
Note that $\l< \bm{\hat{n}} \r>$ has two identical ends, so in the aligned regions, even though we have a cell orientation, we cannot define which direction is positive or negative. 
As a result, we combined the data $(p_n,v_n)$ and $(-p_n,-v_n)$ in Fig.~1f of the main text and in Fig.~\ref{Efig:polarity_velocity}a.   
In an area around a $+1/2$ defect, we reoriented the cells so that the ``comet head'' was on the right-hand side and the ``comet tail'' on the left, and the tail became horizontal, as shown in Fig.~\ref{Efig:polarity_velocity}b. 
We calculated $p_n = \l< p_\text{h} \r>$ and $v_n = \l< v_\text{h} \r>$ in the same way, but now the square box is explicitly chosen in the comet tail region. 
Since the director field of the $+1/2$ defect breaks the spatial symmetry, now we can define the positive direction as pointing to the right, as indicated by the arrow in Fig.~\ref{Efig:polarity_velocity}b. 


\begin{figure*}
  \begin{center}
    \includegraphics[width=0.95\textwidth]{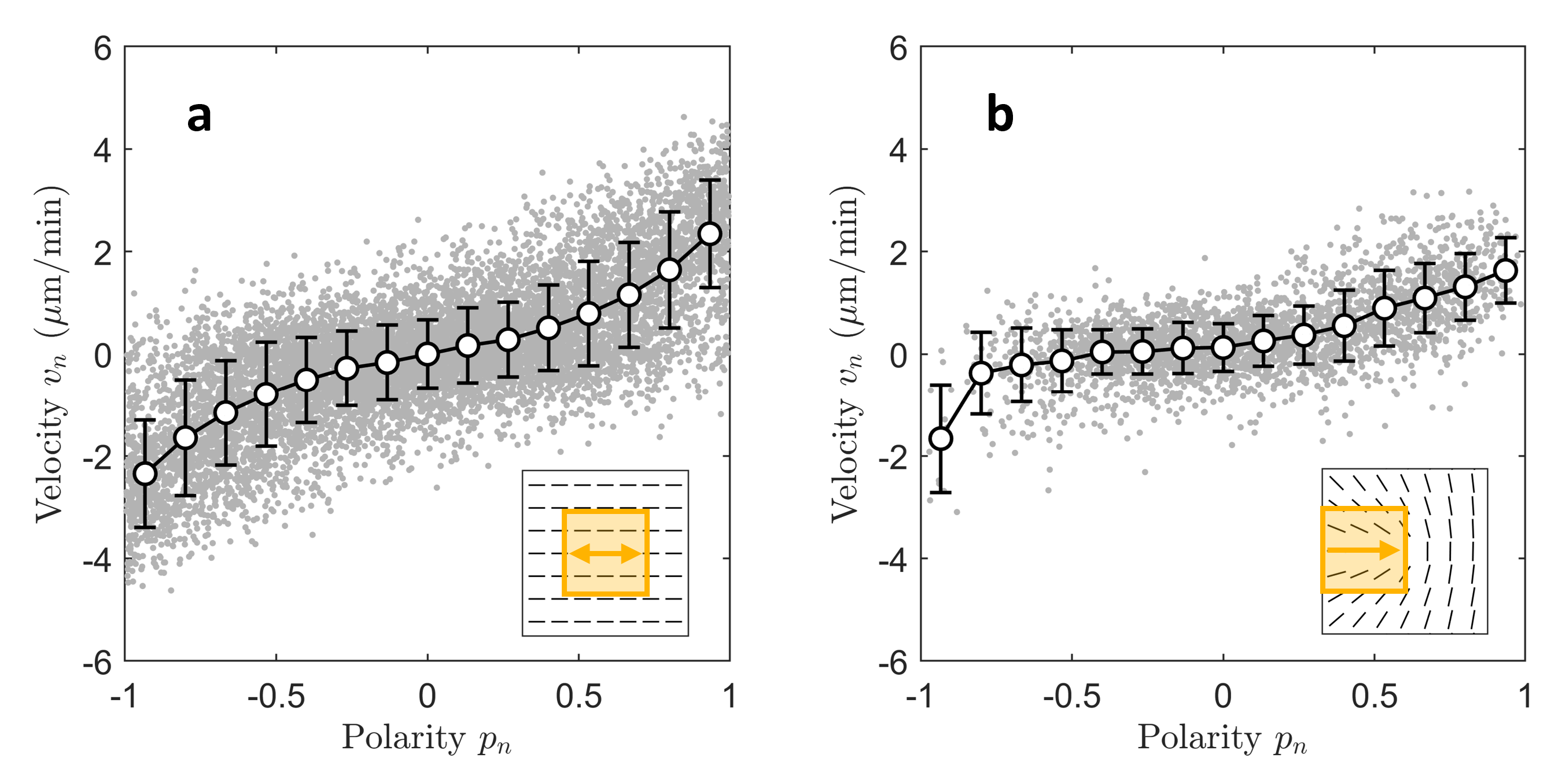}
  \end{center}
  \caption{ \label{SI:polarity_velocity} Relationship between polarity and velocity in ordered regions and near $+1/2$ defects obtained with the polarity assay. 
(a) Relationship between the local mean velocity $v_n$ and polarity $p_n$ in ordered regions, which were calculated in $12 \times 12~\mu$m$^2$ boxes (orange square in the inset). First, we reoriented the area so that the cells aligned horizontally. Then we used the horizontal component of velocity $v_\text{h}$ and polarity $p_\text{h}$ to obtain $v_n = \l< v_\text{h} \r>$ and $p_n = \l< p_\text{h} \r>$, where the mean was calculated within the square. Each gray point in the plot was obtained with one of such squares. Note that the directors $\l< \bm{\hat{n}} \r>$ and $-\l< \bm{\hat{n}} \r>$ are equivalent, so here we show both $(p_n,v_n)$ and $(-p_n,-v_n)$. The black circles and curve show the mean $v_n$ at different $p_n$, and the error bars show the corresponding standard deviation. 
(b) Relationship between the local mean velocity $v_n$ and polarity $p_n$ in the tail region of $+1/2$ defects (orange square in the inset, also $12 \times 12~\mu$m$^2$ in area). The symbols are the same as in (a). Similar to the aligned regions, here we reoriented the areas so that they look like the inset, and then we calculated $v_n = \l< v_\text{h} \r>$ and $p_n = \l< p_\text{h} \r>$. Different from the ordered case, here the defect breaks the spatial symmetry, so we defined a positive direction: $v_n$ and $p_n$ are positive when pointing to the right and negative when pointing to the left based on the orientation of the inset. Besides the orange square, we tried a triangular area in the tail region with the same area when performing these calculations, and the results showed no qualitative difference. 
} 
\end{figure*}

\section{Measuring traction from fluorescence images}
\label{sec:TFM_method}

The fluorescence images captured the fluorescent particles close to the surface of the gel. 
We used particle imaging velocimetry (PIV) to track their motion and obtained the displacement field of the substrate in the $x$-$y$ plane, which is parallel to its surface. 
Then we used the displacement field to reconstruct the traction field. 
In Section~\ref{sec:normal_deformation}, we discuss measuring the deformation of the substrate in the direction normal to its surface (the $z$ direction).

\subsection{Pre-processing fluorescence images}
\label{sec:laser_process}

\begin{figure}[t]
    \begin{center}
        \includegraphics[width=1\textwidth]{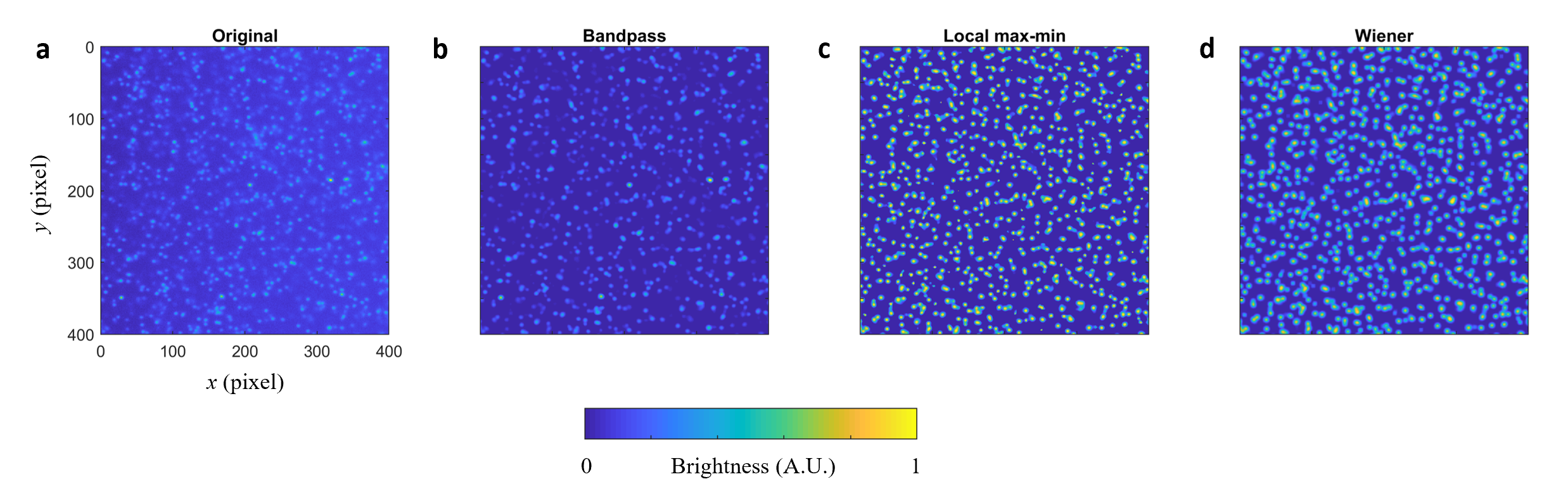}
    \end{center}
\caption{ Images of the 110~nm fluorescent particles at $60 \times$ magnification before (a) and after each step of processing: bandpass filter (b), local max-min filter (c), and Wiener filter (d). The length scale is 0.11~$\mu$m/pixel. 
}
\label{SI_Filters}
\end{figure}

\begin{figure}
    \begin{center}
        \includegraphics[width=1\textwidth]{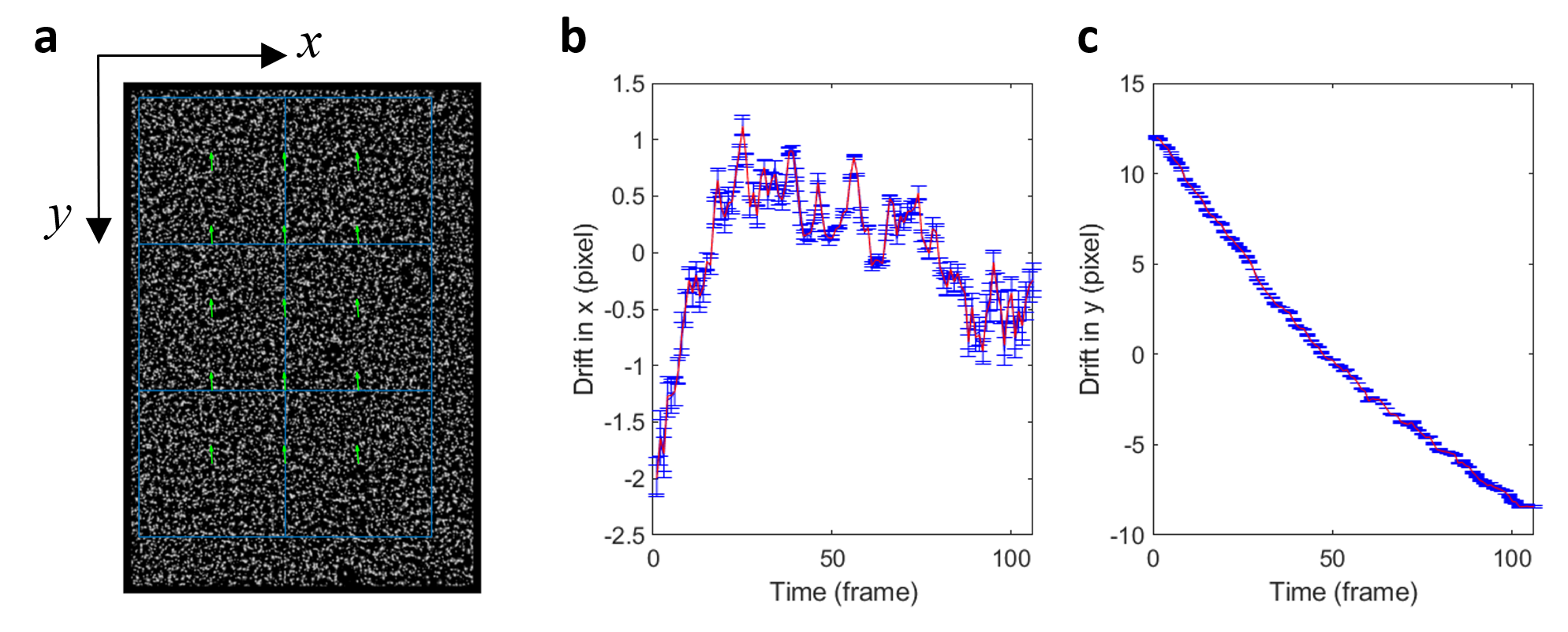}
    \end{center}
\caption{ Global drift in $x$ and $y$ directions of an exemplary video. 
(a) Displacement of frame 11 with respect to the reference frame. The reference frame was frame 53 in this case. The blue lines show the interrogation boxes and the green arrows show the displacement in every box. 
(b) Average displacement in the $x$ direction. The red line shows the relative displacement averaged across all the interrogation boxes, and the blue error bars show the standard deviation. (c) Average displacement in the $y$ direction. The labels are identical to (b). }
\label{SI_drift}
\end{figure}

Before running the PIV algorithm, we processed the images using a band pass filter, a local min-max filter, and a Wiener filter in this order.  
For the bandpass filter, we used the bpass() function in the particle tracking velocimetry (PTV) code written by Blair and Dufresne \cite{PTV}. 
The window sizes of the filters were chosen to be larger than the size of individual particles but smaller than the interrogation box in PIV \cite{Deen_PIV}. 
Exemplary images in Fig.~\ref{SI_Filters} show the original and processed fluorescent images after every step for 110~nm beads at $60 \times$ magnification. 

Even though we sealed the whole sample, the hydrogel still dehydrated slowly from time to time. 
As a result, in some videos, the substrate had a mild drift, where both the cells and the tracer particles moved together slowly in the field of view. 
An example of this global drift is shown in Fig.~\ref{SI_drift}. 
Within 100 frames (about 25~min), the sample drifted approximately 20 pixels (about $1.5~\mu$m) in the $y$ direction, while remaining relatively steady in the $x$ direction. 
We removed such global drifts by performing a low-resolution PIV with 400 pixel $\times$ 400 pixel interrogation boxes, calculating the mean displacement in all the boxes in each image, and cropping the images using a moving window that follows this drift.

\subsection{Measuring displacement at the substrate surface}
\label{sec:displacement}

\begin{figure}[t]
    \begin{center}
        \includegraphics[width=0.8\textwidth]{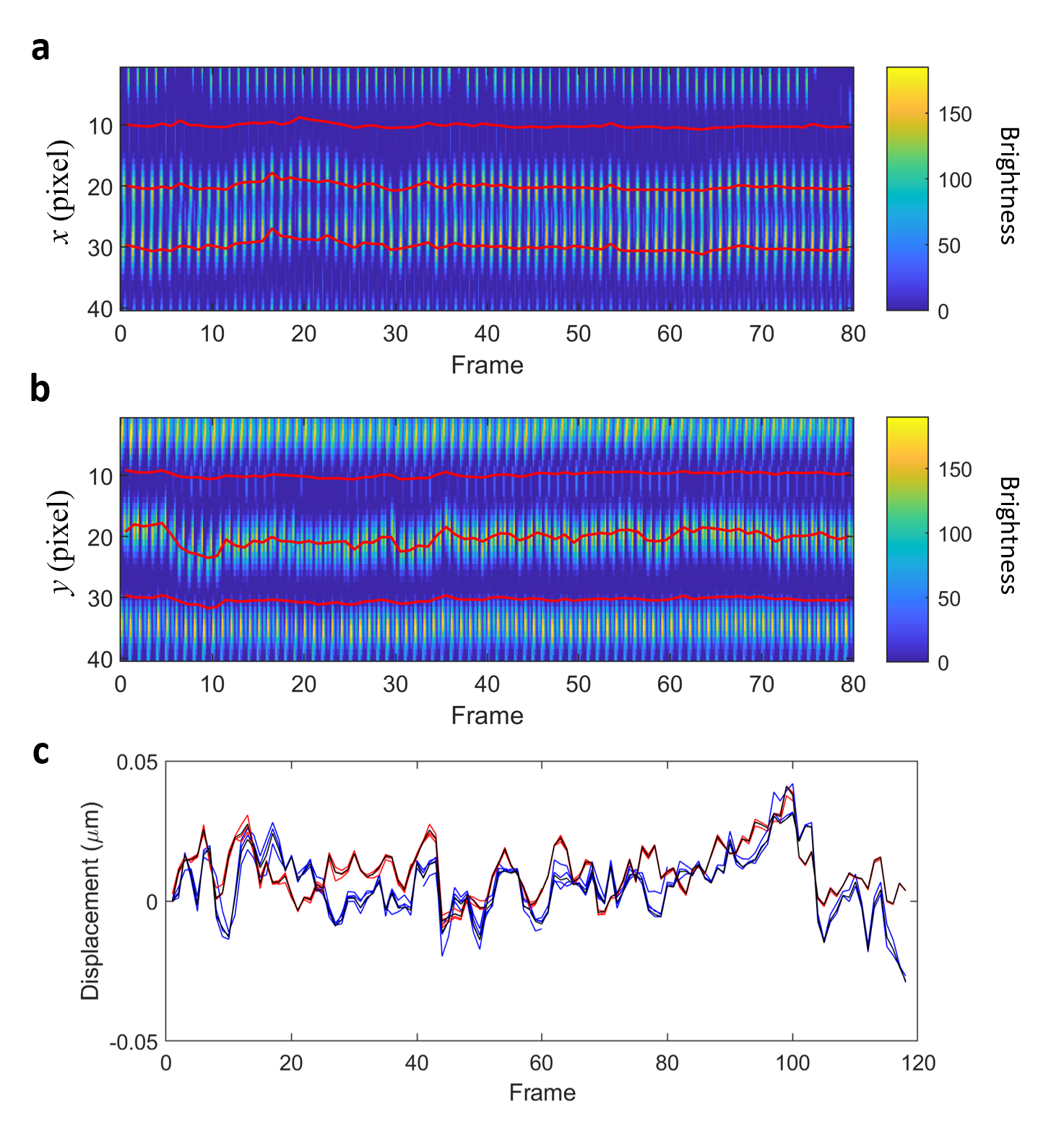}
    \end{center}
\caption{ Tracking the motion of particles. (a, b) Kymographs of speckle motion in the $x$ (a) and $y$ (b) directions. The local displacement calculated with the PIV algorithm is shown by the red curves. For each frame, we show an area of two interrogation boxes tall (40 pixels) and one box wide (20 pixels). The color map represents the brightness of the images. (c) An example of $x$ (blue) and $y$ (red) displacements calculated with different reference frames and stitched together. The black curves show the average displacements. }
\label{SI_PIV_Kymo}
\end{figure}

The displacement of the substrate parallel to its surface was calculated using a custom PIV algorithm. 
In the PIV analysis, we used an interrogation box of $20 \times 20$ pixels with an overlap (oversampling) of $50\%$ (10 pixels). 
Fig.~\ref{SI_PIV_Kymo}a and b present exemplary kymographs showing the motion of the speckles in a column of interrogation boxes in the $x$ and $y$ directions, respectively. 
The corresponding PIV results are overlaid to show a reasonably good agreement. 
To further reduce the noise in the measured displacement, we used multiple reference images while performing PIV. 
The gap between adjacent reference images was 20 frames. 
For example, for a video with 117 images, we made frames 1, 21, 41, ... 101 the reference images. 
Each reference image $i$ was used to calculate the displacement field in the frames $i-40$ to $i+40$. 
Consequently, the displacement field in each frame of the video was calculated independently multiple times with different reference images. 
Then we combined the results obtained from two adjacent reference images by matching the mean displacement within the overlapped frames. 
The mean displacement in each frame was then calculated after removing the outliers. 
Lastly, we shifted the displacement of the first frame to zero everywhere, so the displacement calculated was all with respect to the first frame of the video. 
Fig.~\ref{SI_PIV_Kymo}c shows how the $x$ and $y$ displacements obtained with different reference images are aligned and stitched together.

\subsection{Traction reconstruction}
\label{sec:trac_recon}

\begin{figure}[!t]
    \begin{center}
        \includegraphics[width=1\textwidth]{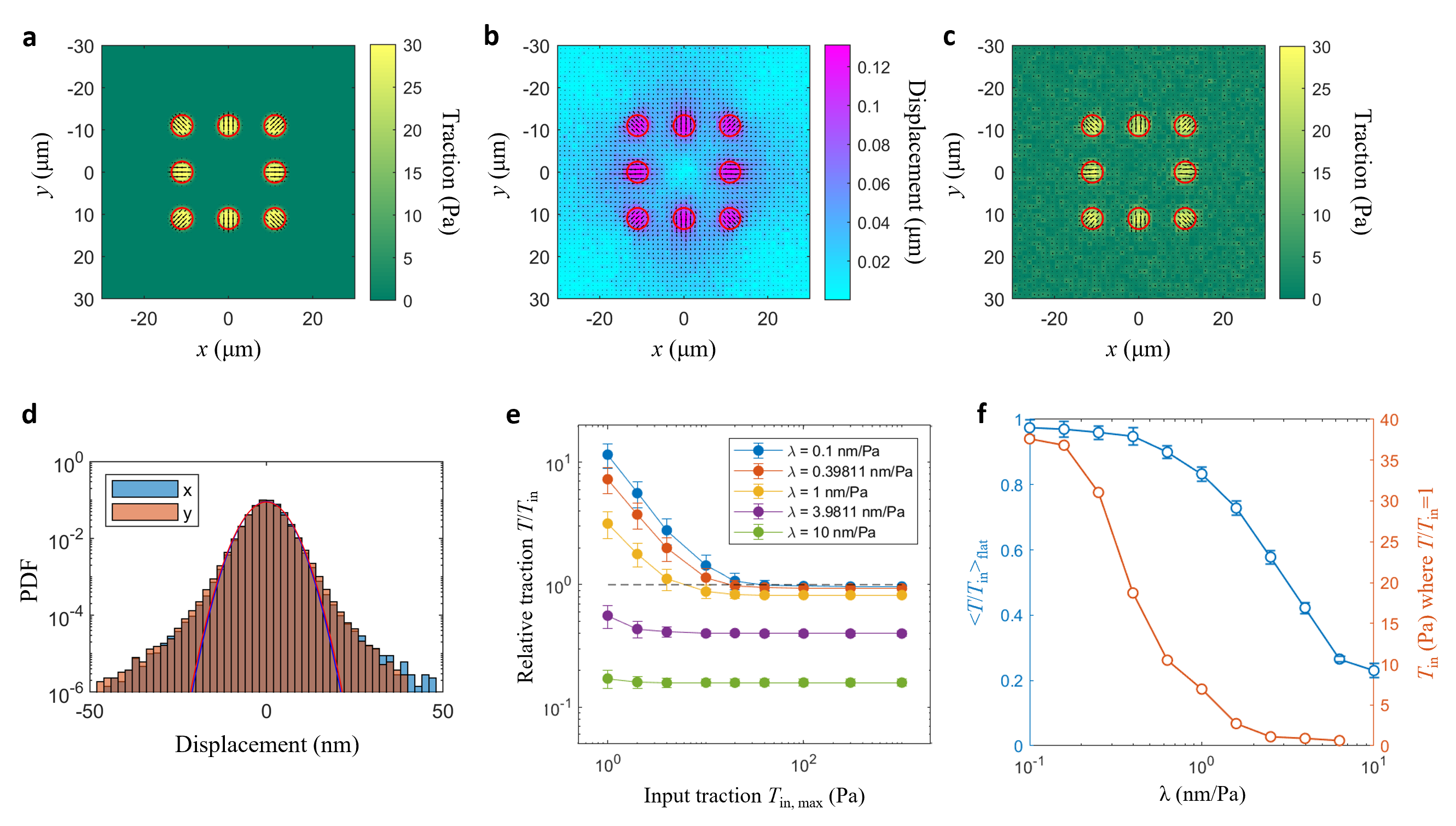}
    \end{center}
\caption{ Find regularization parameter $\lambda$ using artificially generated traction map. 
(a) Input traction map. 
(b) Coarse-grained displacement field with noise. 
(c) Reconstructed traction map with $\lambda = 1$~nm/Pa. 
(d) Distribution of $x$ (blue) and $y$ (orange) displacement when there was no cell on the surface of the gel at $60 \times$ magnification. 
(e) Ratio between the reconstructed and the input traction magnitudes $T/T_\text{in}$ in the regions where $T_\text{in} > 15$~Pa ($50\%$ peak value) as functions of $T_\text{in}$ when reconstructed with different $\lambda$. 
(f) We obtained where $T/T_\text{in} = 1$ in (e). On the right hand side of these intersections, the curves are approximately flat. Here we show the mean and standard deviation of the ratio $T/T_\text{in}$ in the flat regime (blue), and the input $T_\text{in}$ at which $T/T_\text{in} = 1$ (orange). }
\label{SI_regularization}
\end{figure}

Based on the $x$ and $y$ displacement field obtained in the previous step, we reconstructed the traction map following the Green's function based method described in \cite{Sabass_review,Sabass_Chapter,Schwarz_TFM}. 
When a point force $\bm{F} = F_x \hat{x} + F_y \hat{y} + F_z \hat{z}$ ($\hat{x}$, $\hat{y}$, and $\hat{z}$ are unit vectors) is applied on the surface of an elastic medium occupying a half-space, the resulting displacement field $\bm{u} = u_x \hat{x} + u_y \hat{y} + u_z \hat{z}$ is 
\begin{subequations}
\begin{align}
    u_x &= \frac{3}{4\pi E} \left\{ \frac{xz}{r^3} F_z +  \frac{1}{r} F_x + \frac{x}{r^3}(xF_x+yF_y) \right\}, \label{eq:SB_1} \\
    u_y &= \frac{3}{4\pi E} \left\{ \frac{yz}{r^3} F_z +  \frac{1}{r} F_y + \frac{y}{r^3}(xF_x+yF_y) \right\}, \label{eq:SB_2} \\
    u_z &= \frac{3}{4\pi E} \left\{ \left[ \frac{1}{r} + \frac{z^2}{r^3} \right] F_z + \frac{z}{r^3}(xF_x+yF_y) \right\}. \label{eq:SB_3}
\end{align}
\end{subequations}
where $E$ is the Young's modulus, and $r = \sqrt{x^2+y^2+z^2}$ \cite{Landau_book}. 
Since the hydrogel is nearly incompressible, we set the Poisson's ratio $\nu = 0.5$ in these equations. 
When we look at the surface of the gel where $z = 0$, the transverse direction ($x$ and $y$) and the normal direction ($z$) are independent. 
\begin{subequations}
\begin{align}
    u_x &= \frac{3}{4\pi E} \left\{ \frac{1}{r} F_x + \frac{x}{r^3}(xF_x+yF_y) \right\}, \label{eq:SB_z0_1} \\
    u_y &= \frac{3}{4\pi E} \left\{ \frac{1}{r} F_y + \frac{y}{r^3}(xF_x+yF_y) \right\}, \label{eq:SB_z0_2} \\
    u_z &= \frac{3}{4\pi E} \frac{F_z}{r}. \label{eq:SB_z0_3}
\end{align}
\end{subequations}
In two dimensions, the relationship between the tangential traction $\bm{T}(x,y,t)$ (local force divided by the box area, unit: Pa) and the displacement field $\textbf{u}(x,y,t)$ is 
\begin{equation}
    \bm{u} = \bm{G} * \bm{T}, 
    \label{eq:trac_to_disp}
\end{equation}
where $\bm{G}$ is the Green's function and $*$ represents convolution in space. 
In practice, the calculations are performed in Fourier space, so we actually calculated
\begin{equation}
    \tilde{\bm{T}} = \left( \tilde{\bm{G}}^T \tilde{\bm{G}} + \lambda^2 \bm{I} \right)^{-1} \tilde{\bm{G}}^T \tilde{\bm{u}}, 
    \label{eq:traction_reconstruction}
\end{equation}
where $\tilde{\bm{T}}$ and $\tilde{\bm{u}}$ are the Fourier transform of $\bm{T}$ and $\bm{u}$, respectively, $\tilde{\textbf{G}}$ is the Fourier-transformed Green's function
\begin{equation}
    \tilde{G}_{ij}(k_x, k_y) = \frac{2(1+\nu)}{ E(k_x^2+k_y^2)^{3/2} } \left( \begin{array}{cc} (k_x^2+k_y^2) - \nu k_x^2 & -\nu k_x k_y \\
    -\nu k_x k_y & (k_x^2+k_y^2)-\nu k_y^2 \end{array} \right),  
    \label{eq:Green_FFT_matrix}
\end{equation}
$\bm{I}$ is the unit matrix, the Poisson's ratio $\nu \approx 0.5$, and $\lambda$ is the regularization parameter. 
In the end, the inverse Fourier transformation of $\tilde{\bm{T}}$ gives us the traction map $\bm{T}$.

\subsection{Optimizing the regularization parameter}

The optimized regularization parameter $\lambda$ can be estimated using the ratio between the standard deviation of displacement $\sigma_u$ when there is no external force applied and the characteristic traction scale of interest $\sigma_T$ \cite{Sabass_Chapter}. 
As $\lambda$ increases, the noise will be attenuated more significantly, but on the other hand, the magnitude of the actual signal is suppressed. 
This trade-off motivates us to take a closer look at how $\lambda$ affects the final result, so we generated some artificial traction maps (Fig.~\ref{SI_regularization}a), calculated the corresponding displacement field using Eq.~\ref{eq:trac_to_disp}, added Gaussian noise with standard deviation $\sigma_u$ to the displacement (Fig.~\ref{SI_regularization}b), and reconstructed the traction using our TFM analysis algorithm (Fig.~\ref{SI_regularization}c). 
The parameters we chose in the calculation matched those in the experiments. 

We measured the noise level of displacement using bare gels without any bacterium on the surface. 
The distribution of $u_x$ and $u_y$ on such bare gel surfaces are shown in Fig~\ref{SI_regularization}d for $60 \times$ magnification, and they each had a standard deviation of $\sigma_u \approx 4.5$~nm. 
The traction $\bm{T}$ generated by $\Delta$\textit{pilA} mono-layers is discussed in detail in the main text. 
The distribution of its magnitude $|\bm{T}| \equiv T$ has a characteristic width $\sigma_T \approx 10$~Pa, so our expected regularization parameter is $\lambda = \sigma_u / \sigma_T \approx 0.5$~nm/Pa. 
In the calculations shown in Fig.~\ref{SI_regularization}a-c, the input traction field had eight circular regions where forces parallel to the surface were applied. 
Each region had a radius of $r_f = 2.5~\mu$m with constant traction of 30~Pa applied uniformly, and then we smoothed their edges with a moving Gaussian filter (the standard deviation of this Gaussian window was $r_f/4$). 
We varied the peak value of the input traction $T_\text{in, max}$ and calculated the ratio $T / T_\text{in}$ in the regions where $T_\text{in} > 0.5 T_\text{in, max}$ with different $\lambda$, as shown in Fig.~\ref{SI_regularization}e. 
We used two parameters to evaluate the effect of $\lambda$: the input traction at which $T/T_\text{in} = 1$ and the average ratio in the flat region on the right hand side $\left< T/T_\text{in} \right>_\text{flat}$. 
An ideal $\lambda$ should reduce the noise as much as possible while keeping $T/T_\text{in}$ close to 1. 
As shown in Fig.~\ref{SI_regularization}f, $\lambda \in [0.4,1]$~nm/Pa is the preferred range, and we chose $\lambda = 0.6$~nm/Pa as the regularization parameter when analyzing the experimental data.

\vspace{10mm}

\section{Deformation normal to the substrate surface}
\label{sec:normal_deformation}

\subsection{Measuring normal deformation of the substrate surface}

\begin{figure*}[!t]
\vspace{-10pt}
  \begin{center}
    \includegraphics[width=0.9\textwidth]{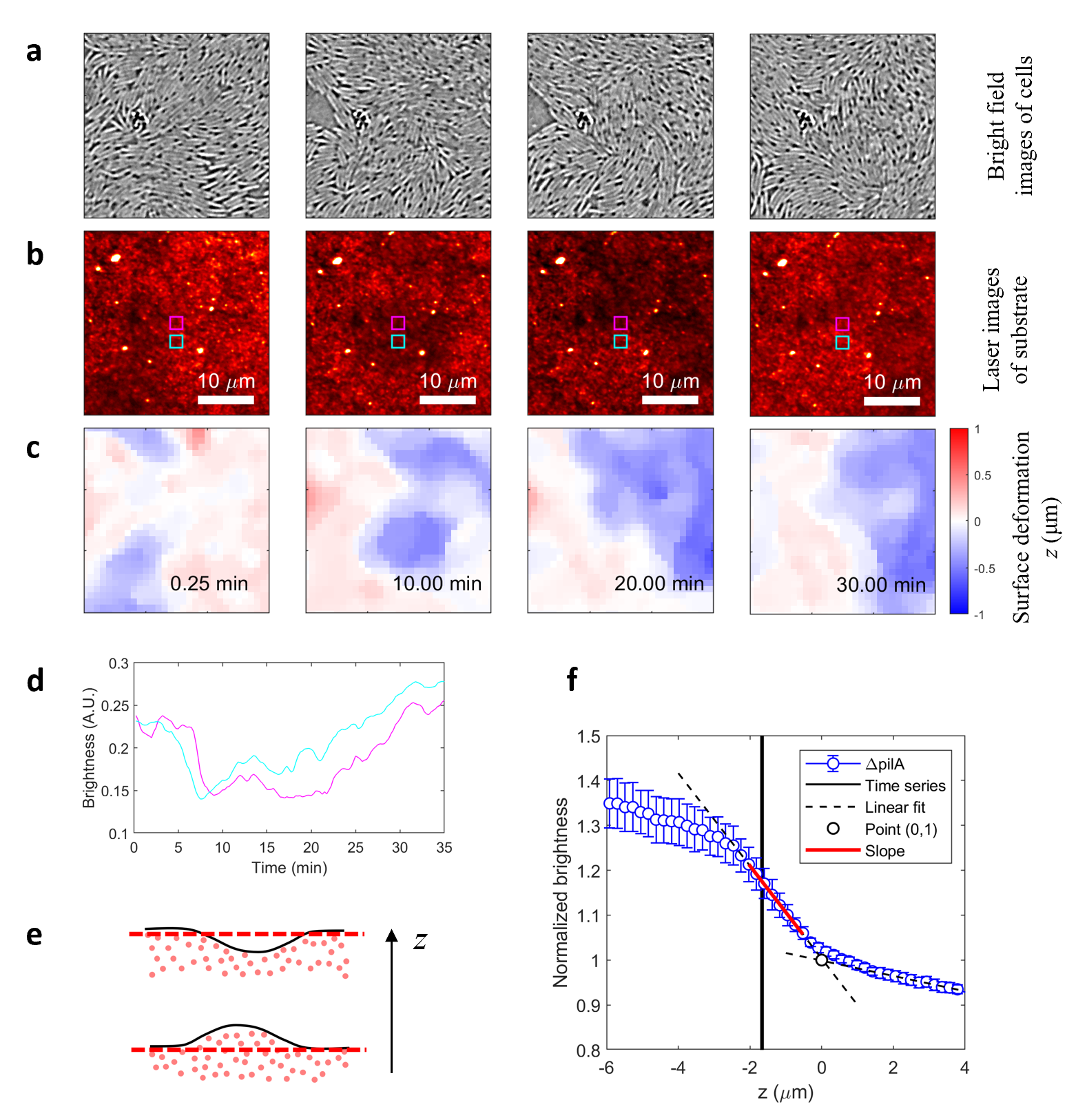}
  \end{center}
  \vspace{-20pt}
  \caption{ \label{SI_surf_deform} Measuring substrate deformation perpendicular to the surface ($z$). 
  (a-c) Bright-field images of the cells (a), fluorescence images of the fluorescent beads (40~nm) in the substrate (b), and measured surface deformation in $z$ at four different times. Images in the same column correspond to the same time as those labeled in (c). The scale bars in (b) indicate 10~$\mu$m. 
  (d) The average brightness as functions of time in the two boxes labeled in (b). 
  (e) Sketches of the relative positions of the fluorescent particles (red dots) and the focal plane (red dashed lines) when the surface of the substrate (black curves) deforms down or up. 
  (f) Normalized brightness of z-stack images of the fluorescent particles as a function of $z$. The blue data were from 12 experiments with $\Delta$\textit{pilA} cells on the substrate. The dashed black lines are the linear fits of the two approximately linear regions near and above the substrate surface. Using the slope of the thick red line, we converted the variation in brightness to the difference in $z$ in the time series, where only one slice in $z$ was imaged. The solid black line shows the average position of this slice in the time series. 
} 
\end{figure*}

Looking at the fluorescent images, we noticed that the brightness of the particles decreased when a thicker layer of cells moved across that area. 
As shown in Fig.~\ref{SI_surf_deform}a and b, a double layer of cells moved across the field of view from the lower left corner to the upper right corner, and it caused a shadow in the fluorescent images that moved with it. 
We used this variation in brightness to measure the deformation of the substrate surface in the normal ($z$) direction (Fig.~\ref{SI_surf_deform}c). 
The fluorescent brightness is controlled by the surface deformation because when imaging the time series, our focal plane was right at the surface of the substrate, as illustrated in Fig.~\ref{SI_surf_deform}d. 
As the surface deformed downward, less light emitted by the fluorescent particles was captured at the focal plane, so this region became darker. 
Similarly, when the surface deformed upward, the light emitted by the particles on both sides of the focal plane was captured, so this region became brighter. 
We divided the fluorescent images with a square lattice with a $20 \times 20$ pixel$^2$ box size and used the average brightness within each box to represent the local brightness. 
The average brightness in two of such boxes (labeled in Fig.~\ref{SI_surf_deform}b) is shown in Fig.~\ref{SI_surf_deform}e, where the colors of the curves and the boxes match. 
The figure shows that the double-layered region both reached and left the cyan box earlier than the magenta box. 

To convert this brightness variation into $z$ displacement, when imaging, we took a z-stack at the same location immediately after taking each time series. 
In the time series, we took consecutive images at the same $z$ (15~s between adjacent frames), while in the z-stack, we imaged different slices in $z$ at a rate of about 5~s per slice. 
Most z-stack images were taken from tens of microns below the substrate surface to several microns above, with a step size of $\Delta z = 0.2~\mu$m. 
We also took z-stacks that covered the whole substrate thickness with $2~\mu$m step size. 
In both the time series and z-stack, we took a bright field image and one or more fluorescent images (one for each necessary fluorescent color) at each time or $z$ step. 
From the z-stack, we calculated the average brightness as a function of $z$ (Fig.~\ref{SI_surf_deform}f) with multiple videos of $\Delta pilA$ cells, in which the cells formed a monolayer with occasional double-layers or holes. 
The brightness decreases as $z$ increases, and its slope steepens near the substrate surface. 
For each experiment, we linearly fitted the section with the steepest slope and the section above the substrate surface (the section on the right), as shown by the black dashed lines in Fig.~\ref{SI_surf_deform}f. 
Their intersection $(z_0,I_0)$ was used to align the curves in the horizontal direction ($z \to z-z_0$) and normalize the brightness ($I \to I/I_0$). 
The normalized $I(z)$ curves align well with each other, and Fig.~\ref{SI_surf_deform}f shows their mean and standard deviation. 
The imaging plane in the time series is indicated by the vertical black line. 
Above this plane (up in $z$), the $I(z)$ curve remains approximately linear across a range of about $1.5~\mu$m. 
The red line is a linear fitting of the data in this regime, and its slope $k_I$ maps the brightness variation in the time series to surface deformation: 
\begin{equation}
    \Delta z = \Delta I / k_I.  
\end{equation}
The resulting measurement of $z$ in the time series is shown in Fig.~\ref{SI_surf_deform}c, where the $z$ position below a monolayer of cells was defined as $z = 0~\mu$m. 
A double layer of cells led to negative $z$, and a hole in the cell layer led to positive $z$. 
As discussed in the main text, this is because the surface tension at the gel-cell-air interface is stronger than the stiffness of the substrate, so the interface remained relatively flat, and a ``second'' layer of cells grew below the ``first'' layer. 

The distributions of $z$ in the time series taken with the $\Delta$\textit{pilA} and $\Delta$\textit{frzE} cells on the substrate are shown in Fig.~\ref{SI_surf_deform}g. 
For $\Delta$\textit{pilA} cells, most cells are within a monolayer (around $z = 0~\mu$m). 
The double layers are at $z \approx -0.4~\mu$m. 
In some experiments, there were more cells within the double layer. 
In some others only small double layers were formed, thus the peaks at $z \approx -0.4~\mu$m were hard to see. 
For $\Delta$\textit{frzE}, more cells formed clusters thicker than a monolayer, and these clusters were less ``layered''. 
As a result, the curves below $z = 0~\mu$m are more smooth, and the widths of these distribution functions are significantly wider than those obtained with $\Delta pilA$ cells. 
We show this in Fig.~\ref{SI_surf_deform}h by measuring the width in $z$ of each individual $P(z)$ curve at half peak height where $P = P_\text{max}$.

\subsection{Zero traction using layer information}

\begin{figure}[t]
    \begin{center}
        \includegraphics[width=0.95\textwidth]{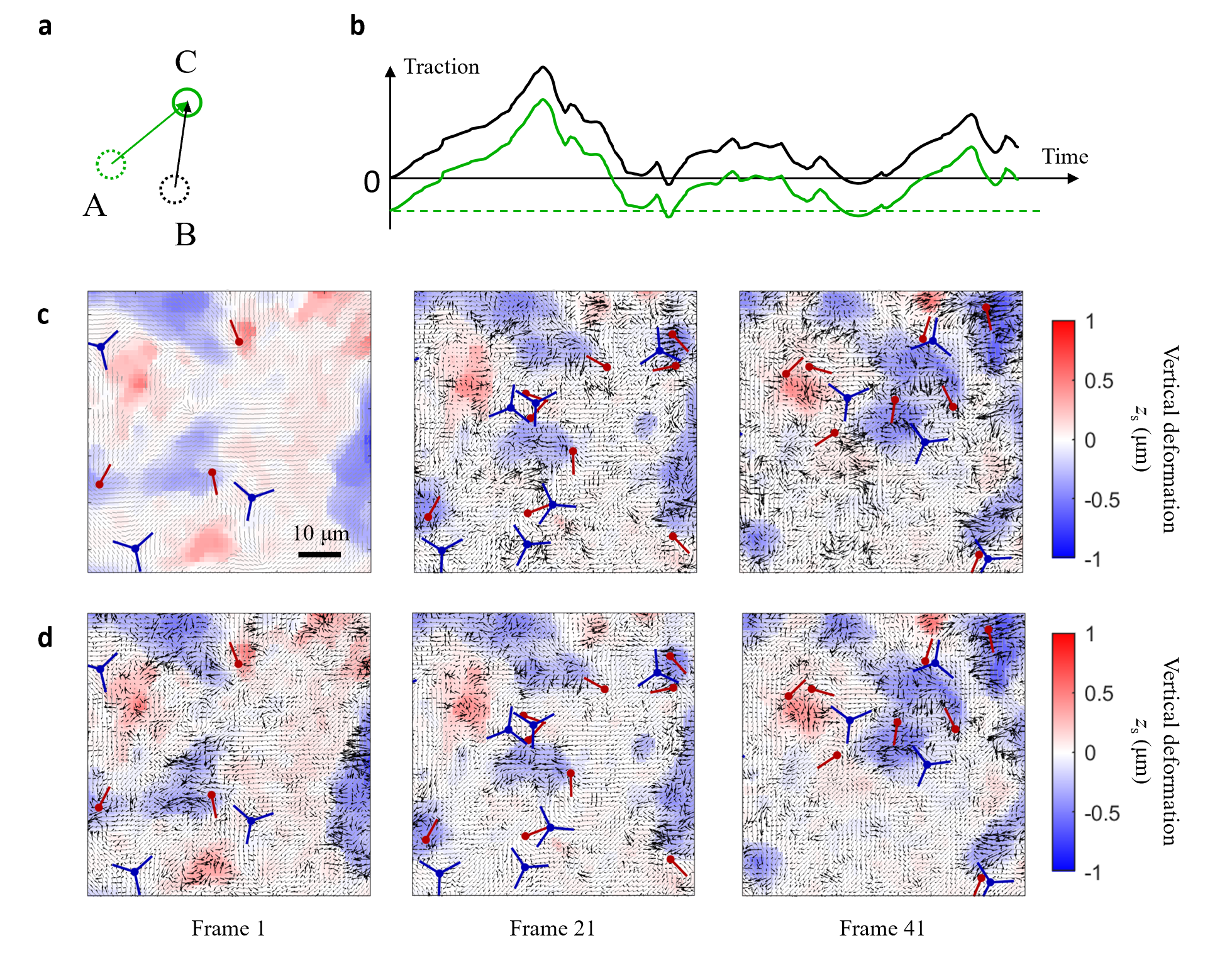}
    \end{center}
\caption{ Separating the DC mode from raw traction. 
(a) A sketch of the problem: when there is no force applied, the particle in the substrate is at point A (dashed green circle). In Frame 1, it is at point B because the substrate is not traction free. In a later frame, the particle moves to point C. The green arrow shows its displacement with respect to A, and the black arrow is with respect to B. At each location, given 
(b) A sketch of the measured traction (black line) and the actual traction (green line) as functions of time. The dashed green line labels the DC component that is removed. 
(c) Raw traction using frame 1 as the reference. The color map shows $z$ deformation of the substrate surface. The scale bar represents $10~\mu$m. 
(d) Traction with the DC component removed. 
}
\label{SI_traction_separation}
\end{figure}

To calculate the traction applied on the substrate, we need to know the location of the particles in the substrate when there is no force applied. 
However, since our system is not driven by cell growth and we could not predict the cell motion, it was difficult for us to start from imaging an area without any cells and wait for the cells to migrate over and form a monolayer. 
Instead, we always imaged regions with cells already in frame one and thus did not have the ``zero displacement'' state of the substrate. 
The consequence is that for each pixel in the reconstructed traction map, there is an unknown constant that represents the traction applied at that location in the first frame. 
As shown in Fig.~\ref{SI_traction_separation}a, the exemplary particle is at position A (dashed green circle) when no force applies on the substrate. 
However, in the first frame of the video, since there is some non-zero traction applied at this location, it has moved to position B (dashed black circle). 
When we calculate its future displacement using frame one as the reference frame, for example, when the particle moves to position C (solid green circle), we obtain the black arrow instead of the green arrow (correct displacement). 
Consequently, at each location, there is an unknown 0~Hz frequency mode (DC component) in the measured time series of traction (Fig.~\ref{SI_traction_separation}b), and the value of this constant varies pixel by pixel. 

One way to remove this unknown DC mode is by setting a threshold frequency and separating the high-frequency traction from the low-frequency component. 
The DC mode is part of the low-frequency component and thus is removed. 
However, according to Fig.~5(g) of the main text, there is no characteristic frequency that justifies the value of this frequency threshold. 
Moreover, removing the low-frequency component flattens the traction jump when a monolayer to double-layer transition happens as shown in Fig.~4 of the main text. 

In this paper, we separated the DC component from the total traction $\bm{T}$ by taking the normal deformation $z$ into consideration as well. 
When the number of cell layers remained constant, the traction fluctuated around a mean value. 
So at pixel $(i,j)$, we found the frames in which $z_{ij} \in [-0.1~0.1]~\mu$m, calculated the mean traction $\l< \bm{T}_{ij} \r>$ in these frames, and removed it from the total traction $\bm{T}_{ij}$ pixel by pixel. 
This way we removed the DC component and kept the traction jump when double layers formed. 
Fig.~\ref{SI_traction_separation}c shows the raw traction $\bm{T}_{ij}$ using frame 1 as the reference frame. 
Fig.~\ref{SI_traction_separation}d shows the traction with the DC components removed $\bm{T}_{ij}-\l< \bm{T}_{ij}(-0.1 \le z_{ij} \le 0.1) \r>$. 
Note that when a hole became a monolayer, or vice versa, there was a traction variation as well. 
Ideally, the traction in a hole is zero, and it should increase when cells migrate into this hole. 
However, when using the traction below monolayers as the reference, we got the opposite, which can be seen in Fig.~\ref{SI_traction_separation}d: in the hole near the bottom edge of frame 1 (red region), the traction was non-zero. 
When the hole was filled later by cells, the local traction nearly vanished. 
Nevertheless, since in this paper, we focus on the monolayer to double-layer transitions, this is not a problem. 



\end{document}